\documentclass[fleqn,usenatbib]{mnras}

\usepackage{newtxtext,newtxmath}
\usepackage[T1]{fontenc}

\DeclareRobustCommand{\VAN}[3]{#2}
\let\VANthebibliography\thebibliography
\def\thebibliography{\DeclareRobustCommand{\VAN}[3]{##3}\VANthebibliography}

\usepackage{graphicx}	% Including figure files
\usepackage[capitalise]{cleveref}
\crefname{equation}{equation}{equations}
\usepackage{amsmath}	% Advanced maths commands
\usepackage{makecell}

\usepackage[normalem]{ulem}
\usepackage{color}
\definecolor{mygreen}{rgb}{0.0, 0.65, 0.20}
\title[Ly$\alpha$ IM forecast for DECaLS/BASS and DESI]{Probing the cosmic web in Ly$\alpha$ emission over large scales: an Intensity Mapping forecast for DECaLS/BASS and DESI}
\author[P. Renard et al.]{
Pablo Renard$^{\rm 1}$,\thanks{E-mail: p.renard.guiral@gmail.com}
Daniele Spinoso$^{\rm 1}$,
Paulo Montero-Camacho$^{\rm 2}$,
Zechang Sun$^{\rm 1}$,
Hu Zou$^{\rm 3}$,
and Zheng Cai$^{\rm 1}$
\\
$^{\rm 1}$Department of Astronomy, Tsinghua University, Beijing 100084, China\\
$^{\rm 2}$Department of Mathematics and Theory, Peng Cheng Laboratory, Shenzhen, Guangdong 518066, China\\
$^{\rm 3}$National Astronomical Observatories, Chinese Academy of Sciences, Beijing 100012, China
}

\date{Accepted XXX. Received YYY; in original form ZZZ}

\pubyear{2015}

\begin{document}
\label{firstpage}
\pagerange{\pageref{firstpage}--\pageref{lastpage}}
\maketitle

% Abstract of the paper. No more than 250 words, no references
\begin{abstract}

Being the most prominent HI line, Ly$\alpha$ permeates the cosmic web in emission. Despite its potential as a cosmological probe, its detection on large scales remains elusive. We present a new methodology to perform Ly$\alpha$ intensity mapping with broad-band optical images, by cross-correlating them with Ly$\alpha$ forest data using a custom one-parameter estimator. We also develop an analytical large-scale Ly$\alpha$ emission model with two parameters (average luminosity $\langle L_{\rm Ly\alpha} \rangle$ and bias $b_{\rm e}$) that respects observational constraints from QSO luminosity functions. We compute a forecast for DECaLS/BASS $g$-band images cross-correlated with DESI Ly$\alpha$ forest data, setting guidelines for reducing images into Ly$\alpha$ intensity maps. Given the transversal scales of our cross-correlation (26.4 arcmin, $\sim$33 cMpc/h), our study effectively integrates Ly$\alpha$ emission over all the cosmic volume inside the DESI footprint at $2.2 < z < 3.4$ (the $g$-band Ly$\alpha$ redshift range). Over the parameter space ($\langle L_{\rm Ly\alpha} \rangle$, $b_{\rm e}$) sampled by our forecast, we find a 3$\sigma$ of large-scale structure in Ly$\alpha$ likely, with a probability of detection of 23.95\% for DESI-DECaLS/BASS, and 54.93\% for a hypothetical DESI phase II with twice as much Ly$\alpha$ QSOs. Without a detection, we derive upper bounds on $\langle L_{\rm Ly\alpha} \rangle$ competitive with optimistic literature estimates ($2.3 \pm 1 \cdot 10^{\rm 41}$ erg/s/cMpc$^3$ for DESI, and $\sim$35\% lower for its hypothetical phase II). Extrapolation to the DESI-Rubin overlap shows that a detection of large-scale structure with Ly$\alpha$ intensity mapping using next-generation imaging surveys is certain. Such detection would allow constraining $\langle L_{\rm Ly\alpha} \rangle$, and explore the constraining power of Ly$\alpha$ intensity mapping as a cosmological probe.

\end{abstract}

\begin{keywords}
large-scale structure of Universe -- intergalactic medium -- cosmology: observations
\end{keywords}

%%%%%%%%%%%%%%%%%%%%%%%%%%%%%%%%%%%%%%%%%%%%%%%%%%

%%%%%%%%%%%%%%%%% BODY OF PAPER %%%%%%%%%%%%%%%%%%

\section{Introduction}\label{sec:Introduction}

The observational study of the Universe consists mostly in the detection and analysis of radiation from cosmic origin, emitted by baryonic matter, and bent by the gravitational potential of matter itself. In the visible spectrum, the vast majority of this radiation is due to thermal and line emission of resolved objects, such as galaxies and quasars. These resolved objects cluster in filaments and nodes, forming a cosmic web that traces the underlying dark matter large-scale structure (LSS). The study of the LSS statistical properties is a cornerstone of observational cosmology, constraining vastly different aspects of cosmological models \citep{Bernardeau2002}, such as inflationary fields \citep{PlanckCOllab2020}, dark matter properties \citep{Kunz2016} or neutrino masses  \citep{Palanque-Delabrouille2015}.

Most of the baryon budget of the universe lies in the intergalactic medium (IGM), in the form of diffuse gas \citep{Shull2012}. One observational technique that traces diffuse baryons in the IGM, as well as objects too faint to be directly resolved, is intensity mapping (IM). IM consists on the analysis of a single emission line across large sky areas, without resolving any objects \citep{Peterson2009}. Since the rest-frame wavelength of the emission line is well-known, observing in a given wavelength range is equivalent to observing in a certain redshift interval. This technique was originally devised for 21cm line emission \citep{Madau1997, Loeb2004, Hu2020}, but its application has also been studied for other emission lines \citep{Bernal2022}, such as CO \citep{Breysse2014, Li2016, Keating2020} and CII \citep{Gong2012, Yue2015, Yang2019} in the infrared, H$\alpha$ in the optical \citep{Parsons2022}, or Ly$\alpha$ in the UV \citep{Silva2013, Pullen2014}. The case for Ly$\alpha$ IM is of particular interest, given that it is the strongest emission line from neutral hydrogen (HI), hence its presence should be ubiquitous in the IGM.

In fact, IGM Ly$\alpha$ absorption has been widely observed in the spectra of high-redshift quasars, where it produces the so-called Ly$\alpha$ forest \citep{Rauch1998}, i.e. a series of absorption lines bluewards of the prominent QSO Ly$\alpha$ emission. Compiling large Ly$\alpha$ forest catalogues has been a key goal of many spectroscopic surveys (e.g., BOSS \citealp{Dawson2013}, eBOSS \citealp{Ahumada2020}, DESI \citealp{Ramirez-Perez2024} or WEAVE-QSO \citealp{Pieri2016}), from which large-scale structure features have been observed, such as Baryon Acoustic Oscillations \citep[BAO, e.g.][]{Bautista2017, Dumasdesbourboux2020, DESICollaboration2024}, and smaller scale features such as the one-dimensional power spectrum \citep{Palanque-Delabrouille2013, Irsic2017, Chabanier2019a}.  The absorption features of the QSOs Ly$\alpha$ forest are due to absorption by neutral gas within relatively-low density regions. This means that the Ly$\alpha$ forest traces a density field more linear than galaxy clustering (i.e., its evolution can be described with reasonable accuracy with linearised equations, \citealp{Bernardeau2002}). However, if the HI density is too high, Ly$\alpha$ absorption saturates, resulting in Lyman limit systems \citep[LLSs,][]{Songaila2010} and Damped Ly$\alpha$ Systems \citep[DLAs,][]{Wolfe2005, Wang2022}. Since the Ly$\alpha$ forest traces the underlying dark-matter distribution in a low-density regime that can primarily be described by linearised equations, it is a particularly useful tracer to place valuable constraints on dark matter models, \citep{Viel2013, Palanque-Delabrouille2020, Irsic2024} neutrino masses \citep{Palanque-Delabrouille2015, Yeche2017}, or even reionisation physics \citep{Montero-Camacho2021}.

On the other hand, Ly$\alpha$ emission traces LSS in a very different manner. UV emission from galaxies, QSOs or the UV background \citep{Chiang2019, Gallego2021} excites HI, which in turn emits in Ly$\alpha$. Being a resonant line, the diffuse Ly$\alpha$ emission scatters into the IGM before reaching the observer \citep{Cantalupo2005, Byrohl2022}. This Ly$\alpha$ emission generates primarily in overdense regions  \citep{Byrohl2022}, which makes it complementary to the Ly$\alpha$ forest absorption tracing the underdense regime. Since it scatters before reaching us, the observed Ly$\alpha$ emission is directly associated to baryons enclosed within the cosmic web. Hence, IGM Ly$\alpha$ emission highly depends on IGM physical properties, such as temperature or ionisation state \citep{Ouchi2020}. Therefore, combining Ly$\alpha$ forest measurements with Ly$\alpha$ emission LSS may help disentangling how both astrophysical and cosmological processes shape the baryon distribution across the cosmic web.

This diffuse Ly$\alpha$ emission has already been well documented up to cMpc scales. At halo scales ($\sim$100 pkpc) diffuse Ly$\alpha$ emission around QSOs and star-forming galaxies has been extensively studied \citep[Ly$\alpha$ blobs,][]{Taniguchi2001, Matsuda2004, Wisotzki2018}, and more recently, larger Ly$\alpha$ nebulae of scales $<$ 1pMpc sourced by QSO UV emission have also been reported \citep[Enourmous Ly$\alpha$ Nebulae, ELANs,][]{Cai2017, Cai2019, Battaia2022}. On even larger scales, filamentary emission between galaxies at scales of few cMpc has been observed \citep{Bacon2021}. Generally, these studies are based on observations on small, ultra-deep fields, since great image depth is usually needed to securely measure faint, extended $\rm Ly\alpha$ emission.

Another possible approach to statistically measure LSS in Ly$\alpha$ emission is to perform IM, i.e., to average observed cosmic emission with a given correlation estimator over large cosmological volumes (i.e., cosmological surveys with much wider but shallower fields than deep, targeted observations). Such studies would allow to detect LSS Ly$\alpha$ emission on scales much larger than those probed by ultra-deep fields \citep[e.g., $\sim$ 5 cMpc field size for][]{Bacon2021}; however, all studies so far have only partially succeeded at this endeavour. For example, \citet{Croft2016} and \citet{Lin2022} detect Ly$\alpha$ emission at scales $<$30 cMpc/h at $z\sim2.5$ by cross-correlating the residuals in galaxy spectra with QSOs in BOSS/eBOSS. However, these cross-correlations only sample the overdense regions surrounding QSOs; a small fraction of the total survey volume that might not be fully representative of the cosmic average. When repeating this cross-correlation with the Ly$\alpha$ forest to sample a much larger volume \citep{Croft2018}, no significant LSS in Ly$\alpha$ emission is found. The detection of a large-scale cross-correlation between Ly$\alpha$ emission in a narrow-band photometric survey \citep[the Physics of the Accelerating Universe Survey, PAUS,][]{Padilla2019} and Ly$\alpha$ forest data is also simulated and evaluated in \citet{Renard2021}. The results of the \citet{Renard2021} forecast were that a detection of large-scale Ly$\alpha$ emission with Ly$\alpha$ forest cross-correlation was seemingly impossible for a projected survey footprint of 100 deg$^2$, but for similar observation times expanding the survey footprint was far more efficient in increasing the SNR than going deeper. It is worth noting that future IM experiments, such as SPHEREx \citep{Dore2014}, also aim to perform IM with Ly$\alpha$ among other lines \citep{Visbal2023}. %However a similar methodology for integrated UV emission over much larger areas \citep[5500 deg$^2$ of GALEX,][]{Chiang2019} detected large-scale structure, but for the entire UV background.

%From a cosmological standpoint, the most useful statistic to evaluate the correlation between two datasets is the two-point correlation function (2PCF), as it can be theoretically modelled and used to constrain cosmological parameters \citep{Bernardeau2002}. However, for the cross-correlation of the Ly$\alpha$ forest and possible Ly$\alpha$ emission present in imaging data, the 2PCF presents some important caveats. First, the bias of the IGM Ly$\alpha$ emission with respect to the dark matter distribution is poorly constrained \citep[e.g.,][]{Croft2016}. Second, the actual bias of a 2PCF between Ly$\alpha$ forest and Ly$\alpha$ emission data would have a degeneracy with the mean value of foreground emission in the cross-correlated images see \citep[e.g.,][]{Renard2021}. Given the impossibility of removing all foreground emission from images (as any unresolved cosmic source that is not the Ly$\alpha$ IGM emission would be considered foregrounds), even if the 2PCF yields a significant signal, it would be difficult to place any meaningful cosmological constraints based on its value.

Here, we present a new methodology to perform Ly$\alpha$ IM by cross-correlating broad-band images and Ly$\alpha$ forest data: being the standard for photometric surveys, most of the night sky has already been observed in broad bands, which potentially allows IM to be applied on very large cosmological volumes. We develop a specific one-parameter correlation estimator to evaluate the fluctuation of Ly$\alpha$ luminosity versus Ly$\alpha$ forest absorption, as well as a simple analytical Ly$\alpha$ emission model for cosmological scales. We perform a forecast for DESI and the $g$ band of its Legacy Imaging Surveys \citep[DECaLS/BASS,][]{Dey2019}, providing realistic expectations for a detection of Ly$\alpha$ LSS, as well as the upper bounds derived from a non-detection.

This work is divided as follows. In \cref{sec:Cosmological surveys}, we describe the relevant specifications of the cosmological surveys used in our forecast. In \cref{sec:Methodology}, we provide an in-depth explanation of the proposed Ly$\alpha$ IM methodology. In \cref{sec:Forecast simulation} the simulation for our forecast is described,  with special emphasis on image reduction and computation of intensity maps; the forecast results are displayed in \cref{sec:Forecast results}. Finally, in \cref{sec:Discussion}, we discuss the implications of our results for upcoming surveys, as well as the lines of research that our analysis may open up. We summarise our conclusions in \cref{sec:Conclusions}. Throughout this paper we assume a flat $\Lambda$CDM cosmology with $h=0.702$, $\Omega_{\rm  m}=0.275$, $\Omega_{\rm  \Lambda}=0.725$, $\Omega_{\rm  b}=0.046$, $n_{\rm  s}=0.968$ and $\sigma_{\rm 8}=0.82$, as this is the parameter set of the hydrodynamic simulation used in \cref{sec:Forecast simulation}.

\section{Cosmological surveys}\label{sec:Cosmological surveys}

Optical cosmological surveys are broadly divided in two categories: imaging/photometric. The former take images of the sky with CCDs covering the field of view of a camera \citep[e.g.,][]{Flaugher2015, Padilla2019}; their wavelength sensitivity is determined by different sets of filters. The latter generally observe spectra of pre-designated targets \citep[e.g.,][]{Eisenstein2011, DESICollaboration2022}, trading off speed and full sky coverage for much higher spectral resolution. Our work consists in detecting the diffuse Ly$\alpha$ emission tracing LSS that must be contained in photometric images, by cross-correlating said images with the Ly$\alpha$ forest contained in QSO spectra. Hence, both spectroscopic and imaging data are required.

\subsection{DESI}\label{sec:DESI}

The Dark Energy Spectroscopic Instrument \citep[DESI, ][]{DESICollaboration2022} is a spectroscopic cosmological survey, currently ongoing at the Mayall telescope in Kitt Peak, Arizona. The survey is expected to cover a total of 14,000 deg$^2$ in its Phase-I, thus becoming the largest spectroscopic survey yet.

DESI targets four different galaxy samples, in ascending redshift order: the Bright Galaxy Sample, Luminous Red Galaxies, Emission Line Galaxies, and quasars (QSO). The latter are split into two sub-samples, according to their redshift. The sample at $z < 2.1$ is only detected via strong QSO emission lines, such as Ly$\alpha$, C-IV, C-III, S-IV typically occurring in QSO spectra \citep[e.g.,][]{VandenBerk2001, Selsing2016}, and thus only QSO positions are used as tracers. On the other hand, the QSOs at z > 2.1 (i.e.,g the Ly$\alpha$ QSOs) show Ly$\alpha$ forest absorption lines in their spectra, in addition to the aforementioned QSO emission lines. This is because at z > 2.1 the spectral range affected by the Ly$\alpha$ forest (bluewards of the Ly$\alpha$ line at the QSO redshift) becomes observable by the DESI spectrograph \citep[3600 to 9800 \AA,][]{DESICollaboration2022}. Our work is focused on the Ly$\alpha$ QSO subsample at $z > 2.1$.

The DESI QSO target selection is based on a colour cut of point-like sources using the optical $grz$ magnitudes from the DESI legacy imaging surveys \citep{Dey2019}, together with the $W1$ and $W2$ infrared magnitudes of the Wide-Field Infrared Survey Explorer satellite \citep[WISE,][]{Wright2010}. This QSO colour selection is expected to yield an approximate completeness of 60$\%$ for a limiting magnitude of $r < 23$ \citep{DESICollaboration2016}. This selection results in a preliminary Ly$\alpha$ QSO density of 40 deg$^{\rm -2}$, and the QSO redshift distribution shown in \cref{fig:qso_vs_z_g_band} (only for the redshift range of interest in this work). We will work with this preliminary estimate, but it is worth noting that current DESI observations reach significantly higher Ly$\alpha$ QSO completeness, yielding a Ly$\alpha$ QSO density of 60 deg$^{\rm -2}$ \citep{Chaussidon2022}, well above the DESI science requirement of 50 deg$^{\rm -2}$. Higher Ly$\alpha$ QSO density on an eventual final DESI phase-I data release means a larger total number of Ly$\alpha$ forests, and thus higher statistical power and SNR for our Ly$\alpha$ IM methodology.

\subsection{DECaLS-BASS}\label{sec:DECaLS-BASS}

Spectroscopic surveys require preliminary target selection, usually based on already existing photometric data. For DESI, three imaging surveys where carried out for this purpose \citep{Dey2019}: the Dark Energy Camera Legacy Survey (DECaLS, $grz$ bands) the Beijing-Arizona Sky Survey \citep[BASS, $gr$ bands,][]{Zou2017} and the Mayall $z$-band Legacy Survey (MzLS, $z$ band). Since this work uses the $g$ band to perform an intensity mapping forecast, we focus only on the first two. The combination of DECaLS and BASS covers the entirety of the DESI footprint on the $gr$ bands.

DECaLS has been carried out at the Blanco Telescope, with the Dark Energy Camera \citep[DECam][]{Flaugher2015}, a large camera with an hexagonal field of view of approximately 3.2 deg$^2$. BASS was observed at the Bok Telescope with the 90Prime camera \citep{Williams2004}, which provides a square field of view of 1.08 x 1.03 deg$^2$. Both surveys reach similar magnitudes: for a point source, the detection limit at 5$\sigma$ is $g=23.95$ for DECaLS and $g=23.65$ for BASS \citep[][see Table 4]{Dey2019}.

For the purposes of our forecast, the survey features we are interested in are the survey depth and the field of view (FOV) of the camera. A deeper survey means a fainter magnitude detection limit, and lower photometric noise in the images. The field of view will determine the maximum angular area in which sky subtraction can be performed without removing clustered diffuse Ly$\alpha$ emission (see \cref{sec:Photometric noise: BASS intensity map and image reduction}), and thus the largest scales we can sample. DECaLS both has a larger field of view and slightly higher survey depth than BASS \citep{Dey2019}, therefore, we will be conservative and model our forecast only after BASS, even though DECaLS covers most of the DESI footprint (9850 deg$^2$ versus 5500 deg$^2$).

\begin{figure}
 	\includegraphics[width=\columnwidth]{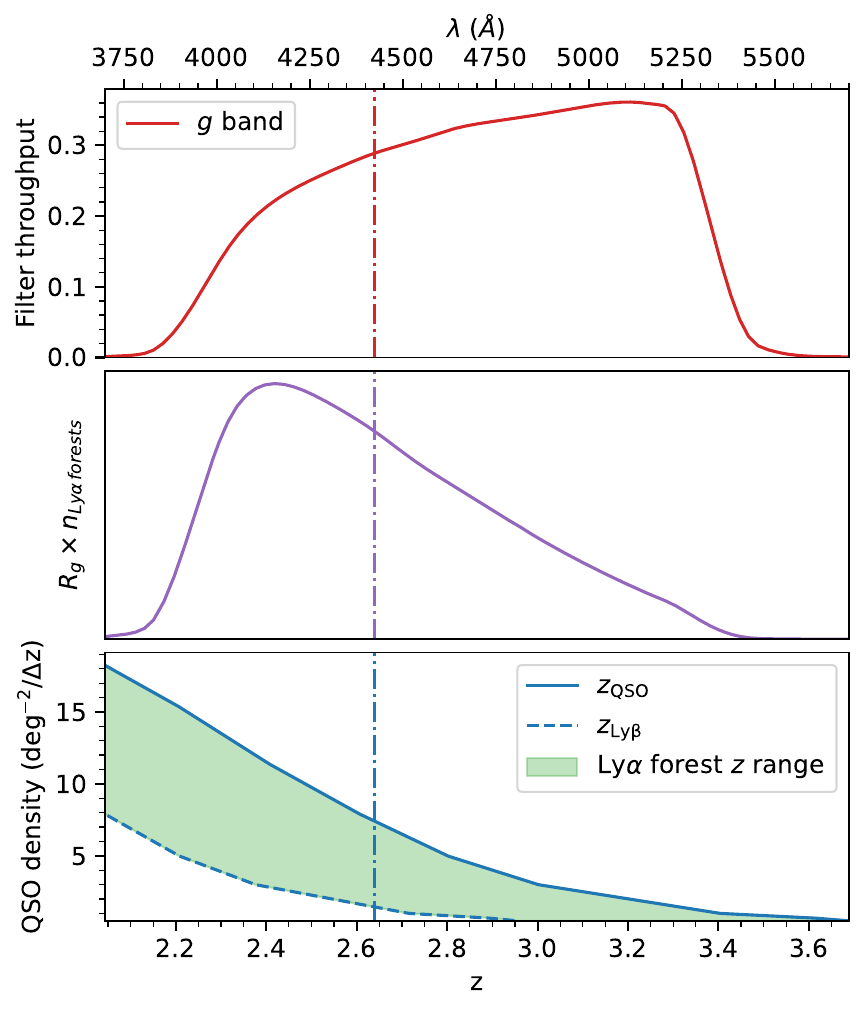}
     \caption{\textit{Upper panel:} standard SDSS $g$-band response function. The $g$-band wavelength in the upper x-axis matches the redshift for the Ly$\alpha$ line in the x-axis of the lower panel. \textit{Middle panel:} Product of the $g$-band response function times the number density of Ly$\alpha$ forest sightlines per redshift, normalised. This is the redshift kernel of our cross-correlation study; the dash-dotted line (in all three panels) represents the average redshift of our convolution kernel. \textit{Lower panel:} DESI projected QSO angular density versus redshift (solid blue line). The accompanying dashed blue line represents the lower redshift limit of the observable Ly$\alpha$ forest, marked by the start of the Ly$\beta$ forest. Therefore, the green-shaded region is the redshift range covered by the Ly$\alpha$ forest for a given QSO redshift.}
     \label{fig:qso_vs_z_g_band}
\end{figure}

For the sake of homogeneity, we use in this work the original standard $g$ band from the Sloan Digital Sky Survey \citep[SDSS,][]{York2000}; hypothetical deviations in the response function of the $g$-band filters used in DECaLS/BASS are expected to be negligible. We show in \cref{fig:qso_vs_z_g_band} the response function of the SDSS $g$ band, with the corresponding redshift range for the Ly$\alpha$ line.  Together with the DESI QSO redshift distribution, this figure gives an  idea of the redshift range sampled by our study: the DECalS/BASS images contain an intensity map of Ly$\alpha$ emission convolved in redshift with the $g$ band, and the Ly$\alpha$ forest of DESI samples Ly$\alpha$-sightlines with a spatial density given by the QSO distribution. Both are convolved in the middle panel of \cref{fig:qso_vs_z_g_band} to show the redshift kernel of our cross-correlation study, whose average redshift is $\langle z \rangle = 2.64$.

\section{Methodology}\label{sec:Methodology}

The optical depth of the Ly$\alpha$ forest is a tracer of the HI density along the line of sight; hence regions with higher optical depths corresponding to higher HI densities. This relationship between Ly$\alpha$ absorption and HI column density depends on both physical parameters of the IGM (e.g., temperature, photoionisation rate, line-of-sight velocity) and cosmological parameters (Hubble constant, baryon density). For an analytical model valid on linear scales, see the Fluctuating Gunn-Peterson Approximation \citep[e.g.,][]{Weinberg1997, Kooistra2022}. The relationship between HI density and Ly$\alpha$ emission is also convoluted, given that it is a resonant line \citep{Dijkstra2017, Ouchi2020}, and the bulk of Ly$\alpha$ emission seems to be sourced from faint Ly$\alpha$ emitters (LAEs) and their surrounding circumgalactic medium (CGM) \citep{Bacon2021, Byrohl2022}.

%Even if the actual physics relating emission to absorption are complex, averaging over cosmic volumes large enough (such as DESI) must yield an absorption/emission correlation that depends only on the dark matter density field, since both are biased tracers of the underlying large-scale LSS.

In order to evaluate the correlation between the Ly$\alpha$ absorption in the DESI QSO spectra and the Ly$\alpha$ emission contained in $g$-band images we will not use the conventional two-point correlation function (2PCF) and its Fourier space analogue, the power spectrum monopole \citep{Bernardeau2002}, but a custom estimator based on stacking with Bayesian inference the fluctuations of Ly$\alpha$ emission binned in bins of Ly$\alpha$ forest absorption.

The 2PCF/power spectrum and its higher-order counterparts (e.g., bispectrum), are the backbone of the study of LSS \citep[e.g.,][]{Bernardeau2002}, including IM \citep{Bernal2022}. However, alternative summary statistics have also been described in the literature and applied to constrain the properties of different LSS tracers. For example, the wavelet transform has seen use to extract non-gaussian information \citep{MaksoraTohfa2024} or capture environmental information \citep{Wang2022a}, among other works. Another example of alternative statistic is the k-estimator \citep{Adelberger2005}, which evaluates clustering strength and constrains linear tracer bias in small fields, using only fine redshift bins \citep[e.g.,][]{Diener2017, HerreroAlonso2021}. In the context of IM, the voxel intensity distribution, i.e., the histogram of voxel fluxes, has also been studied as a complementary statistic to the power spectrum \citep[e.g.,][]{Ihle2019, Sato-Polito2022}.

Following these examples, in this work we define a custom estimator instead of using the 2PCF/power spectrum. Our main reason is the degeneracy between the absolute value of the 2PCF and the foreground value in the Ly$\alpha$ intensity map \citep{Renard2021}.  Due to this degeneracy, we would need a significant detection of the 2PCF in at least two distance bins to place any constraints on the combined bias of diffuse Ly$\alpha$ emission as a LSS tracer, by constraining the shape of the 2PCF. Unfortunately, the SNR we expect with DESI-DECaLS/BASS is very likely to not be enough (\cref{sec:Forecast results}). Moreover, the bias of the Ly$\alpha$ emission field is not properly constrained (see \cref{sec:Lya emission model} for a short discussion), which adds an extra degeneracy to potential constraints obtained through the 2PCF.

Therefore, to fully exploit the low SNR of the Ly$\alpha$ emission intensity map, we use a single-parameter estimator with the fewest assumptions possible, namely: the cosmological principle. We assume both the Ly$\alpha$ absorption and emission fields to be isotropic, which should be sufficient as a first approximation (especially given the very large redshift convolution kernel, see \cref{fig:qso_vs_z_g_band}). We assume a correlation monotonically decreasing with comoving distance between two arbitrary points, and both Ly$\alpha$ emission and absorption fields being virtually homogeneous on large enough scales (the homogeneity scale $\chi_{\rm h}$, e.g. see \citealt{Sarkar2009, Pandey2015, Goncalves2018}). Despite the use of a custom estimator, we consider our work an IM study, since we constrain the properties of an emission line tracing LSS by integrating all observed emission, without resolving any objects (as in the general definitions given in e.g., \citealt{Peterson2009, Kovetz2017, Bernal2022}). Moreover, the transversal scales we sample are large enough to be considered LSS (\cref{sec:Optimal patch radius}), and our Ly$\alpha$ emission model is defined in the linear regime, where the effects of gravitational collapse become negligible (\cref{sec:Lya emission model}).

\begin{figure*}
 	\includegraphics[width=0.8\textwidth]{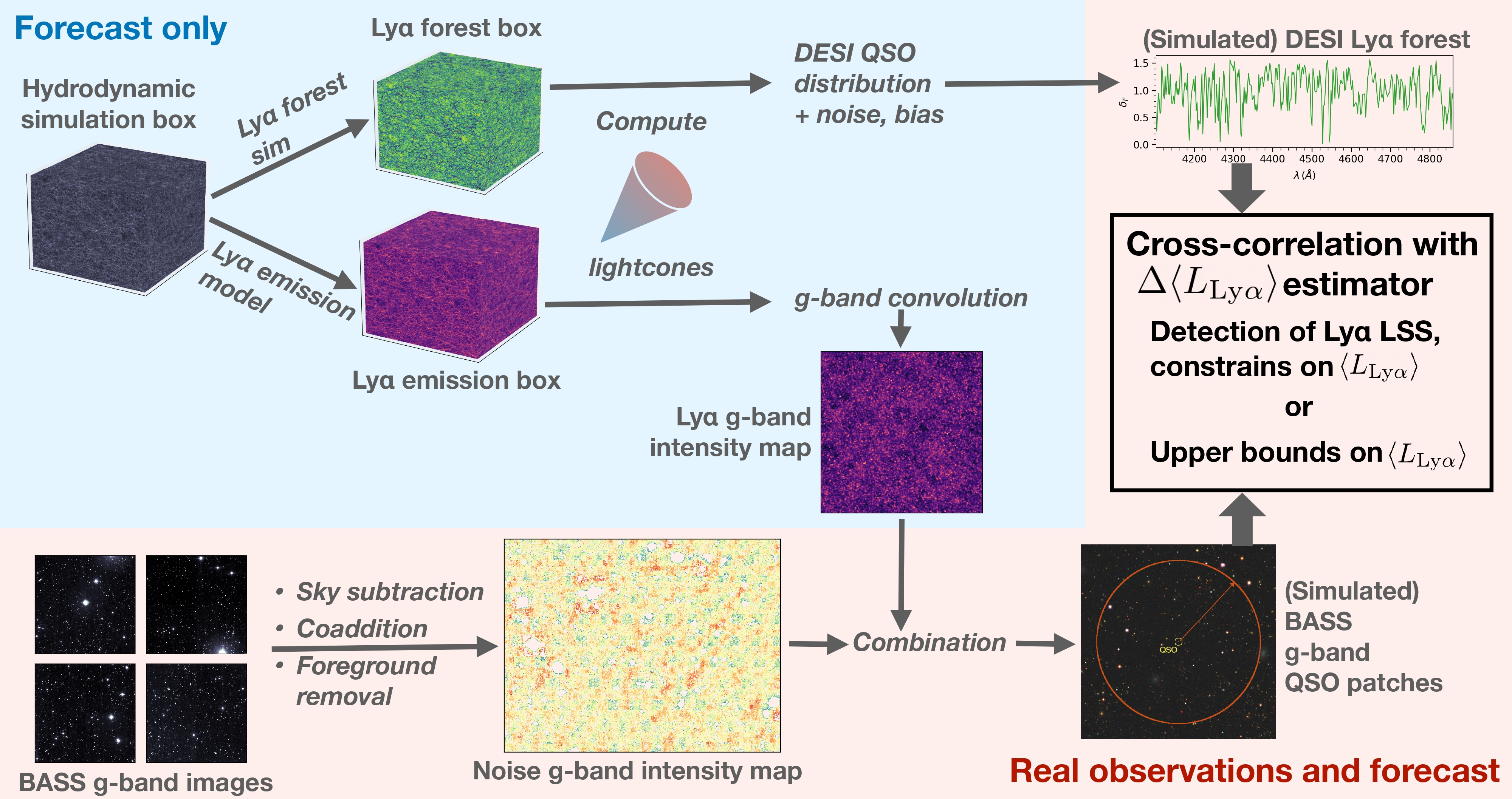}
     \caption{Diagram of the Ly$\alpha$ IM forecast presented in this paper. Blue region corresponds to forecast-only steps (i.e., simulation of the Ly$\alpha$ absorption/emission signals). Red region corresponds to steps carried out in this forecast, but also expected to be performed on an eventual observational study (e.g., image reduction, cross-correlation).}
     \label{fig:lya_im_diagram}
\end{figure*}

In \cref{fig:lya_im_diagram}, we provide a diagram which illustrates the overall structure of our forecast: how different simulated and observed data are combined to simulate DESI and DECaLS/BASS, and perform Ly$\alpha$ IM by cross-correlating them. This is intended just as a visual reference; the cross-correlation estimator is fully described in this section, and in \cref{sec:Forecast simulation} the simulation of our forecast is thoroughly laid out.

\subsection{Imaging data: QSO patches}\label{sec:Imaging data: QSO patches}

The $g$-band images are a 2D projection of the Ly$\alpha$ emission field, convolved over a large redshift range. Given our assumption that the correlation between Ly$\alpha$ emission and absorption decreases with distance and that isotropy holds, our analysis must concentrate on circular regions centred on the Ly$\alpha$ DESI quasars. This approach will maximise the correlation and consequently yield the highest SNR. Throughout the paper, by distance $r$ we will refer to transversal angular distance, projected onto the sky. We will assume that the flat-sky approximation holds true on the scales we sample (of the order of arcmin), and thus our assumptions derived from the cosmological principle also hold true.

Hence, we define a \textit{patch} of a given quasar $i$ as a circular aperture of radius $r$ centred on such quasar $i$, on the image of a given broad-band filter (the $g$ band in our case). The \textit{patch flux} in the $g$ band, $f_{\rm i}^{\rm g}$, will be the average integrated flux in such patch, minus the flux of all resolved sources inside the patch (below a given magnitude threshold):    

\begin{equation}\label{eq:patch_flux}
    f_{\rm i}^{\rm g} \equiv \frac{1}{\pi r_{\rm patch}^2} \left[ \int\limits_0^{\rm 2\pi} \int\limits_0^{\rm r_{\rm patch}} f^{\rm g}(r, \theta) r^2 dr d\theta  - \sum\limits_{\rm \substack{g<g_{\rm lim} \\ r<r_{\rm patch}}} 
    f^{\rm g}_{\rm src} \right],
\end{equation} 

where $f^{\rm g}(r, \theta)$ is the observed flux in the $g$ band image, in polar coordinates centred on the respective quasar, $r_{\rm patch}$ the patch radius, and $f^{\rm g}_{\rm src}$ the flux of resolved sources in the image, brighter than the foreground threshold $g_{\rm lim}$. This source flux is summed over all sources inside the patch ($r < r_{\rm patch}$) and with magnitude brighter than the foreground cut magnitude limit $g< g_{\rm lim}$. Even if the QSO at the centre of the patch itself has $g < g_{\rm lim}$, it is not included in this sum, since we do not consider it foregrounds (a significant part of its integrated emission will be Ly$\alpha$).

The only criterion we have adopted to choose to remove a given resolve source is $g <g_{\rm lim}$. Therefore, if any QSO with available Ly$\alpha$ forest data is inside another Ly$\alpha$ QSO patch, it will only be removed if its $g$ magnitude is brighter than $g_{\rm lim}$. However, we will choose a conservative foreground cut ($g_{\rm lim} = 19$) that preserves the vast majority of DESI Ly$\alpha$ QSOs, as discussed in \cref{sec:Photometric noise: BASS intensity map and image reduction}.

In a more general sense, we could define a patch as 

\begin{equation}\label{eq:patch_flux_kernel}
    f_{\rm i}^{\rm g} \equiv \frac{\int\limits_0^{\rm 2\pi} \int\limits_0^{\infty} W(r) f^{\rm g}(r, \theta) r^2 dr d\theta  - \sum\limits_{\rm \substack{g<g_{\rm lim} \\ r<r_{\rm patch}}} 
    f^{\rm g}_{\rm src}}{2\pi \int\limits_0^{\infty} W(r) dr},
\end{equation} 

with $W(r)$ being a radial kernel such that $\lim_{r \to \infty} W(r) = 0$, normalised by its integral. Under this definition, \cref{eq:patch_flux}
represents the particular case where $W(r)$ is a step function with $W(r) = 1$ for $r < r_{\rm patch}$ and $W(r) = 0$ elsewhere.  A different radial kernel $W(r)$ that weights more the more correlated (closer) regions might result in a higher SNR of our estimator (i.e., a Gaussian kernel, or any monotonically decreasing function). However, since we will perform our forecast for the $r_{\rm patch}$ that yields the maximum SNR for our estimator (\cref{sec:Optimal patch radius}), we argue that the gains in SNR we may derive from using a different $W(r)$ must be fairly limited.

Moreover, we also choose a step kernel and follow \cref{eq:patch_flux} to ensure that we reach radial scales large enough to sample the whole cosmological volume defined by the DESI footprint and the $g$-band Ly$\alpha$ redshift range. A $W(r)$ preferentially weighting closer regions could lead to an overestimation the cross-correlation radial length, since larger radial scales would be unfavourably weighted, regardless of the correlation length we define (e.g., the FWHM of $W(r)$ or its second moment). However, with a step function, $r_{\rm patch}$ is unequivocally the maximum radial scale that contributes to our cross-correlation.

Being able to sample the entire survey cosmological volume is an important feature of our methodology. Previous Ly$\alpha$ IM works cross-correlating QSO positions with residuals in galaxy spectra \citep{Croft2016, Lin2022} only sample $\sim$20-30 cMpc around QSOs in both transversal and line-of-sight directions, which is a small fraction of the total survey volume: overdense regions affected by the QSO proximity effect that may not be representative of the whole survey volume.

Throughout the paper we will adopt the convention of applying a superscript $^{\rm g}$ to cosmic emission/absorption fields that have been observed and convolved in redshift/wavelength with the $g$ band. Moreover, we will assume flux units of erg/s/cm$^2$/\AA\, through this work, and erg/s/cm$^2$/\AA/arcmin$^2$ units for integrated patch fluxes.

Most of the flux $f^{\rm g}$ observed inside a given QSO patch will not come from Ly$\alpha$ itself ($f^{\rm g}_{\rm Ly\alpha}$), but contamination of various origins ($f^{\rm g}_{\rm noise}$): instrumental (CCD thermal noise), atmospheric (telluric lines, airglow), solar/lunar (moonlight, zodiacal light), and cosmic (unresolved/non-removed emission from other rest-frame wavelengths than Ly$\alpha$). For this reason, it is crucial to understand how the emitted Ly$\alpha$ luminosity of cosmic origin, $L_{\rm Ly\alpha}$, is transformed into the signal we observe, $f^{\rm g}_{\rm Ly\alpha}$. If we consider an infinitesimal element of volume in redshift and celestial equatorial coordinates ($dz d\alpha d\delta$), the Ly$\alpha$ luminosity emitted in such element of volume will be $L_{\rm  Ly\alpha}(z, \alpha, \theta)dz d\alpha d\delta$. Therefore, the contribution of each volume-element to the total observed Ly$\alpha$ flux density would be:

\begin{equation}\label{eq:lya_observed_flux}
    f_{\rm Ly\alpha}(z,\alpha,\delta) dz d\alpha d\delta = 
    \frac{L_{\rm Ly\alpha}(z, \alpha, \delta)\, dz d\alpha d\delta}
    {4\pi D_L(z)^2 \Delta\lambda^{\rm  obs}},
\end{equation}
where $D_L(z)$ is the luminosity distance to said volume element, and $\Delta \lambda^{\rm  obs}$ the wavelength width of the $Ly\alpha$ line emitted by this element. Neglecting line broadening, this wavelength width is:

\begin{equation}\label{eq:lya_wavelength_width}
    \Delta \lambda^{\rm  obs} = 
    \lambda_{\rm  Ly\alpha} dz.
\end{equation}

Given that any line broadening due to thermal motion, turbulence or line resonance \citep{Ouchi2020} is at least an order of magnitude smaller than the FWHM of any broad-band filter ($\sim$1000 \AA), this assumption fully holds for $g$-band images. Finally, to determine the total Ly$\alpha$ flux of cosmic origin observed with the $g$ band, we just need to convolve \cref{eq:lya_observed_flux} with the $g$-band transmission curve/response function:

\begin{equation}\label{eq:lya_emission_g_band}
    f^{\rm g}_{\rm Ly\alpha}(\alpha, \delta)d\alpha d\delta = 
    \frac{\int_0^\infty f_{\rm  Ly\alpha}(z,\alpha, \delta)d\alpha d\delta R_{\rm g}((1+z)\cdot\lambda_{\rm  Ly\alpha})dz}
    {\int_0^\infty R_{\rm g}((1+z)\cdot\lambda_{\rm  Ly\alpha})dz},
\end{equation}

where $R_{\rm g}$ is the response function of the $g$ band displayed in the upper panel of \cref{fig:qso_vs_z_g_band}. The integration limits have been simply been set to $z=0$ and $z=\infty$, as the redshift interval is naturally bounded by $R_{\rm g}$.

\begin{figure*}
 	\includegraphics[width=\textwidth]{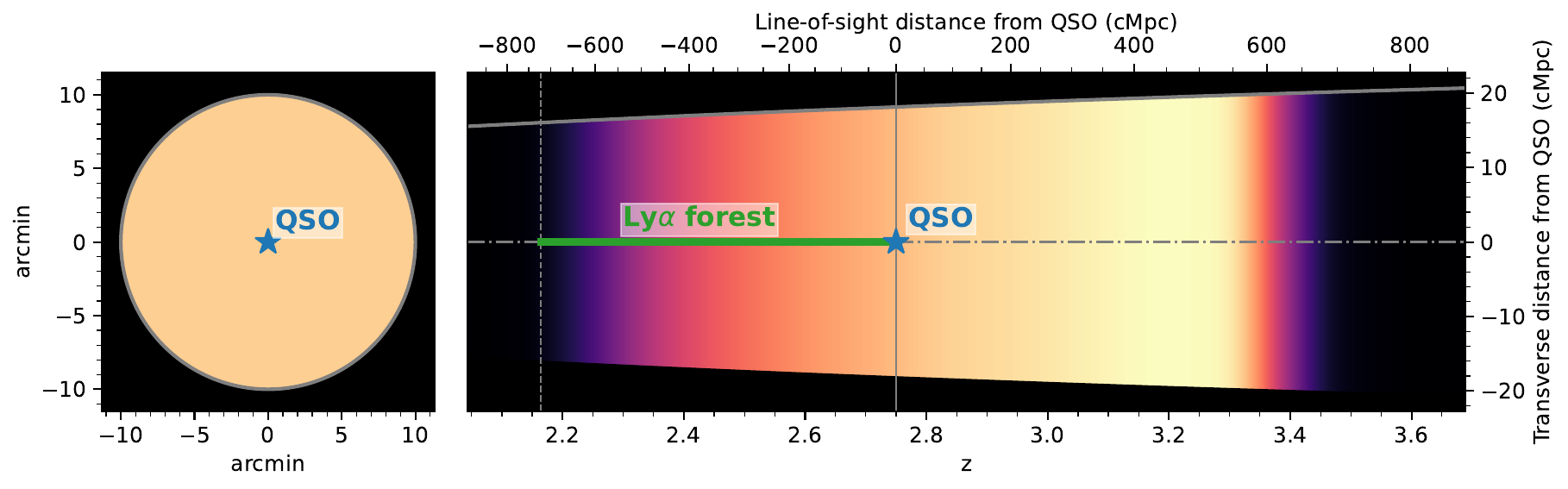}
     \caption{\textit{Left panel:} Cartoon of a $g$-band QSO patch with a radius of 10 arcmin, centered around a $z=2.75$ QSO (not to scale). \textit{Right panel:} Deprojection of the patch along the redshift direction ($g$-band deconvolution). The lower x-axis shows the redshift of the Ly$\alpha$ signal in linear scale, and the upper x-axis the corresponding comoving line-of-sight distance from the QSO. The y-axis shows the comoving transverse distance from the line of sight, and the solid grey lines the comoving limits of the patch, which evolve with redshift (not to scale with the line-of-sight distance). The coloured region inside the patch cylinder is the sensitivity of the $g$ band response to Ly$\alpha$ emission, with brighter regions inside the volume being more sensitive, and black regions not observing any Ly$\alpha$ emission. Finally, both the QSO line of sight (dash-dotted line)  and the redshift range covered by its Ly$\alpha$ forest (green solid line) are represented along the central axis of the patch.}
     \label{fig:patch_diagram}
\end{figure*}

This $f^{\rm g}_{\rm Ly\alpha}(\alpha, \delta)$ flux is all the information we receive from the IGM Ly$\alpha$ emission in $g$-band images; proper integration over angular coordinates will yield the patch fluxes $f_i^{\rm g}$ expressed in \cref{eq:patch_flux}. Each one of these patches can be seen as the redshift projection of a narrow cylinder of Ly$\alpha$ emission centred on the QSO (see \cref{fig:patch_diagram}), plus all the noise from different sources. Given that we expect noise to be dominant over the Ly$\alpha$ emission signal, the Ly$\alpha$ emission contained in patches should be impossible to observe on an individual patch basis, so a very large sample and/or a signal to cross-correlate with are needed for a detection.

\subsection{Spectroscopic data: Convolved forest probabilities}\label{sec:Spectroscopic data: Convolved forest probabilities}

The ideal dataset to cross-correlate the QSO patches with is the Ly$\alpha$ forest, as it contains high-SNR information of the Ly$\alpha$ absorption field in the line of sight of the QSO (centre of the patch). Moreover, its redshift range totally or partially overlaps with the Ly$\alpha$ emission observed by the $g$ band (\cref{fig:patch_diagram}). Nevertheless, to properly correlate both quantities, the absorption field of the Ly$\alpha$ forest must undergo the same transformation imprinted by the $g$ band on the Ly$\alpha$ emission field (equations \ref{eq:lya_observed_flux} to \ref{eq:lya_emission_g_band}). First, we define the flux absorption contrast (i.e. $\delta_{\rm F}$) for the Ly$\alpha$ forest in a given QSO spectrum $i$ as:

\begin{equation}\label{eq:delta_flux}
    \delta(z)_{\rm F\, i} \equiv \frac{e^{\rm -\tau_{\rm QSO\, i}(z)}}{\langle e^{\rm -\tau} \rangle},
\end{equation}

where $\tau(z)$ is the optical depth of the Ly$\alpha$ forest at redshift $z$. We note that $\delta_{\rm F}$ is a cosmological scalar field in spherical coordinates, similar to $L_{\rm Ly\alpha}$ in \cref{eq:lya_observed_flux}. Therefore, the redshift $z$ in \cref{eq:delta_flux} acts as radial coordinate along the sightline; angular coordinates are omitted since these are fixed by the respective QSO whose spectra contains the Ly$\alpha$ forest. 

We can apply to $\delta(z)_{\rm F\, i}$ from \cref{eq:delta_flux} the same transformation we have applied to convert Ly$\alpha$ luminosity to observed $g$-band Ly$\alpha$ flux (equations \ref{eq:lya_observed_flux} to \ref{eq:lya_emission_g_band}) as follows:

\begin{equation}\label{eq:delta_g}
    \delta_{\rm F\, i}^{\rm g} = \frac{1}{\langle\delta_{\rm F}^{\rm g}\rangle} \frac{\int_{\rm z_{\rm min\, i}}^{\rm z_{\rm max\, i}} \frac{\delta(z)_{\rm F\, i}}{4\pi D_L(z)^2 \Delta\lambda^{\rm  obs}} R_{\rm g}((1+z)\cdot\lambda_{\rm Ly\alpha})dz} {\int_{\rm z_{\rm min\, i}}^{\rm z_{\rm max\, i}} R_{\rm g}((1+z)\cdot\lambda_{\rm Ly\alpha})dz}
    ,
\end{equation}
where we have divided the convolved $\delta_{\rm F\, i}^{\rm g}$ by its average for all QSOs, $\langle \delta_{\rm F}^{\rm g} \rangle$, to keep it dimensionless\footnote{Without the $\langle \delta_{\rm F}^{\rm g} \rangle^{\rm -1}$ factor, $\delta_{\rm F}^{\rm g}$ would have dimensions of cm$^{\rm -2}$\AA$^{\rm -1}$, and an extremely small order of magnitude ($\sim10^{\rm -61}$).}. We also note that we adopted a definition of the $\delta_{\rm F}$ flux contrast in \cref{eq:delta_flux} without the -1 term \citep[which is the standard convention, e.g. ][]{Cisewski2014, Ozbek2016, Agathe2019}. This choice has been made to prevent $\langle \delta_{\rm F}^{\rm g} \rangle \sim 0$ and keep the $\langle \delta_{\rm F}^{\rm g} \rangle^{\rm -1}$ factor finite and numerically stable.

Notably, the integration ranges in \cref{eq:delta_g} have been left specified as $z_{\rm min\, i}$ and $z_{\rm max\, i}$, instead of going from 0 to $\infty$, since the redshift range in which the Ly$\alpha$ forest can be observed is limited. The observational redshift limits are given by

\begin{gather}\label{eq:zmax}
    z_{\rm max\, i} = z_{\rm QSO\, i} \\
    z_{\rm min\, i} = \max(2.1, z_{\rm Ly\beta\, i}); \quad z_{\rm  Ly\beta} \equiv \frac{\lambda_{\rm  Ly\beta}}{\lambda_{\rm  Ly\alpha}}
    (1+z_{\rm QSO\, i}) - 1,\label{eq:zmin}
\end{gather}

where the numerical value 2.1 is simply the lowest redshift at which Ly$\alpha$ can be observed with the DESI spectrograph, while $z_{\rm Ly\beta}$ is the redshift at which the Ly$\beta$ forest starts. Since $\lambda_{\rm  Ly\beta}=1025.722$ \AA\, and $\lambda_{\rm  Ly\beta}<\lambda_{\rm  Ly\alpha}$, $z_{\rm  Ly\beta}$ is necessarily a lower redshift limit for the observation of the Ly$\alpha$ forest. At $z<z_{\rm Ly\beta}$, the observed forest will be the superimposition of Ly$\alpha$ and Ly$\beta$ lines, which are produced by largely uncorrelated structure, given the large redshift offset. While some works use the Ly$\beta$-contaminated zone of the Ly$\alpha$ forest \citep[e.g., to measure BAO in][]{Blomqvist2019}, or even perform power spectrum measurements on the Ly$\beta$ forest itself \citep[e.g.,][]{Wilson2022}, we will be conservative and exclude the Ly$\beta$ regions from our forecast.

Knowing that the redshift range covered by the Ly$\alpha$ forest data is only a fraction of the $g$-band redshift coverage, and that it will also vary for each QSO, a question arises: what is the optimal way to make use of the limited redshift range? To address this question, we first introduce a nomenclature distinction. We define $\delta_{\rm F}^{\rm g\, obs}$ as the $\delta_{\rm F}$ of the Ly$\alpha$ forest convolved over the observable redshift range (\cref{eq:delta_g} with the integration limits defined in equations \ref{eq:zmax} to \ref{eq:zmin}), and $\delta_{\rm F}^{\rm g\, full}$ as the convolution with the $g$ band over the whole redshift range covered by the $g$ band (i.e., \cref{eq:delta_g} with $z_{\rm min} = 0$ and $z_{\rm max} = \infty$, as in \cref{eq:lya_emission_g_band}). Following this definition, $\delta_{\rm F}^{\rm g\, full}$ can not be directly observed. We could just cross-correlate $\delta_{\rm F}^{\rm g\, obs}$ with $f^{\rm g}$ (the patch fluxes), but $f^{\rm g}$ samples a significantly larger redshift range, and thus it will only be partially correlated with $\delta_{\rm F}^{\rm g\, obs}$. Moreover, due to the $D_L(z)^{\rm -2}$ term in \cref{eq:delta_g}, the value of $\delta_{\rm F}^{\rm g\, obs}$ has a strong correlation with $z_{\rm QSO}$ (the higher $z_{\rm QSO}$, the smaller the $D_L(z)^{\rm -2}$ term, and thus the average $\delta_{\rm F}^{\rm g\, obs}$). This is a trend uncorrelated to $f^{\rm g}$ (since it should be independent of $z_{QSO}$) that would be an undesired systematic in our analysis.

We may instead adopt a Bayesian approach and determine the probability of having a certain $\delta_{\rm F}^{\rm g\, full}$ on a QSO sightline, having observed a given $\delta_{\rm F}^{\rm g\, obs}$. This can be done in a model-independent way by resampling observational data, if we assume that all $\delta_{\rm F\, i}(z)$ values (i.e., all redshift bins/spectral pixels) for a given forest in a QSO spectrum $i$ are uncorrelated. This approximation holds for bins of $\delta_{\rm F}(z)$ above the scale of homogeneity, $\chi_{\rm h}$; to do so, we must smooth the Ly$\alpha$ forest extracted from the spectra in bins much larger than the pixels of the DESI spectrograph \citep{DESICollaboration2022} (or the resolution of the simulation in a forecast like this work, see \cref{sec:Forecast simulation}).

For this work, we will assume $\chi_{\rm h} = 150$ cMpc ($\sim$105 cMpc/h), a scale similar to values reported in the literature \citep[e.g.,][]{Pandey2015, Goncalves2018}. It is worth noting that the homogeneity scale itself depends on the criterion used to measure it, as the correlation tends to zero asymptotically. For example, \citet{Pandey2015} find a transition to homogeneity between 120 and 140 cMpc/h applying Shannon entropy to SDSS galaxy samples at $z<0.2$, while \citet{Goncalves2018} find $\chi_{\rm h}< 100$ cMpc/h for most redshift bins of SDSS quasars at $z>0.8$ using the fractal dimension (with a threshold for homogeneity $D_2>2.97$). They also note that, when correcting for quasar bias, $\chi_{\rm h}$ shows a clear decreasing trend with redshift; hence our assumption of $\chi_{\rm h} = 150$ cMpc at $z > 2.2$ is quite conservative.

\begin{figure}
 	\includegraphics[width=\columnwidth]{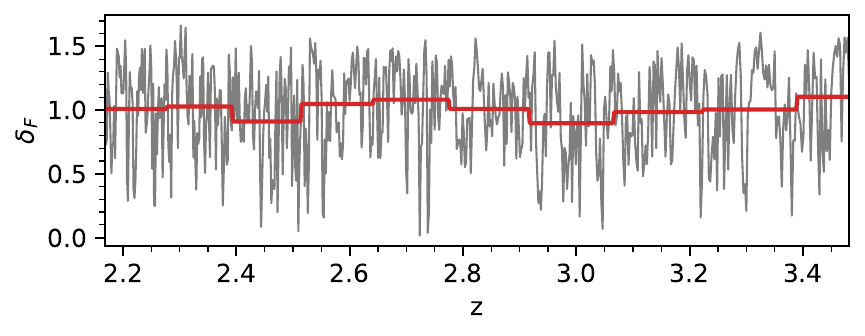}
     \caption{$\delta_{\rm F}(z)$ from a full Ly$\alpha$ forest from our simulation (see \cref{sec:Forecast simulation}) in grey, together with the same forest, smoothed in comoving bins of width 150 cMpc, in red. The bins are not uniform in wavelength/redshift space.}
     \label{fig:smoothed_forest_example}
\end{figure}

Consequently, if we smooth the forest data into wide redshift bins of comoving size $\Delta \chi = \chi_{\rm h}$, (as in \cref{fig:smoothed_forest_example}), the correlation of one bin with their adjacent bins will be close to negligible. By performing this smoothing over all our Ly$\alpha$ forest sample, we can generate new smoothed forests by randomly drawing the $\delta_{\rm F}(z)$ values of each redshift bin from all observed values of said bin. While any information on scales smaller than the smoothing scale $\Delta \chi$ is lost, the scale of the $g$ band convolution is one order of magnitude larger than $\chi_{\rm h}$ ($\sim$1 cGpc, see upper x-axis on \cref{fig:patch_diagram}), so the effect of this forest smoothing is close to negligible for our study (see \cref{sec:Validation of forest smoothing assumptions} for further proof of the validity of this assumption).

With this method, we can generate an indefinite amount of smoothed forests covering the whole $g$-band redshift range just by resampling from observed forest spectra, provided that they have been previously smoothed in comoving bins of size $\Delta \chi_{\rm h} = \chi_{\rm h}$. This procedure allows us to integrate these generated forests using \cref{eq:delta_g} (with $z_{\rm min}=0$ and $z_{\rm max}=\infty$) to draw likely values of $\delta_{\rm F}^{\rm g\, full}$ based on real data, despite $\delta_{\rm F}^{\rm g\, full}$ not being directly observable. For a large enough sample of generated smoothed forests, the empirical distribution function (EDF) of $\delta_{\rm F}^{\rm g\, full}$ will be akin to $P(\delta_{\rm F}^{\rm g\, full})$.

\begin{figure}
 	\includegraphics[width=\columnwidth]{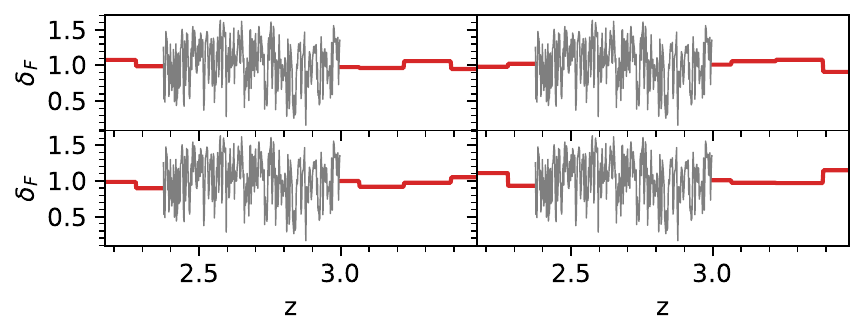}
     \caption{Four examples of generated forests (in red) with the insertion of an observed forest with $z_{\rm QSO}=3$ (in grey).}
     \label{fig:injected_forest_example}
\end{figure}

Observed forest data can then be inserted into the generated smoothed forest sample by replacing the randomly drawn value of $\delta_{\rm F}$ by the real observed $\delta_{\rm F}$, in the redshift range of the observed forest (dependant on $z_{\rm QSO}$, see equations \ref{eq:zmax} to \ref{eq:zmin}). \Cref{fig:injected_forest_example} shows the distribution of $\delta_{\rm F}^{g\,obs}$ for 4 generated smoothed forests with the insertion of the same observed forest at $z_{\rm QSO} = 3$. We note that the difference in spectral resolution/comoving bin size of the generated and inserted forest in \cref{fig:injected_forest_example} does not affect significantly the resulting $\delta^{\rm g\, full}_{\rm F}$ value (see \cref{sec:Validation of forest smoothing assumptions} for details).

\begin{figure}
 	\includegraphics[width=\columnwidth]{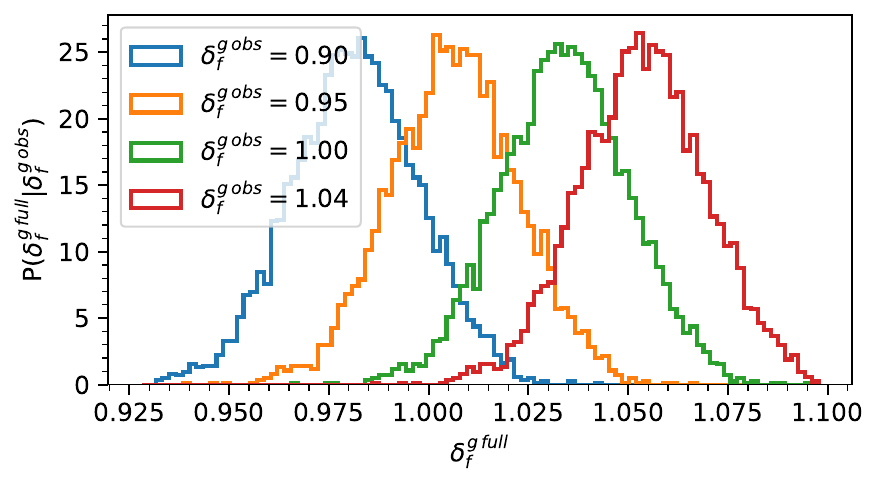}
     \caption{The EDF $P(\delta_{\rm F}^{\rm g\, full} | \delta_{\rm F}^{\rm g\, obs})$ for four different QSOs of different $\delta_{\rm F}^{\rm g\, obs}$, all of them at approximately $z_{\rm QSO}\sim2.7$. All data extracted from our full simulation of DESI Ly$\alpha$ forest data \cref{sec:Spectra simulation: Lya forest probabilities}.}
     \label{fig:edf_forest_example}
\end{figure}

If we compute the EDF of $\delta_{\rm F}^{\rm g\, full}$ for the generated sample of smoothed forests, all of them with an insertion of a given observed forest, we obtain the distribution of $\delta_{\rm F}^{\rm g\, full}$ having observed a given $\delta_{\rm F}^{\rm g\, obs}$. In a Bayesian sense, when we insert the observed forest data we update a prior that contains no information of a specific QSO ($P(\delta_{\rm F}^{\rm g\, full})$) to a posterior with the observed information of a specific QSO ($\delta_{\rm F}^{\rm g\, obs}$), which yields $P(\delta_{\rm F}^{\rm g\, full} | \delta_{\rm F}^{\rm g\, obs})$, i.e., the probability distribution of $\delta_{\rm F}^{\rm g\, full}$ updated with the information observed for a given QSO ($\delta_{\rm F}^{\rm g\, obs}$). \cref{fig:edf_forest_example} displays an example of four different $P(\delta_{\rm F}^{\rm g\, full} | \delta_{\rm F}^{\rm g\, obs})$ for four QSOs at $z_{\rm QSO}\sim2.7$ with different $\delta_{\rm F}^{\rm g\, obs}$ values. We can see that the EDFs shift towards higher values of $\delta_{\rm F}^{\rm g\, full}$ when $\delta_{\rm F}^{\rm g\, obs} $ increases.

\subsection{Correlation estimator: Ly$\alpha$ average fluctuation}\label{sec:Correlation estimator: Lya average fluctuation}

For each Ly$\alpha$ QSO in DESI we have defined two observable quantities. First, the $g$-band patch flux surrounding the QSO, $f^{\rm g}$, which contains the information of the LSS Ly$\alpha$ emission field. Second, the probability distribution of $\delta_{\rm F}$ convolved through the whole $g$ band, having a given observed Ly$\alpha$ forest: $P(\delta_{\rm F}^{\rm g\, full} | \delta_{\rm F}^{\rm g\, obs})$. This is a Bayesian estimation of the LSS Ly$\alpha$ absorption field based on the redshift range we can observe. Both these quantities must be correlated, as these Ly$\alpha$ emission and absorption fields trace the same LSS. Now we want to define a correlation estimator with maximal SNR, which also maximises our possibilities of detecting the Ly$\alpha$ emission in the $g$ band images, despite the very low SNR expected from the diffuse Ly$\alpha$ emission.

First, let us define a $\delta_{\rm F}^{\rm g\, full}$ bin $j$, comprising all values $\delta_{\rm F\, j\, min}^{\rm g\, full} <\delta_{\rm F}^{\rm g\, full}<\delta_{\rm F\, j\, max}^{\rm g\, full}$. If we consider a QSO $i$, with a patch flux $f_i^{\rm g}$, and an associated forest observed value $\delta_{\rm F\, i}^{\rm g\, obs}$, we can express the probability of such patch $i$ having a $\delta_{\rm F\, i}^{\rm g\, full}$ value inside the bin $j$ as:

\begin{equation}\label{eq:patch_weight}
    w_{\rm ij} = \int_{\rm \delta_{\rm F\, j\, min}^{\rm g\, full}}^{\rm \delta_{\rm F\, j\, max}^{\rm g\, full}} P(\delta_{\rm F}^{\rm g\, full} | \delta_{\rm F\, i}^{\rm g\, obs})\, d\delta_{\rm F}^{\rm g\, full}
\end{equation}

Since this probability $w_{\rm ij}$ is a degree of belief, we can use it as a weight in the determination of the average patch flux in the forest bin $j$, $\langle f^{\rm g} \rangle_j$. In this way, for each $i$-th patch $w_{\rm ij}$ will be proportional to the probability of $\delta^{\rm g, full}_{\rm F\, i}$ being within the $j$-th bin. The average patch flux for a given $\delta_{\rm F}^{\rm g\, full}$ bin $j$ will be

\begin{equation}\label{eq:weighted_patch_average}
     \langle f^{\rm g} \rangle_j =  \langle f^{\rm g} \rangle (\delta_{\rm F\, j\, min}^{\rm g\, full}, \delta_{\rm F\, j\, max}^{\rm g\, full}) = \frac{\sum_{\rm i} w_{\rm ij} f_i^{\rm g}}{\sum_{\rm i} w_{\rm ij}}.
\end{equation}

This average patch flux $\langle f^{\rm g} \rangle_j$ can be understood as a \textit{bayesian stacking} of observed flux $f_{\rm g}$ in bins $j$ of the Ly$\alpha$ forest absorption convolved with the entire $g$ band ($\delta_{\rm F}^{\rm g\, full}$).

If the $g$-band images contain unresolved Ly$\alpha$ emission tracing LSS (and this is correlated to the Ly$\alpha$ forest), the average patch fluxes $\langle f^{\rm g} \rangle_j$ must evolve for different $\delta^{\rm g\, full}_{\rm F}$ bins $j$. In other words, integrating the weights in \cref{eq:patch_weight} for different ranges of $\delta_{\rm F}^{\rm g\,full}$ should yield different values of $\langle f^{\rm g} \rangle$. If this evolution of $\langle f^{\rm g} \rangle_j$ versus $\delta^{\rm g\,full}_{\rm F\, j}$ is statistically significant and physically meaningful (e.g., pointing towards more overdense regions having brighter Ly$\alpha$ emission), then we will have detected LSS in Ly$\alpha$ emission. Such a detection would not be a \textit{cross-correlation} in the strict sense usually applied in cosmology (i.e., the 2PCF computed using two different tracers), but it will still be a \textit{cross-correlation} in the broadest sense (i.e., an estimator that proves that a correlation exists between two different datasets).

We will not assume a given functional form to fit to the $\langle f^{\rm g} \rangle_j$ versus $\delta^{\rm g\,full}_j$ relation. Instead, we will start from the most agnostic assumption possible: on average, overdense regions will have brighter integrated Ly$\alpha$ emission than underdense regions. Therefore, our correlation estimator will be the difference in average Ly$\alpha$ patch flux between underdense and overdense regions, i.e,

\begin{equation}\label{eq:lya_diff_avg}
      \Delta \langle L_{\rm Ly\alpha} \rangle \equiv \langle f^{\rm g} \rangle_{\rm over} - \langle f^{\rm g} \rangle_{\rm under} = \langle f^{\rm g} \rangle (0, 1) - \langle f^{\rm g} \rangle (1, \infty),
\end{equation}

where $\langle f^{\rm g} \rangle_{\rm over}$ is the average patch flux for overdense regions (for $\delta^{\rm g\,full} < 1$, i.e., regions where the transmitted flux fraction in the Ly$\alpha$ forest is smaller than average), and $\langle f^{\rm g} \rangle_{\rm under}$ is its counterpart for underdense regions. Hence, we expect $\Delta \langle L_{\rm Ly\alpha} \rangle > 0$, and the significance of $\Delta \langle L_{\rm Ly\alpha} \rangle > 0$ will be the significance of the detection of LSS in Ly$\alpha$ emission. The behaviour of the estimator versus properties of the Ly$\alpha$ emission field, as well as its reasonable values, will be explored and discussed in \cref{sec:Estimator behaviour and SNR optimisation} and \cref{sec:Modelling the Lya emission field}. Since the estimator in \cref{eq:lya_diff_avg} does not make any assumption of underlying physics (other than overdense regions being brighter than underdense), we expect it to effectively detect LSS in Ly$\alpha$ regardless of the true nature of the Ly$\alpha$ emission field, as long as SNR is high enough. However, given the large redshift convolution of the estimator (\cref{fig:qso_vs_z_g_band}), we do not expect it to be able to detect very overdense/underdense local features in specific regions, such as protoclusters or large voids.

The error of $\Delta \langle L_{\rm Ly\alpha} \rangle$ in \cref{eq:lya_diff_avg} is not trivial to propagate, since the probability distributions of $\langle f^{\rm g} \rangle_{\rm \rm over}$ and $\langle f^{\rm g} \rangle_{\rm \rm under}$ are not independent. While both are expected to have Gaussian distributions simply by being the weighted average of a large number of independent patch fluxes, the weights in \cref{eq:weighted_patch_average} for $\langle f^{\rm g} \rangle_{\rm \rm over}$ and $\langle f^{\rm g} \rangle_{\rm \rm under}$ are correlated. For any QSO, the sum of the two weights used for $\langle f^{\rm g} \rangle_{\rm \rm over}$ and $\langle f^{\rm g} \rangle_{\rm \rm under}$ must be 1, as that sum would be the full integral of $P(\delta_{\rm F}^{\rm g\, full} | \delta_{\rm F\, i}^{\rm g\, obs})$ over the set of all real numbers $\mathbb{R}$ (equation \ref{eq:patch_weight}). Therefore, we will compute the errors of $ \Delta \langle L_{\rm Ly\alpha} \rangle$ via bootstrapping, resampling our data by randomly drawing QSOs subsamples from our simulation with replacement (see \cref{sec:Forecast results} for more details).

\section{Forecast simulation}\label{sec:Forecast simulation}

In this section, we will briefly explain how both the DECaLS/BASS images and the DESI Ly$\alpha$ forest spectra are simulated for our forecast. First, the Ly$\alpha$ emission and absorption field (i.e., our signal) is modelled by employing a hydrodynamic simulation specifically designed for Ly$\alpha$ forest studies. This simulation was performed with the P-GADGET code (\citealt{Springel2005, DiMatteo2012}), using the cosmology specified in \cref{sec:Introduction}, for a comoving box size of 400 cMpc/h (571.4 cMpc). This is a volume larger than the current publicly available, state-of-the-art hydrodynamic simulations \citep[e.g., IllustrisTNG,][with a maximum comoving size of 300 Mpc]{Nelson2019}, so some approximations had to be made to reduce computation time.

The simulation contains $2 \cdot 4096^2$ particles, with masses of $1.19\times 10^7\, M_{\rm \rm \odot}$/h and $5.92 \times 10^7 \, M_{\rm \rm \odot}$/h for baryons and dark matter respectively, calculated with a gravitational softening length of 3.25 kpc/h. A density threshold for star formation lower than usual \citep[e.g.,][]{Springel2003, Pillepich2018a} was adopted (just 1000 times the average gas density), to allow for gas particles to be more easily turned into collisionless star particles and hence reduce the computational cost of following the gas hydrodynamics for longer periods. Moreover, black hole formation and stellar feedback were not applied. All these approximations simplify calculations and reduce computation time, but result in inaccurate stellar properties for galaxies. While AGN feedback \citep{Bird2023} or star formation \citep{Sorini2020} may have a limited effect on the the IGM, and thus its associated Ly$\alpha$ forest, we consider these effects subdominant and thus do not model them in our study, as in \citet{Viel2004}.

\begin{figure*}
 	\includegraphics[width=\columnwidth]{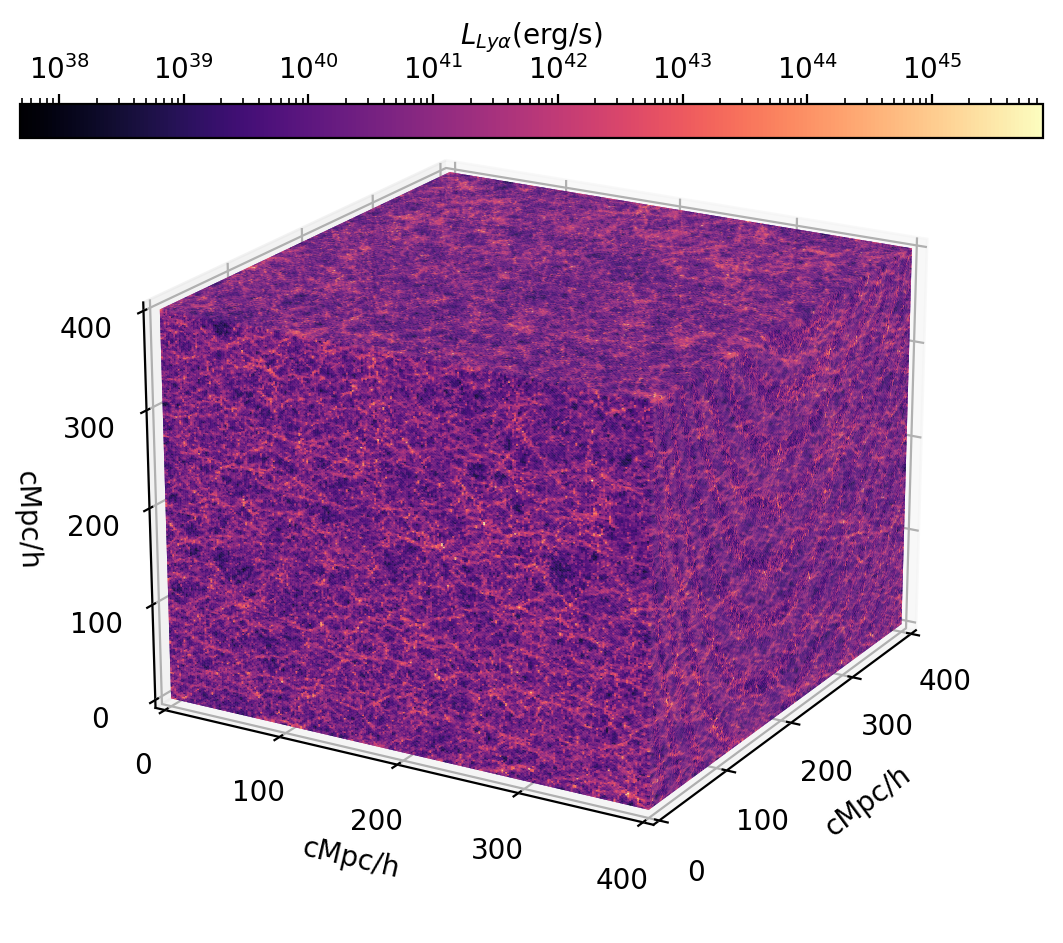}
 	\includegraphics[width=\columnwidth]{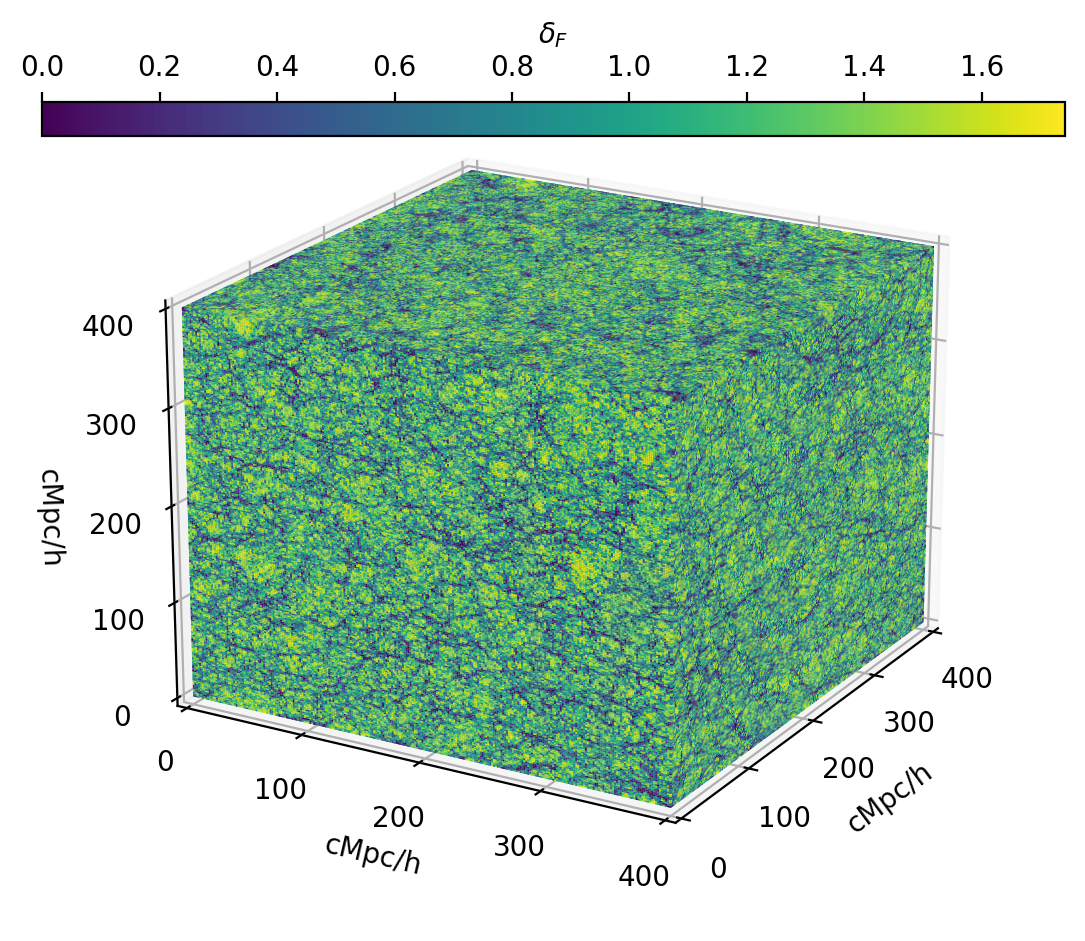}
     \caption{\textit{Left panel:} Simulation box for the Ly$\alpha$ emission field, with the parameters of an intermediate case ($\langle L_{\rm Ly\alpha} \rangle = 7.5 \cdot 10^{\rm 40}$ erg/s/cMpc$^3$, $b_{\rm e}=2.5$). \textit{Right panel:} Simulation box for the Ly$\alpha$ absorption field.}
     \label{fig:simulation_boxes}
\end{figure*}

This simulation box was used in \citet{Cisewski2014} and \citet{Ozbek2016} in order to test different approaches to interpolate the three-dimensional IGM distribution based on Ly$\alpha$ forest observations. The same simulation was again used by \citet{Croft2018} to place constraints on Ly$\alpha$ emission models by comparing it to observational data, and by \citet{Renard2021} to perform a Ly$\alpha$ IM forecast with narrow-band photometry over a smaller footprint. A voxel plot of the simulation snapshot at $z=3$ (the one used for this work) is displayed in \cref{fig:simulation_boxes}, for both the Ly$\alpha$ emission (with intermediate parameters of our Ly$\alpha$ emission model, see \cref{sec:Lya emission model}) and Ly$\alpha$ absorption fields. For our analysis, this snapshot of the simulation box is divided into 256 equal bins per side, which results in cubic voxels of 1.56 cMpc/h (2.23 cMpc) per side (613 pkpc at $\langle z \rangle=2.64$).

\subsection{Ly$\alpha$ emission model}\label{sec:Lya emission model}

The Ly$\alpha$ forest is a well-understood and extensively studied cosmological probe, and thus, its simulation has already been addressed in a multitude of works \citep[e.g.,][]{Peirani2014, Ramirez-Perez2022}. In this work, the optical depth of the Ly$\alpha$ absorption field was computed along sightlines of the hydrodynamic simulation box, following \citet{Hernquist1996}.

In contrast, the Ly$\alpha$ emission is far less constrained, lacking straightforward and standardised simulation methods apart from computationally expensive radiative transfer implementations \citep[e.g.,][]{Cantalupo2005, Byrohl2022}. Given the modeling constraints and the observational limitations, we have opted for a simple analytical model that determines the Ly$\alpha$ luminosity at any given simulation voxel as a function of the baryonic density. The model has two free parameters which we will explore in our forecast: $\langle L_{\rm Ly\alpha} \rangle$, the average Ly$\alpha$ luminosity per unit of volume (in erg/s/cMpc$^3$), and $b_{\rm e}$, the linear bias of the power spectrum of Ly$\alpha$ emission. In detail, below a given luminosity threshold $L_{\rm Ly\alpha\, cut} \sim 10^{\rm 45}$ erg/s, the Ly$\alpha$ luminosity follows the power law

\begin{equation}\label{eq:lya_emission_cosmo}
    L_{\rm Ly\alpha}(\delta) = C_L \cdot (1+\delta)^{b_{\rm e}},
\end{equation}
where $\delta$ is the dark-matter overdensity field, $b_{\rm e}$ the bias and $C_L$ a normalisation constant computed numerically so that the average luminosity per unit volume in our simulation box matches the input value, $\langle L_{\rm Ly\alpha} \rangle$.  While this simple model successfully reproduces any desired average luminosity and bias in the linear regime, at $\langle L_{\rm Ly\alpha} \rangle > 10^{\rm 40}$ erg/s/cMpc$^3$ the voxels with highest $\delta$ may result in excessively bright Ly$\alpha$ luminosities \citep[a problem already discussed in][]{Renard2021}. It is also worth noting that no physics are assumed in this Ly$\alpha$ emission model: just a normalisation constant derived from $\langle L_{\rm Ly\alpha} \rangle$, and a linear bias with respect to the dark-matter LSS  \citep{Bernardeau2002}. More sophisticated alternatives implementing the physics behind the generation and scattering of Ly$\alpha$ emission will be discussed in \cref{sec:Modelling the Lya emission field}.

To explore the wide range of average Ly$\alpha$ luminosities and biases described in the literature, we impose an \textit{ad hoc} power-law cutoff for $L_{\rm Ly\alpha} > L_{\rm Ly\alpha\mathrm{\, cut}}$. This observationally motivated cutoff ensures that the brightest voxels have $L_{\rm Ly\alpha}$ values within approximately 1 dex of the observed luminosity functions (LFs) for bright quasars \citep[e.g.,][]{Spinoso2020, Liu2022, Torralba-Torregrosa2023}. Additionally, it sets the maximum total $L_{\rm Ly\alpha}$ in a voxel to around $10^{45.5}$ erg/s, consistent with the integrated luminosity of the largest observed Ly$\alpha$ nebulae \citep{Cantalupo2014, Cai2017}. The details of this bright-end cutoff and the whole Ly$\alpha$ emission model are thoroughly discussed in \cref{sec:Lya emission model cutoff}.

\begin{figure}
 	\includegraphics[width=\columnwidth]{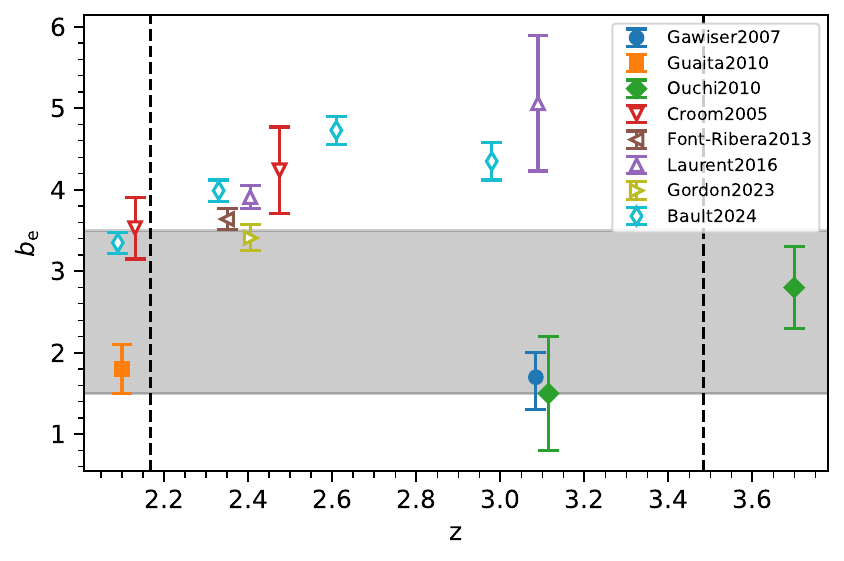}
     \caption{Bias values found in the literature for QSOs (empty markers) and LAEs (solid markers). The grey interval represents the range considered in our model. Dashed vertical line are the sensitivity limits of the $g$ band to Ly$\alpha$ (5\% of maximum sensitivity).}
     \label{fig:bias_values}
\end{figure}

\begin{figure}
 	\includegraphics[width=\columnwidth]{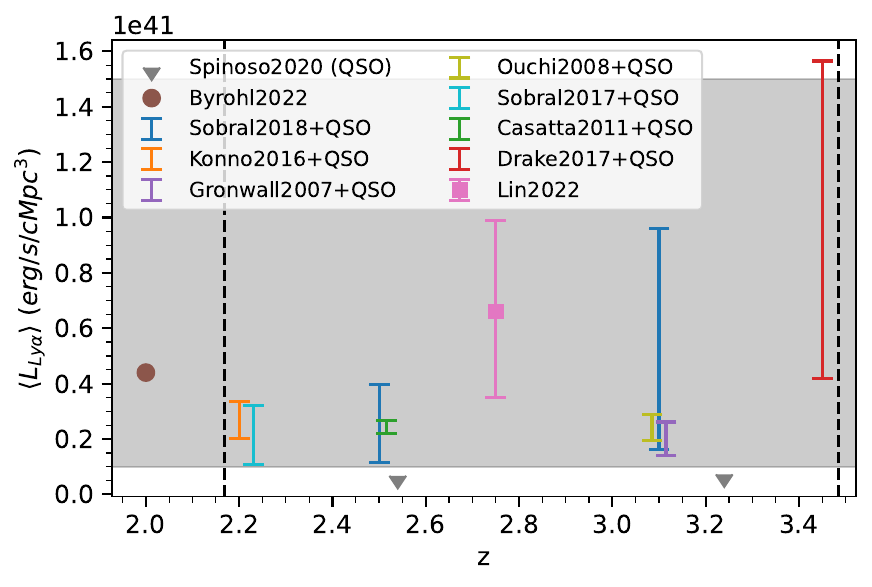}
     \caption{$\langle L_{\rm Ly\alpha} \rangle$ values found in the literature. The grey triangles are the lower limits obtained by integrating the bright QSO LFs from \citet{Spinoso2020}, recomputed after applying the $g < 19$ foreground cut from \cref{sec:Photometric noise: BASS intensity map and image reduction}. All luminosity intervals without central point correspond to integrals of LAE LFs + \citet{Spinoso2020} QSO LFs. The brown point \citep{Byrohl2022} is derived from simulations, and the pink point \citep{Lin2022} from QSO-galaxy spectra cross-correlation. The grey interval represents the interval considered as realistic in our model. Dashed vertical line are the sensitivity limits of the $g$ band to Ly$\alpha$ (5\% of maximum sensitivity).}
     \label{fig:lya_avg_values}
\end{figure}

Our model encompasses a range of bias values from $1.5 \leq b_{\rm e}\leq 3.5$. For values outside this range, the model's inner parameters diverge (see \cref{sec:Lya emission model cutoff}). Within this bias range, the model can adopt any value of $\langle L_{\rm Ly\alpha} \rangle$. However, we consider realistic values of $\langle L_{\rm Ly\alpha} \rangle$ to lie within the interval $1.5 \cdot 10^{40}, {\rm erg/s/cMpc^3} \leq \langle L_{\rm Ly\alpha} \rangle \leq 1.5 \cdot 10^{41}, {\rm erg/s/cMpc^3}$, based on current literature estimates \citep[e.g.,][]{Ouchi2008, Drake2017, Lin2022}. We will explore larger values of $\langle L_{\rm Ly\alpha} \rangle$ in \cref{sec:Forecast results} (up to $5 \cdot 10^{41}, {\rm erg/s/cMpc^3}$), but we consider these unlikely to represent the average luminosity of the real Ly$\alpha$ emission field. We also caution that in our Ly$\alpha$ emission model, such high $\langle L_{\rm Ly\alpha} \rangle$ values may yield non-physically high luminosities in the brightest Ly$\alpha$ voxels, even if only by a factor of a few. However, we do not consider this an issue since it only happens for $\langle L_{\rm Ly\alpha} \rangle$. values we already deem unphysical. These ranges of $b_{\rm e}$ and $\langle L_{\rm Ly\alpha} \rangle$ are supported by the literature, as shown in \cref{fig:bias_values} and \cref{fig:lya_avg_values}.

\Cref{fig:bias_values} displays different values of bias found in the literature for LAEs \citep{Gawiser2007, Guaita2010, Ouchi2010} and QSOs \citep{Croom2005, Font-Ribera2013, Laurent2016, Gordon2023, Bault2024}, in the the approximate redshift range of the $g$ band for Ly$\alpha$. The upper limit of the bias parameter in our model is close to the lowest value of the QSO bias measured to date \citep[][at $z \sim 2.1$]{Croom2005}. On the other hand, the lower limit corresponds to the lowest LAE bias \citep[][at $z \sim 3.1$]{Ouchi2010}. Since the $g$-band IM comprises all Ly$\alpha$ emission, from unresolved sources to all QSOs (except the ones removed by our foreground cut, see \cref{tab:g_cut_sigma}), it is reasonable to assume that the bias $b_{\rm e}$ of the whole Ly$\alpha$ emission field lies in-between that of QSOs and that of LAEs, as show with the grey area in 
\cref{fig:bias_values}. Even if unresolved LAEs or diffuse IGM emission were significant components of the total Ly$\alpha$ emission field, according to \citet{Byrohl2022}, the smallest dark matter halos with significant LAEs have $M_{\rm halo} = 10^9 M_{\rm \odot}$. Such halos have a bias $b_{\rm e} \sim 1$ at $z=3$ \citep[see e.g., fig. 7 of][]{Ouchi2018}. Given this absolute lower boundary to Ly$\alpha$ emission bias, and given that bright QSOs may reach $b_{\rm e} \sim 5$ and represent approximately from 3\% to 50\% of total Ly$\alpha$ emission in our model (\cref{fig:lya_avg_values}), it is highly unlikely that the true bias of the integrated Ly$\alpha$ emission field is significantly lower than 1.5.

\Cref{fig:lya_avg_values} shows the range of $\langle L_{\rm Ly\alpha} \rangle$ of our study, together with several values from the literature. First, the grey triangles are the $\langle L_{\rm Ly\alpha} \rangle$ obtained by integrating the QSO LFs from \citep{Spinoso2020} (recomputed after applying the foreground cut $g \leq 19$ to their sample, as specified in \cref{sec:Photometric noise: BASS intensity map and image reduction}). All the luminosity intervals without central points \citep{Gronwall2007, Ouchi2008, Cassata2011, Drake2017, Sobral2017, Sobral2018a} correspond to integrated LAE LFs. For all cases, the lower limit is the respective LF integrated to the observational limit specified in each work, plus the bright QSO LF integral from \citet{Spinoso2020} (as none of these works sample sources as bright as \citealp{Spinoso2020}). The upper limit is the LF integrated to the absolute lower limit $L_{\rm Ly\alpha}=10^{\rm 37}$ erg/s proposed in \citet{Bacon2021} and \citet{Alonso2023}, plus the contribution from the QSO LF integral. The remaining points are derived from a theoretical model of Ly$\alpha$ emission applied to the IllustrisTNG simulation \citep{Byrohl2022}, and a cross-correlation study in eBOSS of residuals of galaxy spectra with QSOs \citep{Lin2022}.

It is worth commenting briefly on the highest estimate of $\langle L_{\rm Ly\alpha} \rangle$ in \cref{fig:lya_avg_values}, namely: the upper limit presented by \citet{Drake2017}. According to \citet{Herenz2019}, most studies in the literature underestimate the faint end of the Ly$\alpha$ luminosity function (LF) by a factor of 2.5 due to correcting for completeness by treating LAEs as point sources rather than extended emitters. Notably, \citet{Drake2017} is the sole study in \cref{fig:lya_avg_values} that accounts for LAEs as extended sources, thereby avoiding the underestimation of LAE density at the faint end. Consequently, even though the upper limit reported by \citet{Drake2017} is 50\% higher than other estimates in \cref{fig:lya_avg_values}, recent literature supports its credibility. This justifies the upper limit of our $\langle L_{\rm Ly\alpha} \rangle$ range.

\subsection{Image simulation: QSO Patches}\label{sec:Image simulation: QSO patches}

The integrated $g$-band flux of the patches around each QSO has two uncorrelated components: the Ly$\alpha$ signal and the noise. Here we explain how both are simulated and combined in our forecast. The steps we follows are: i) the construction of a noise intensity map using real BASS $g$-band images, ii) the generation of a Ly$\alpha$ intensity map from our simulation and iii) the final combination of these two elements, plus the simulation of all relevant systematics.

\subsubsection{Photometric noise: BASS intensity map}\label{sec:Photometric noise: BASS intensity map and image reduction}

Any signal in the $g$-band images that does not correspond to Ly$\alpha$ emission of cosmic origin is noise for the purposes of our study. This noise encompasses various sources such as cosmic background/foregrounds, zodiacal light, airglow, moonlight, and thermal noise. In \citet{Renard2021}, cosmic foregrounds and all other noise sources were modelled as two distinct components (a mock galaxy catalog and noise extrapolations from real PAUS images, respectively). However, here we do model all noise sources with one single element. In order to do so, we employ real BASS $g$-band images to model both cosmic foregrounds and other noise sources, following the same image reduction strategy we propose for an eventual observational study with full imaging surveys. This procedure ensures that cosmic foregrounds are real and not simulated, as they are already contained in the BASS images, and also allows us to develop and test an image reduction strategy.

It is of utmost importance that the diffuse Ly$\alpha$ signal is preserved throughout the entire data reduction of the $g$-band images. This imposes a hard limit on how aggressive can the foreground and noise removal procedures be when applied to these images. Although in this forecast the Ly$\alpha$ signal is simulated, and we only use the real $g$-band images as noise, we still make sure that the real Ly$\alpha$ emission contained in them is preserved. To do so, we take all BASS reduced $g$-band science-images in an arbitrary patch of 20 deg$^2$ centred on RA = 235$^\circ$, dec = 45$^\circ$. Since the standard BASS pipeline only performs background subtraction when computing the photometric catalogue \citep{Zou2017a}, these images do still contain the diffuse Ly$\alpha$ emission, and can be used as provided for our forecast (and an eventual observational study). However, if any sort of background subtraction by performing a fit to the images themselves was part of the standard image reduction, a custom reduction without it would be required, as the unresolved Ly$\alpha$ emission that constitutes the bulk of our signal would have been removed.

We apply additional reduction steps on the selected images as follows. First, we subtract the SKY value of each observation to all CCD images in the camera. For each observation, we take all CCD images (4 for BASS), compute the mean of their median SKY values, and subtract it from the 4 CCD images. This procedure greatly homogenises the $g$-band intensity map, mitigating the variability due to atmospheric and moon conditions, which is one of the main sources of noise in our study. However, by subtracting the median SKY value at the camera level, we effectively erase any signal of structure in Ly$\alpha$ at scales larger than the FOV of the camera itself. 90Prime, the camera used by BASS, has an effective approximately square FOV of $1.08 \times 1.03$
deg$^2$ \citep{Williams2004}, which corresponds to 120 cMpc at $\langle z \rangle = 2.64$. This imposes a hard limit to the transversal scales we can sample with this study. For DECaLS, the DECam has a significantly larger FOV. Thus, by assuming the BASS specifications we are being again conservative in our forecast. If larger transversal scales needed to be sampled, more sophisticated methods of sky subtraction could be implemented. For instance, modeling sky emission without averaging out all observed flux over a given area \citep[e.g.,][]{Kimeswenger2015, Han2023}.

Second, we crop the image borders (12 pixels) to avoid edge effects, and coadd all selected images in a low-resolution intensity map with \textsc{swarp} \citep{swarp}, with the same pixel size as our Ly$\alpha$ $g$-band simulated image (66.125 $\times$ 66.125 arcsec$^2$, see \cref{sec:Photometric signal: Lya intensity map}). We then deproject the resulting intensity map into a simple equal-area projection \citep[Sanson-Flamsteed projection, see e.g.][]{Calabretta2002}. This projection is adequate for the 20 deg$^2$ used in this forecast. However, a study with the full DECaLS/BASS surveys may require either the computation of several small intensity maps, or the use of more sophisticated projections such as HEALPix \citep{Gorski2005}.

Third, we apply extra reduction steps on the intensity map to remove noise sources. In particular, we employ $\sigma$-clipping on the image at the 3$\sigma$ level; this mostly removes very bright stars, mimicking the effect of a survey mask (in a eventual observational study, the DECaLS/BASS survey masks could be applied instead). Then, we perform a foreground cut by removing all sources with $g < 19$. Since photometry has already been performed over the entire BASS survey, we sum the flux of all sources with $g < 19$ for each pixel of the intensity map and subtract it, instead of masking the foregrounds in each individual image. The $g < 19$ threshold has been chosen to remove the brightest sources, while keeping the great majority of bright QSOs from \citet{Spinoso2020} (and thus, the great majority of DESI Ly$\alpha$ QSOs). Indeed, considering a pessimistic scenario, these bright QSOs may represent up to 50\% of the total Ly$\alpha$ luminosity (\cref{fig:lya_avg_values}). In \cref{tab:g_cut_sigma}, we show the $\sigma$ of the BASS $g$-band intensity map for different foreground cuts, together with the $\%$ of QSOs in \citet{Spinoso2020} after applying the same foreground cut. We  checked that cuts more stringent than $g < 19$ would remove a significant fraction of the Ly$\alpha$ QSOs while decreasing the $\sigma$ of the intensity map by less than 5\%.

\begin{table}
\centering
	\caption{Percentage of Ly$\alpha$ QSOs remaining in the \citet{Spinoso2020} sample for different $g$-band magnitude cuts, together with $\sigma$, the standard deviation of the observed pixel flux in our BASS $g$-band intensity map (\cref{fig:bass_im_lya_img}), after foreground subtraction with the same $g$-band magnitude cuts.}
 	\label{tab:g_cut_sigma}
 	\begin{tabular}{rccc}
 		\hline
 		 Foreground cut& $g<19$ &$g < 20$ &$g < 21$\\
 		\hline
 		QSO fraction (\%)& 93.353&49.245&3.001\\
        $\sigma$ ($10^{\rm -16}$ erg/s/cm$^2$/\AA)& 1.606&1.559&1.540\\
		\hline
 	\end{tabular}
 \end{table}

\begin{figure}
 	\includegraphics[width=1.01\columnwidth]{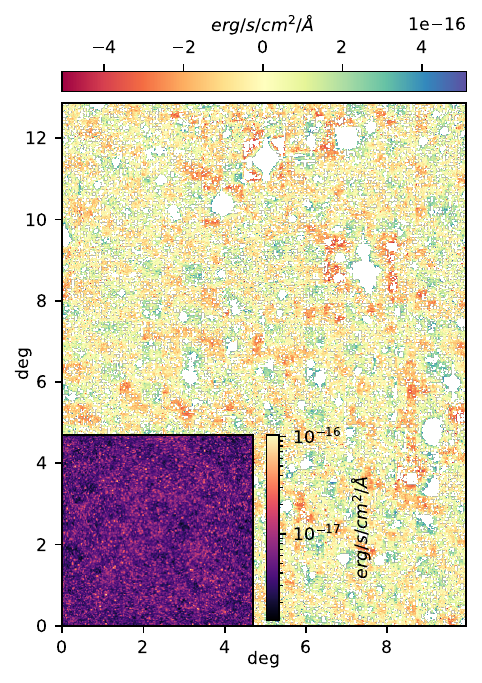}
     \caption{Intensity map of 20 deg$^2$ of BASS $g$-band images, overlaid with the simulated $g$-band Ly$\alpha$ emission intensity map, to scale (lower left corner). The Ly$\alpha$ emission model parameters were $\langle L_{\rm Ly\alpha} \rangle = 7.5 \cdot 10^{\rm 40}$ erg/s/cMpc$^3$, $b_{\rm e}=2.5$. The white regions on the BASS intensity map represent masked areas.}
     \label{fig:bass_im_lya_img}
\end{figure}

Finally, we apply once more a $\sigma$-clipping at 3$\sigma$ to the intensity map instead of masking very bright sources. Indeed, at this stage, the latter may be contiguous-pixels areas in the map image, so the effect of this last $\sigma$-clipping is to mask very bright isolated pixels (hence removing outliers and making the intensity map smoother) without significantly increasing the fraction of masked pixels\footnote{Only $\sim$0.3\% of pixels are masked, since it is a 3$\sigma$-clipping and large contiguous areas were already masked with the previous 3$\sigma$-clipping.}. The resulting intensity map, with the mean flux subtracted, is shown in \cref{fig:bass_im_lya_img} 

It is worth noting that, despite the image reduction,  a chequered pattern on the BASS intensity map is still clearly noticeable by visual inspection. This pattern is composed of squares of approximately 0.5 x 0.5 deg$^2$; roughly the size of the BASS camera CCDs \citep{Williams2004}.Thus, the primary source of noise for our forecast is not foregrounds of cosmic origin, but rather variations between CCDs and exposures. These variations likely arise from either the instrumental noise of the camera or changes in flux due to atmospheric or lunar sources (e.g., moonlight, airglow). We will further discuss the ramifications of this for our forecast and future Ly$\alpha$ IM studies in \cref{sec:Discussion}.

\subsubsection{Photometric signal: Ly$\alpha$ intensity map}\label{sec:Photometric signal: Lya intensity map}

To simulate the Ly$\alpha$ emission intensity map, we apply the Ly$\alpha$ emission model described in \cref{sec:Lya emission model} to our simulation box (\cref{fig:simulation_boxes}). This allows us to obtain a Ly$\alpha$ emission simulation box from which we then compute a lightcone. For this, we replicate and rotate twice (by 90$^\circ$) the resulting luminosity box in the redshift direction, to fully cover the Ly$\alpha$ redshift range associated to the $g$-band . The simulation voxels are converted from transversal comoving to angular coordinates, interpolating to the grid at lowest redshift, i.e., the lowest angular resolution. This angular resolution limit constrains the pixel size of our forecast: 66.125 $\times$ 66.125 arcsec$^2$. Finally, the Ly$\alpha$ luminosity within each voxel is converted to observed flux following \cref{eq:lya_observed_flux}. The result is a lightcone of Ly$\alpha$ observed flux, which is convolved with the $g$ band to obtain the Ly$\alpha$ intensity map displayed in the inset of \cref{fig:bass_im_lya_img}.

\subsubsection{Photometric signal and noise combination}\label{sec:Photometric signal and noise combination}

The simulated Ly$\alpha$ intensity map plus the real $g$-band BASS intensity map need to be combined into a single dataset simulating real $g$-band patches around DESI Ly$\alpha$ QSOs.  There are more projected DESI Ly$\alpha$ QSOs (700,000) than pixels in our intensity maps (65,536 for the Ly$\alpha$ simulation, 379,241 for the BASS intensity map), so some repetition of Ly$\alpha$ emission/noise patches is unavoidable. To solve this issue, we separately select pixels from the Ly$\alpha$ emission and BASS noise intensity maps as QSOs, in a random and independent fashion.

For each DESI Ly$\alpha$ QSO, we randomly draw with replacement a pixel from the Ly$\alpha$ simulated map and the BASS real map. In other words, a pixel that has been drawn once can be drawn again; the probability of drawing a given pixel remains constant regardless of previous outcomes. We integrate the fluxes for a Ly$\alpha$ patch (from the Ly$\alpha$ map) and noise patch (from the BASS map) centred around the drawn pixels, and add them to simulate the $g$-band patch flux around a given QSO, containing both the Ly$\alpha$ emission signal and noise. By drawing random pairs of both intensity maps, we ensure that no artificial correlations arise from the periodicity of the hydrodynamic simulation combined with the periodic CCD-sized patterns in the BASS intensity map (\cref{fig:bass_im_lya_img}). Moreover, the amount of possible combinations of Ly$\alpha$ patches + BASS noise patches largely outnumbers the expected number of DESI Ly$\alpha$ QSOs (by 4 orders of magnitude). This implies that the impact of the repetition of Ly$\alpha$ signal + noise combinations is negligible, even if we include all expected DESI Ly$\alpha$ QSOs in our study.

The most important systematics affecting the observed flux are modelled before combining the Ly$\alpha$ + noise patches. Zero-point calibration error is included in the Ly$\alpha$ patch fluxes by converting the fluxes to magnitudes, adding a random magnitude offset drawn from a Gaussian of $\mu = 0$ and $\sigma=7$ mmag \citep{Dey2019}, and then converting back to fluxes. Since BASS noise patches are computed on real data, the calibration error is already present, so no extra offset is added. It is worth noting that calibration techniques with zero-points varying across a given CCD exist \citep[e.g., J-PLUS/J-PAS,][]{Lopez-Sanjuan2019}, instead of per-CCD calibration as in DECaLS/BASS \citep{Dey2019}. These calibration methods may significantly reduce our calibration error, since when computing patch fluxes we integrate relatively large fractions of a given CCD.

Fluxes in the noise patches are integrated by summing the flux in all pixels inside each patch, barring masked pixels (according to the mask computed in \cref{sec:Photometric noise: BASS intensity map and image reduction} and displayed as white regions in \cref{fig:bass_im_lya_img}). For the Ly$\alpha$ patches, however, we do not replicate the same mask, as this would require generating a single Ly$\alpha$ cutout with the respective mask for every QSO patch we need to simulate, dramatically increasing computation time. Instead, we first pre-compute all possible Ly$\alpha$ patches of the desired radius in our simulated Ly$\alpha$ image, and determine the flux percentiles for each possible patch, $P^\%_{\rm Ly\alpha}$. Then, we simulate the error stemming from masking by shifting the flux percentile as follows:

\begin{equation}\label{eq:lya_flux_masked}
    P^\%_{\rm Ly\alpha\, masked} = P^\%_{\rm Ly\alpha} + U(-\%_{\rm masked}/2, \%_{\rm masked}/2),
\end{equation}
where $U$ is the uniform distribution, and $\%_{\rm masked}$ is the percentage of area masked in the respective BASS noise patch. For example, a Ly$\alpha$ patch with a flux in the 50th percentile, that is superimposed with a BASS noise patch that has 10\% of its area masked, will have a Ly$\alpha$ flux after making between the 45th and 55th percentile (if the result was to be below the 0th or above the 100th percentile, it is capped at 0 or 100 respectively). This method results in a flux scatter similar to what would be obtained by the Ly$\alpha$ image with the BASS mask, at a negligible computational cost. After the zero-point calibration error and the mask offset following \cref{eq:lya_flux_masked} is added to the Ly$\alpha$ patch fluxes, and the BASS noise patch fluxes have been integrated with their actual mask, we sum the BASS noise patch fluxes to the Ly$\alpha$ patch fluxes. The result is a set of simulated $g$-band patches for each DESI QSO, each one of them with a (very likely) unique Ly$\alpha$ signal+noise combination.

\subsection{Spectra simulation: Ly$\alpha$ forest probabilities}\label{sec:Spectra simulation: Lya forest probabilities}

\begin{figure}
 	\includegraphics[width=\columnwidth]{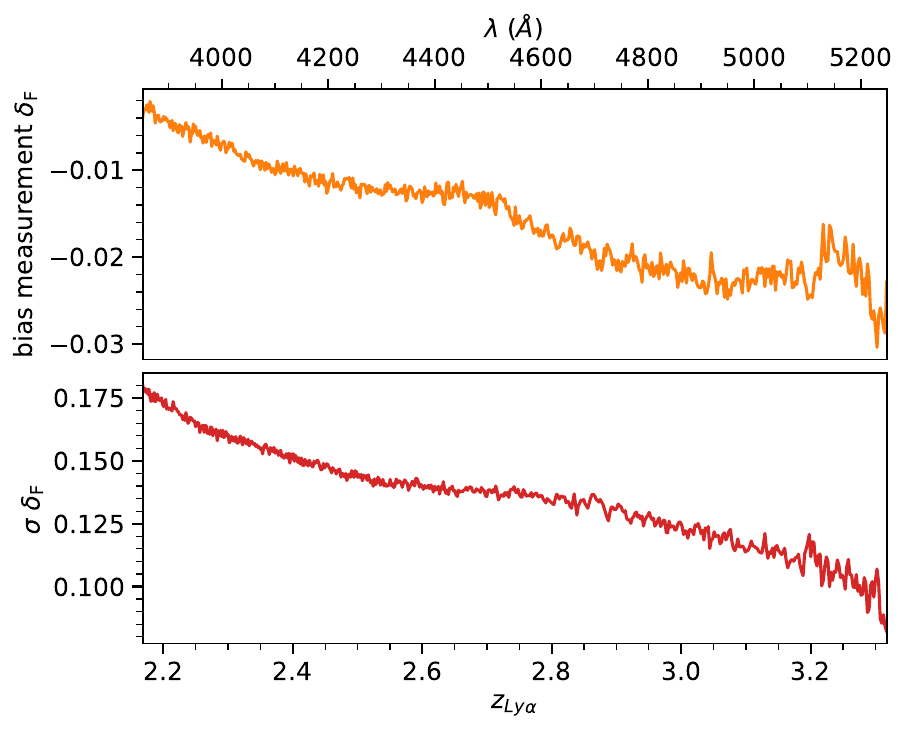}
     \caption{\textit{Upper panel}: Bias resulting from applying the quasar continuum-subtraction algorithm QFA \citep{Sun2022} to the Ly$\alpha$ forest mocks of \citet{Bautista2015}. \textit{Lower panel}: Average observational error of the \citet{Bautista2015} Ly$\alpha$ forest mocks.}
     \label{fig:forest_bias_error}
\end{figure}

To simulate the Ly$\alpha$ forest data, first the Ly$\alpha$ absorption simulation box (\cref{fig:simulation_boxes}) is replicated and rotated in the redshift direction, as well as interpolated to angular coordinates, following the same procedure carried out for the Ly$\alpha$ emission box (\cref{sec:Photometric signal: Lya intensity map}). We add observational noise to this Ly$\alpha$ forest lightcone and continuum-subtraction bias as a function of redshift/observed wavelength. The observational noise is extracted from the SDSS BOSS mocks in \citet{Bautista2015}, and the bias is derived from the continuum-subtraction algorithm presented in \citet{Sun2022}. \Cref{fig:forest_bias_error} displays both the bias and error as a function of Ly$\alpha$ redshift/observed wavelength.

After adding noise and bias, we duplicate the lightcone, and smooth the duplicate in redshift bins of  width 150 cMpc. This duplicate, smoothed version of the lightcone will be used to randomly generate smoothed forests to fill the redshift range not covered by the observed forests, as described in \cref{sec:Spectroscopic data: Convolved forest probabilities}. 

We simulate the observed forests from the unsmoothed lightcone by randomly drawing a QSO redshift for each angular bin (i.e., each possible sightline), following the DESI QSO redshift distribution (\cref{fig:qso_vs_z_g_band}). Indeed, the latter sets the observed redshift/wavelength range in each lightcone sightline. Damped Ly$\alpha$ Absorbers \citep[DLAs][]{Wolfe2005} are added to the observed forests approximately following the redshift distribution and width of the DLAs in the \citet{Bautista2015} mocks. Since Ly$\alpha$ absorption is saturated in DLAs, the underlying HI density field is significantly more difficult to measure, thus we mask them. However, we fill the masked DLA regions with the average Ly$\alpha$ forest absorption $\delta_{\rm F}$ at its redshift. This is done because in the convolution of $\delta_{\rm F}$  with the $g$-band ($\delta_{\rm F}^{\rm g}$, as expressed in \cref{eq:delta_g}), not all the redshift pixels contribute equally (because of the $D_L(z)^{\rm -2}$ factor). Therefore, omitting lower redshift regions from integration results in lower values of $\delta^{\rm g}_{\rm F}$, introducing an undesired bias in our sample.

After computing a simulation of full smoothed forests and another of unsmoothed observed forests, we follow the procedure outlined in \cref{sec:Spectroscopic data: Convolved forest probabilities} to obtain, for each sightline in our simulation, the posterior distribution of full convolved forests given an observed forest, $P(\delta^{\rm g\, full}_{\rm F} | \delta^{\rm g\, obs}_{\rm F})$. To do so, for each observed lightcone sightline we generate 5,000 smoothed forests over the whole $g$-band redshift range by drawing with replacement values for each redshift bin from all the smoothed lightcone sightlines. We inject the observed forests into the 5,000 generated smoothed forests, convolve all of them with \cref{eq:delta_g}, and compute their EDF, which is analogous to  $P(\delta^{\rm g\, full}_{\rm F} | \delta^{\rm g\, obs}_{\rm F})$.

Therefore, every pixel in our simulated Ly$\alpha$ intensity map (\cref{fig:bass_im_lya_img}) has an associated posterior of the full Ly$\alpha$ forest convolution, $P(\delta^{\rm g\, full}_{\rm F} | \delta^{\rm g\, obs}_{\rm F})$. When randomly selecting a given pixel in the Ly$\alpha$ intensity map as a DESI Ly$\alpha$ QSO to compute the $g$-band patch around it, its correspondent $P(\delta^{\rm g\, full}_{\rm F} | \delta^{\rm g\, obs}_{\rm F})$ is also selected. %By randomly drawing simulation pixels for the 700,000 projected DESI Ly$\alpha$ QSOs, the result is a simulation of both BASS $g$-band patches and its associated DESI Ly$\alpha$ forest posteriors.

\section{Forecast results}\label{sec:Forecast results}

\subsection{Estimator behaviour and SNR optimisation}\label{sec:Estimator behaviour and SNR optimisation}

Before simulating the actual forecast for DESI-DECaLS/BASS, we need to evaluate the behaviour of the $\Delta \langle L_{\rm Ly\alpha} \rangle$ estimator and, more importantly, its error. We acknowledge that we have made the conservative choice of using BASS images to model our noise intensity map (\cref{sec:Photometric noise: BASS intensity map and image reduction}); however, we compute our forecast for the combined DECaLS/BASS footprint (i.e., the entirety of DESI). Therefore, we will refer to the cross-correlation simulated in this work as DESI-DECaLS/BASS. The expected changes we may expect in our results by using actual DECaLS images to simulate its respective fraction of DESI are discussed in \cref{sec:Extrapolation to number of observed QSOs}; however these anticipated changes do not modify our overarching conclusions.

Since we expect the SNR to be low, and a detection of Ly$\alpha$ LSS to be uncertain, it is crucial that the errors provided for $\Delta \langle L_{\rm Ly\alpha} \rangle$ are properly justified, and the observational parameters are chosen to maximise the SNR. Unless specified otherwise, in this subsection we will use a "fiducial" Ly$\alpha$ emission model with $\langle L_{\rm Ly\alpha} \rangle = 7.5 \cdot 10^{\rm 40}$ erg/s/cMpc$^3$ and $b_{\rm e}=2.5$. 

We will assume that all results can be extrapolated to any other combination of $\langle L_{\rm Ly\alpha} \rangle$ and $b_{\rm e}$ of our model. Since $\langle L_{\rm Ly\alpha} \rangle$ is just a multiplicative constant of our model (equation \ref{eq:lya_emission_cosmo}), we have no reason to believe the optimal parameters yielding the highest SNR will be sensitive to $\langle L_{\rm Ly\alpha} \rangle$. Moreover, since our estimator $\Delta \langle L_{\rm Ly\alpha} \rangle$ evolves symmetrically with respect to $\langle L_{\rm Ly\alpha} \rangle$ and $b_{\rm e}$ (\cref{fig:lya_diff_map_with_isolines}), we assume the same invariance with respect to $b_{\rm e}$.

\subsubsection{Error estimation}\label{sec:Error estimation}

As specified in \cref{sec:Correlation estimator: Lya average fluctuation}, we will compute the errors via bootstrapping. Having a given number of QSOs $N_{\rm QSO}$, we randomly draw with replacement $N_{\rm QSO}$ QSOs $N_B$ times, creating $N_B$ bootstrap resamplings. For each one of these resamples, we compute $\Delta \langle L_{\rm Ly\alpha} \rangle_B$, the value of the estimator in a given bootstrap resampling. The value and error of $\Delta \langle L_{\rm Ly\alpha} \rangle$ will be the average and standard deviation of all $\Delta \langle L_{\rm Ly\alpha} \rangle_B$, respectively.

\begin{figure}
 	\includegraphics[width=\columnwidth]{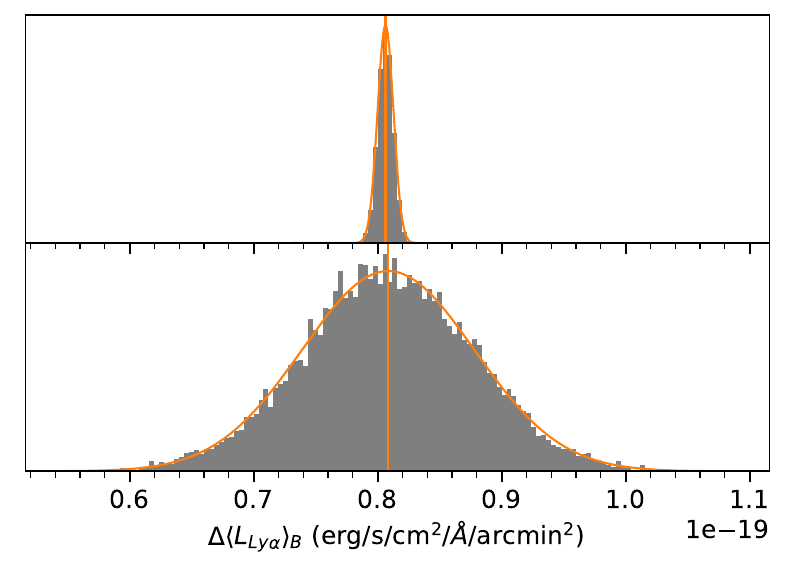}
     \caption{Histograms of $\Delta \langle L_{\rm Ly\alpha} \rangle_B$ for 10,000 bootstrap resamplings of our forecast (grey), together with a Gaussian distribution with the mean and $\sigma$ of the bootstrap resampling (orange). Upper panel for noiseless forecast, lower panel for forecast with 5\% of expected noise.}
     \label{fig:lya_diff_bootstrap}
\end{figure}

\Cref{fig:lya_diff_bootstrap} shows the distribution of $\Delta \langle L_{\rm Ly\alpha} \rangle_B$ for 10,000 bootstrap resamplings, for a noiseless case and for a case with 5\% of the expected BASS noise. In detail, for the former, the BASS noise patches were not combined with the Ly$\alpha$ patches from the Ly$\alpha$ image (i.e., the BASS intensity map was not used in the forecast at all); for the latter, we added the BASS noise patches, after multiplying their fluxes by 0.05. Additional systematics such as masking and flux calibration errors were added only in the 5\% noise case. For both cases, the histogram of $\Delta \langle L_{\rm Ly\alpha} \rangle_B$ values closely follows a Gaussian distribution; adding the 5\% BASS noise and systematics widens the Gaussian and increases the error, but it does not significantly alter the average value of $\Delta \langle L_{\rm Ly\alpha} \rangle_B$ (approximately at $8.16\cdot 10^{\rm -20}$ erg/s/cm$^2$/\AA/arcmin$^2$). Hence, \cref{fig:lya_diff_bootstrap} shows that the error estimated via bootstrapping is well-behaved, and the introduction of noise and systematics from a real imaging survey like BASS does not bias the results. 

Beyond the intrinsic cosmic variance in the hydrodynamic simulation and the noise/systematics broadening the $\Delta \langle L_{Ly\alpha} \rangle_B$ distribution in \cref{fig:lya_diff_bootstrap}, we must consider an additional source of variability. Our forecast involves randomly selecting pixels from the Ly$\alpha$ simulated image as QSOs (patch centres) and pairing them with randomly chosen pixels from the BASS intensity map. In other words, we randomly pair each patch of integrated signal with a patch of integrated noise. Depending on how these signal-noise patches are paired, the value of  $\Delta \langle L_{\rm Ly\alpha} \rangle$ may artificially be increased (e.g., if areas with high Ly$\alpha$ emission happen to be paired with exposures with a particularly bright sky) or decreased (e.g., if a large masked area caused by a bright star overlaps with a bright Ly$\alpha$ region). In the following, we will call a given set of randomly chosen signal-noise patches a \textit{realisation} of our forecast.

\begin{figure}
 	\includegraphics[width=\columnwidth]{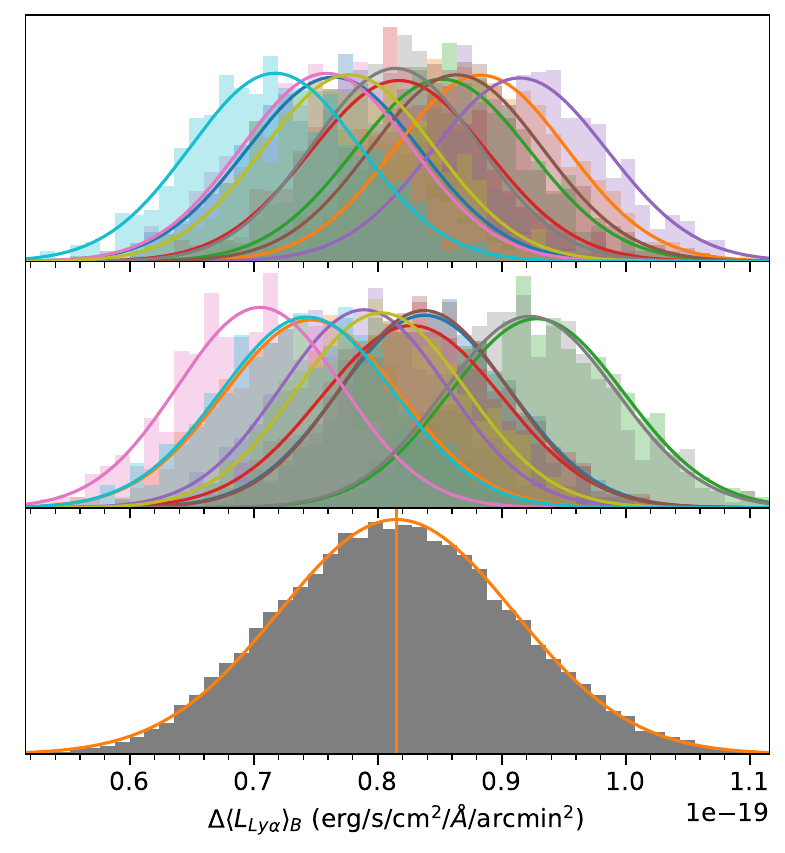}
     \caption{\textit{Upper and middle panel:} Histograms of $\Delta \langle L_{\rm Ly\alpha} \rangle_B$ for 20 realisations of Ly$\alpha$ patches-noise patches, with 1,000 bootstrap resamplings each, together with their respective derived Gaussian distributions. The realisations have been randomly divided in two panels, each one displaying 10. \textit{Lower panel:} Histogram and Gaussian distributions for all the $\Delta \langle L_{\rm Ly\alpha} \rangle_B$ from the 20 realisations combined. The x-axis has the same scale as \cref{fig:lya_diff_bootstrap}.}
     \label{fig:n_realizations_distributions}
\end{figure}
In \cref{fig:n_realizations_distributions}, we show how $\Delta \langle L_{\rm Ly\alpha} \rangle$ varies for 20 different realisations of our forecast; each one obtained with 1,000 bootstrap resamplings and 5\% of BASS noise. The values of $\Delta \langle L_{\rm Ly\alpha} \rangle$ (the mean of $\Delta \langle L_{\rm Ly\alpha} \rangle_B$) vary between realisations, the error (the $\sigma$ of $\Delta \langle L_{\rm Ly\alpha} \rangle_B$) remains approximately constant. For example, in \cref{fig:n_realizations_distributions}, the average error and its standard deviation for all 20 realisations is $6.89 \pm 0.15 \cdot 10^{-21}$ erg/s/cm$^2$/\AA/arcmin$^2$ (i.e., the $\sigma$ of the error is $\sim$2\% of its average value).

If all the realisations are combined into a single one (i.e., combining the 20 realisations of 1,000 bootstraps into a single one with 20,000), the result is a wider Gaussian distribution for $\Delta \langle L_{\rm Ly\alpha} \rangle_B$ than the distributions for separate realisations (lower panel of \cref{fig:n_realizations_distributions}). The standard deviation for this combined distribution is $9.48 \cdot 10^{-21}$ erg/s/cm$^2$/\AA/arcmin$^2$; an increase of almost 40\% compared to individual realisations. In fact, some of the Gaussian distributions corresponding to different realisations in \cref{fig:n_realizations_distributions} are separated by more than 1$\sigma$, indicating that this variation between realisations might be significant in some cases (larger than the error of $\Delta \langle L_{\rm Ly\alpha} \rangle$).

This variability of results between realisations is not a limitation of our forecast, but an additional systematic that is also expected to arise when carrying out the Ly$\alpha$ IM study with real data. Indeed, just as in our forecast, the correlated noise (chequered pattern in \cref{fig:bass_im_lya_img}) and masking over the full DECaLS/BASS $g$-band intensity map may artificially enhance or worsen the structure of the real Ly$\alpha$ emission field, resulting in a shift of the true value of  $\Delta \langle L_{\rm Ly\alpha} \rangle$. Moreover, in the real Universe there is only one realisation of the sky positions of the QSOs targeted by DESI; hence the variability between realisations is not the error of the $\Delta \langle L_{\rm Ly\alpha} \rangle$ measurement, but the variability of outcome for a measurement of $\Delta \langle L_{\rm Ly\alpha} \rangle$ with real DESI-DECaLS/BASS data.

Consequently, the variability between forecast realisations imposes a hard limit on the precision with which we can determine $\Delta \langle L_{\rm Ly\alpha} \rangle$, in a similar way cosmic variance limits any cosmological statistics \citep[e.g.,][]{Marra2013, Ma2024}. In fact, the intrinsic variability of $\Delta \langle L_{\rm Ly\alpha} \rangle$ can be seen as the combined effect of cosmic variance, the limited space sampling of the Ly$\alpha$ forest, and the correlated noise in the $g$-band intensity map. Hence, all error intervals and regions shown hereafter, both for $\Delta \langle L_{\rm Ly\alpha} \rangle$ and its SNR, correspond to their variability between forecast realisations.

\subsubsection{Optimal $r_{\rm patch}$}\label{sec:Optimal patch radius}

The most relevant free parameter in our Ly$\alpha$ IM methodology is $r_{\rm patch}$; in principle we can choose any $r_{\rm patch}$ as long as it remains below the angular scales of the BASS camera. Hence, we should evaluate how our $\Delta \langle L_{\rm Ly\alpha} \rangle$ (and more importantly, its SNR), evolve versus $r_{\rm patch}$, with the aim of finding its optimal value. By \textit{optimal} $r_{\rm patch}$, we exclusively refer to the $r_{\rm patch}$ that yields the maximum $\Delta \langle L_{\rm Ly\alpha} \rangle$ SNR, not the $r_{\rm patch}$ that may provide tighter constraints for other parameters (e.g., $\langle L_{\rm Ly\alpha} \rangle$ or $b_{\rm e}$).

\begin{figure}
 	\includegraphics[width=\columnwidth]{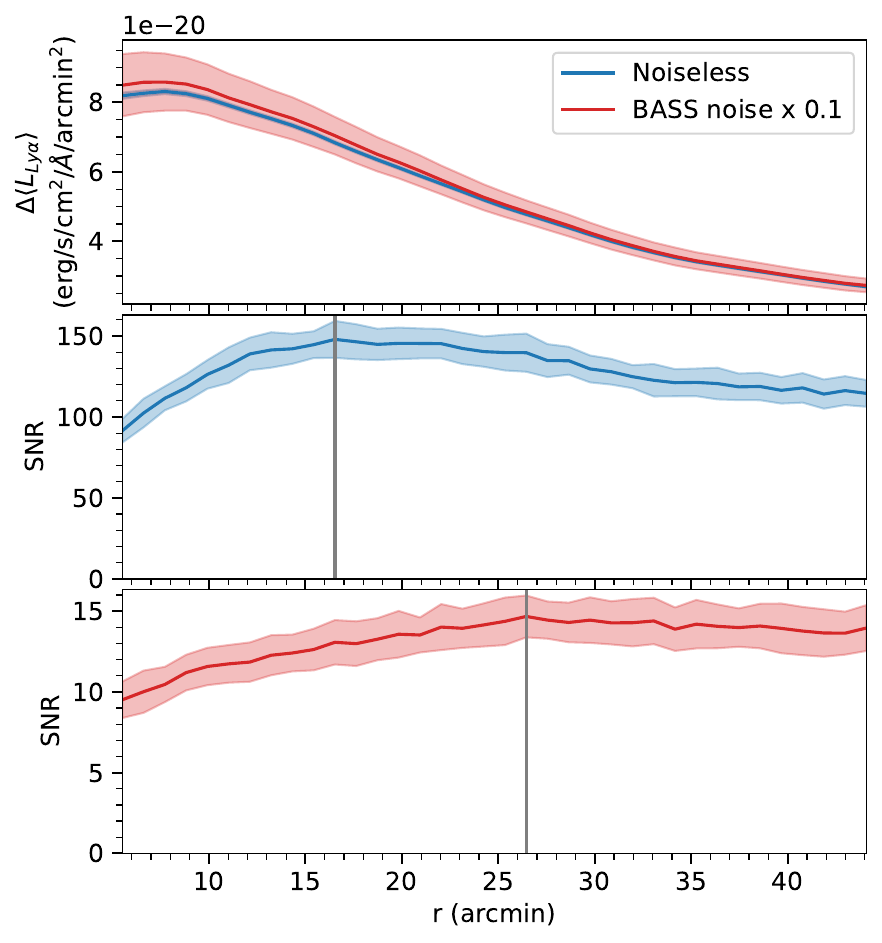}
     \caption{\textit{Upper panel:} Evolution of $\Delta \langle L_{\rm Ly\alpha} \rangle$ versus patch radius, for a noiseless simulation (blue) and a forecast with 10\% of BASS noise (red). \textit{Middle panel:} SNR versus patch radius, for the noiseless case; vertical line indicates the maximum SNR radius. \textit{Lower panel:} SNR versus patch radius, for the 10\% noise case.}
     \label{fig:snr_vs_radius}
\end{figure}

In \cref{fig:snr_vs_radius}, we display how the value of $\Delta \langle L_{\rm Ly\alpha} \rangle$ and, more importantly, its SNR evolve versus patch radius. We show both a noiseless forecast and one with 10\% of BASS noise (for both cases, 100 forecast realisations with 100 bootstrap resamplings were computed). The value of $\Delta \langle L_{\rm Ly\alpha} \rangle$ monotonically decreases with radius, as would be expected from integrating regions further away from the Ly$\alpha$ forest sightline. The SNR, however, increases with radius up to a given maximum, and then smoothly decreases. This SNR maximum is caused by two opposing trends: the integration of more Ly$\alpha$ emission versus the smaller correlation between Ly$\alpha$ emission and Ly$\alpha$ forest absorption as the patch radius increases. The SNR maximum is localised at $r_{\rm patch}=16.5$ arcmin for the noiseless forecast, but when we add noise the optimal radius shifts to $r_{\rm patch}=26.4$ arcmin (approximately 33 cMpc/h at $\langle z \rangle=2.64$). We have verified that this radius remains constant for increasing noise levels, so we will adopt $r_{\rm patch}=26.4$ arcmin for our forecast. Moreover, we have also verified that smaller trends such as the apparent minimum at $r_{\rm patch}\sim35$ arcmin in the lower panel of \cref{fig:snr_vs_radius} or the apparent SNR increase at the largest $r_{\rm patch}$ on the same panel are not significant. They stem from the variability between realisations, and they do not consistently appear if we recompute \cref{fig:snr_vs_radius} with a different set of realisations.

Moreover, by choosing such a large $r_{\rm patch}$, we ensure that all Ly$\alpha$ emission inside the cosmic volume defined by the DESI footprint and the $g$-band redshift range ($2.2 < z < 3.4$)  is integrated in our study. A circular patch of radius 26.4 arcmin has an area of 0.61 deg$^2$; according to the DESI QSO density distribution (\cref{fig:qso_vs_z_g_band}), on average such an area should contain at least one Ly$\alpha$ QSO (and thus a patch should be integrated around it) up to $z = 3.75$. Since this redshift is higher than the maximum Ly$\alpha$ redshift observed by the $g$ band, we can state that on average all Ly$\alpha$ emission in the cosmic volume defined by the DESI footprint and the $g$ band redshift convolution is sampled by at least one patch.

Three additional remarks are worth mentioning regarding \cref{fig:snr_vs_radius}. First, at $r_{\rm patch}>26.4$ the SNR decreases with radius very slowly, to the point that the decrease until $r_{\rm patch}=45$ is barely significant taking into account the aforementioned variability between forecast realisations (shaded regions). Hence, the SNR is almost insensitive to patch radius at $r_{\rm patch}>26.4$, which means that a real observational study does not necessarily have to use $r_{\rm patch}=26.4$ arcmin, and that a significant detection of $\Delta \langle L_{\rm Ly\alpha} \rangle > 0$ would very likely imply a detection at larger patch radii, and thus larger scales.

Second, it is possible that the SNR peak at $r_{\rm patch}=26.4$ and the slow decline at larger radii are partially or totally caused by the limited size of our simulation box. The maximum scales at which the 2PCF/power spectrum can be sampled in a given simulation are limited by the simulation size \citep{Bagla2005}. In \citet{Renard2021}, using the same hydrodynamic simulation (with a box size of 400 cMpc/h, \cref{sec:Forecast simulation}) a sharp decrease in the 2PCF and its respective SNR at scales of $>35$ cMpc/h could be seen. Direct extrapolations of \citet{Renard2021} results cannot be made here to understand the impact of the box size. This is because \citet{Renard2021} used fine redshift slices to compute the 2PCF while here we define a custom estimator over a much larger lightcone. Nevertheless, given their results, it is likely that the limited size of the simulation box artificially decreases the SNR of our forecast after a given $r_{\rm patch}$. Regardless, this limitation of our forecast would only mean that the  $r_{\rm patch}$ for true maximum SNR is larger than 26.4 arcmin, and thus by choosing $r_{\rm patch}=26.4$ arcmin we might be being conservative on our SNR estimates. Based on this reasoning, while we suggest $r_{\rm patch}=26.4$ arcmin for an actual observational study based on \cref{fig:snr_vs_radius}, we warn that repeating said study with larger $r_{\rm patch}$ may yield higher SNR.

Third, the value of  $r_{\rm patch}=26.4$ is close to the CCD size of 90Prime, the BASS camera. It is also possible that the slow SNR decrease at $r_{\rm patch}>26.4$  is due to the chequered CCD-sized pattern in \cref{fig:bass_im_lya_img} at similar scales. Regardless, even if we used a DECaLS intensity map for the forecast instead, the transversal scales we can sample are still limited by the size of our hydrodynamic simulation; without a significantly larger simulation box we cannot disentangle those effects. Hence, our conclusions remain the same: it is possible that a larger $r_{\rm patch}$ yields a higher SNR with a real data study, and it is also possible that the real optimal $r_{\rm patch}$ is limited by CCD or FOV size. However, since the $\Delta \langle L_{\rm Ly\alpha} \rangle$ SNR vs $r_{\rm patch}$ in \cref{fig:snr_vs_radius} fluctuates by less than 50\%, we expect the predicted SNR to have the same order of magnitude as our forecast even if the optimal $r_{\rm patch}$ for real data ends up being significantly different than the $r_{\rm patch}$ we employ.

\subsubsection{Optimal $z_{\rm min \, QSO}$}\label{sec:Optimal z qso min}

In addition to $r_{\rm patch}$, another parameter we may explore to optimise SNR is $z_{\rm min\, QSO}$; the minimum redshift at which we take the Ly$\alpha$ QSOs from DESI into our Ly$\alpha$ IM study. Following \citet{DESICollaboration2016}, all QSOs with $z>2.1$ have Ly$\alpha$ forest that can be observed with DESI, but the Ly$\alpha$ redshift range of the $g$ band is approximately $2.2 < z < 3.4$ (\cref{fig:qso_vs_z_g_band}). Hence, we can choose a higher value of $z_{\rm min\,QSO}$ to increase the minimum redshift overlap between the Ly$\alpha$ forest data and the integrated Ly$\alpha$ emission in the $g$-band patches. However, by choosing a higher $z_{\rm min\,QSO}$ cut, we also select a smaller total number of QSOs, reducing the size of our sample.

\begin{figure}
 	\includegraphics[width=\columnwidth]{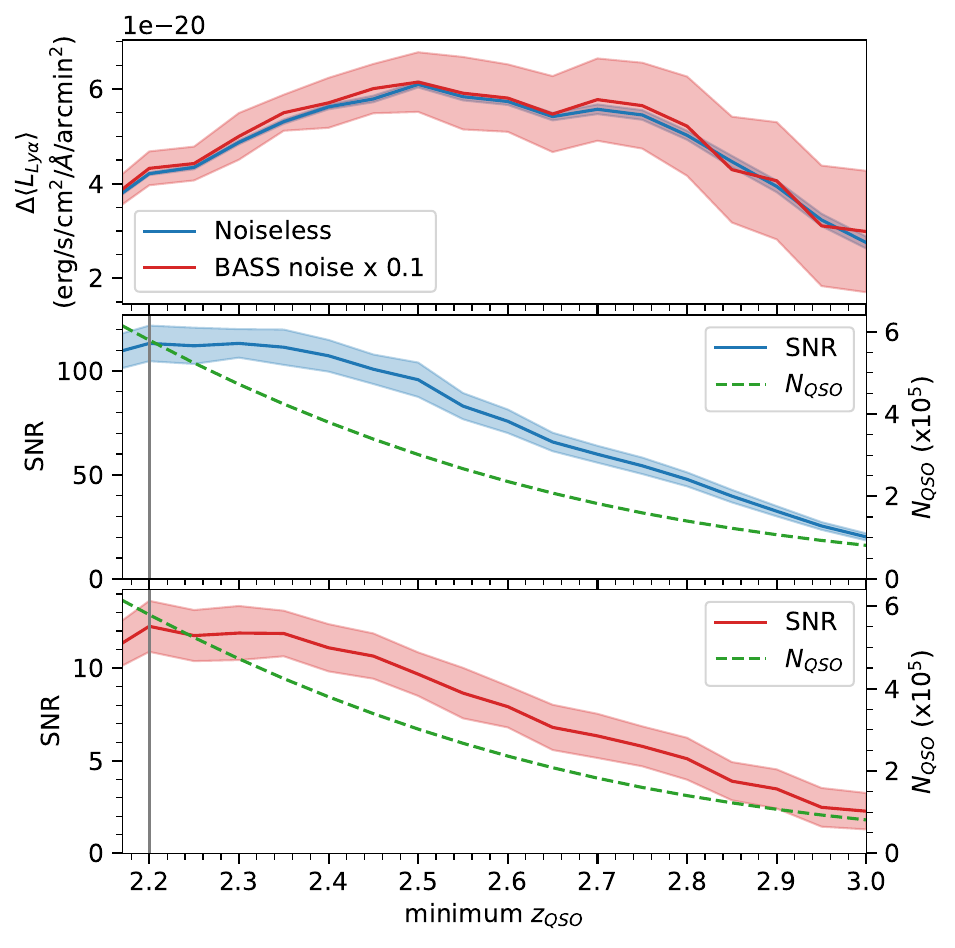}
     \caption{\textit{Upper panel:} Evolution of $\Delta \langle L_{\rm Ly\alpha} \rangle$ versus minimum QSO redshift in our study, for a noiseless case (blue) and 10\% of BASS noise case (red). \textit{Middle panel:} SNR versus minimum QSO redshift (left y axis) for the noiseless case, together with the total DESI QSOs for a given minimum redshift (right y axis, green dashed line). The vertical line indicated the minimum QSO redshift of maximum SNR. \textit{Lower panel:} Analogous plot, for the 10\% noise case.}
     \label{fig:snr_vs_z}
\end{figure}

This trade-off between number of QSOs and redshift overlap is explored in \cref{fig:snr_vs_z}, which shows the evolution of $\Delta \langle L_{\rm Ly\alpha} \rangle$ and its SNR versus  $z_{\rm min\,QSO}$, for our fiducial Ly$\alpha$ emission model and 100 realisations with 100 bootstrap resamplings in each case. The QSO number has been computed following \citet{DESICollaboration2016}, which assumes a total of 700,000 QSOs\footnote{This number of QSOs is smaller than the number of Ly$\alpha$ QSOs used for the first DESI BAO analysis by almost 10,000 \citep{DESICollaboration2024}, although these comprise a lower redshift range ($1.77 < z < 4.16$).} for the full DESI survey at $z_{\rm min\,QSO} = 2.1$. The $\Delta \langle L_{\rm Ly\alpha} \rangle$ values in \cref{fig:snr_vs_z} increase with $z_{\rm min\,QSO}$ up until $z_{\rm min\,QSO} = 2.5$ (since we remove the QSOs with the least redshift overlap with the $g$-band, and thus the least correlated to patch flux). At higher $z_{\rm min\,QSO}$, the $\Delta \langle L_{\rm Ly\alpha} \rangle$ value steadily decreases (because at that point, increasing $z_{\rm min \, QSO}$ removes the most overlapping/correlated QSOs). 

The SNR, however, peaks just at $z_{\rm min\,QSO} = 2.2$, both in the noiseless case and for 10\% BASS noise. At $z_{\rm min\,QSO} > 2.2$, removing the least overlapping/correlated QSOs in exchange for reducing our sample does not increase  $\Delta \langle L_{\rm Ly\alpha} \rangle$ SNR. This behaviour is to be expected, since the QSO redshift distribution is monotonically decreasing in our redshift range, and by removing QSOs starting by the lowest redshift we significantly reduce our sample, and thus the SNR of $\Delta \langle L_{\rm Ly\alpha} \rangle$. Hence, we choose $z_{\rm min\,QSO} = 2.2$ for our study, which results in 579,183 DESI QSOs  \citet{DESICollaboration2016}.

\subsubsection{Evolution of $\Delta \langle L_{\rm Ly\alpha} \rangle$ versus $\langle L_{\rm Ly\alpha} \rangle$ and $b_{\rm e}$}\label{sec:Evolution of estimator vs lya parameters}

\begin{figure}
 	\includegraphics[width=\columnwidth]{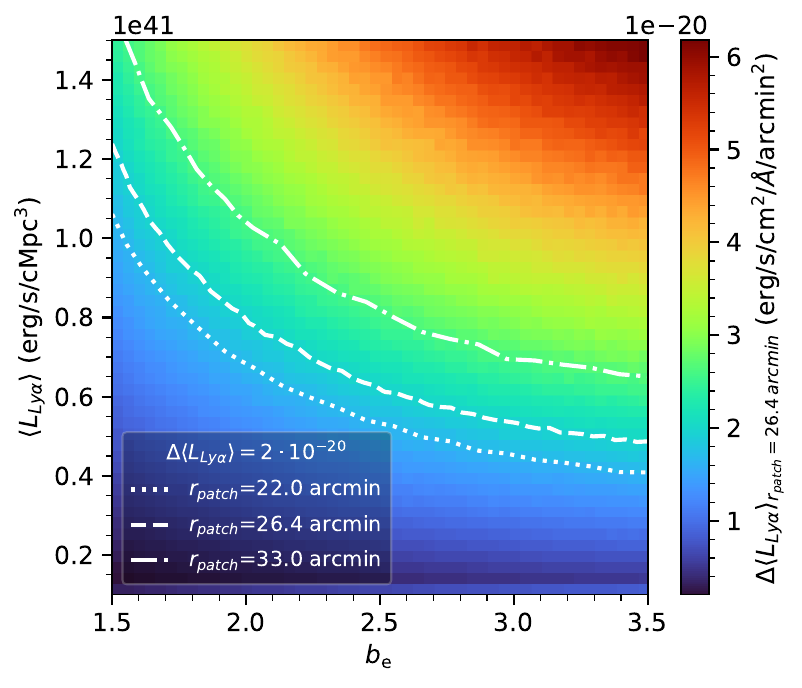}
     \caption{Colormap of $\Delta \langle L_{\rm Ly\alpha} \rangle$ values for the ranges of $b_{\rm e}$ of our model and  $\langle L_{\rm Ly\alpha} \rangle$ we deem realistic, for $r_{\rm patch}=26.4$ arcmin. The isolines are for the value $\Delta \langle L_{\rm Ly\alpha} \rangle = 2\cdot 10^{\rm -20}$ erg/s/cm$^2$/\AA/arcmin$^2$ for $r_{\rm patch}=22$ arcmin (dotted line), $r_{\rm patch}=26.4$ arcmin (dashed line) and $r_{\rm patch}=33$ arcmin (dash-dotted line).}
     \label{fig:lya_diff_map_with_isolines}
\end{figure}

After determining the optimal $r_{\rm patch}$ and $z_{\rm QSO\, min}$ to maximise SNR, we show in \cref{fig:lya_diff_map_with_isolines}  how our statistic $\Delta \langle L_{\rm Ly\alpha} \rangle$ evolves in the $\langle L_{\rm Ly\alpha} \rangle$, $b_{\rm e}$ parameter space of our Ly$\alpha$ emission model. Here, for every pixel on the colormap we computed our forecast without BASS noise nor systematics, using the $\langle L_{\rm Ly\alpha} \rangle$, $b_{\rm e}$ parameter values corresponding to the pixel coordinates (shown on \cref{fig:lya_diff_map_with_isolines} axis), instead of the "fiducial" model used so far. For all $\langle L_{\rm Ly\alpha} \rangle$, $b_{\rm e}$ pairs we used the same QSO realisation, by choosing the exact same pixels of our Ly$\alpha$ and noise intensity maps for all cases. This was done to control that source of variability. Furthermore, we did not resample via bootstrapping because the SNR is high enough (SNR$\sim 100$) for the bootstrapping variability to become negligible.

From \cref{fig:lya_diff_map_with_isolines}, we can see that $\Delta \langle L_{\rm Ly\alpha} \rangle$ increases for both higher values of $\langle L_{\rm Ly\alpha} \rangle$ and $b_{\rm e}$. These trends are in line with our Ly$\alpha$ emission model (\cref{eq:lya_emission_cosmo}):  $\langle L_{\rm Ly\alpha} \rangle$ is simply a multiplicative factor, and the bias $b_{\rm e}$ increases the contrast between underdense and overdense regions. The \cref{fig:lya_diff_map_with_isolines} colormap also exhibits an almost radial symmetry centred at the highest $\langle L_{\rm Ly\alpha} \rangle$, $b_{\rm e}$ corner (upper right). Therefore, a given measured value of $\Delta \langle L_{\rm Ly\alpha} \rangle$ just constrains a curve in $\langle L_{\rm Ly\alpha} \rangle$, $b_{\rm e}$ space. Nevertheless, it is worth examining if we can break the $\langle L_{\rm Ly\alpha} \rangle$-$b_{\rm e}$ degeneracy by measuring the value of $\Delta \langle L_{\rm Ly\alpha} \rangle$ for different patch radii.

This question can be answered by examining the isolines plotted over the colormap in \cref{fig:lya_diff_map_with_isolines}. These correspond to the arbitrary value of $\Delta \langle L_{\rm Ly\alpha} \rangle = 2\cdot 10^{\rm -20}$ erg/s/cm$^2$/\AA/arcmin$^2$, for the optimal $r_{\rm patch}$ of 26.4 arcmin, as well as for patch radii of 22 and 33 arcmin. Increasing (decreasing) patch radius just offsets the $\langle L_{\rm Ly\alpha} \rangle$-$b_{\rm e}$ curve away (towards) the upper right corner (the "centre" of radial symmetry). Interestingly, curves corresponding to different $r_{\rm patch}$ do not intersect. Consequently, modifying $r_{\rm patch}$ only offsets our  $\Delta \langle L_{\rm Ly\alpha} \rangle$ by a given value, without providing additional constraints for either  $\langle L_{\rm Ly\alpha} \rangle$ or $b_{\rm e}$.

Regardless, this degeneracy is not an issue, since we have developed the $\Delta \langle L_{\rm Ly\alpha} \rangle$ estimator assuming a low SNR scenario with a significant detection for a single $r_{\rm patch}$ at most. If the  $\Delta \langle L_{\rm Ly\alpha} \rangle$ SNR with an eventual observational study was high enough to constrain $\Delta \langle L_{\rm Ly\alpha} \rangle$ in several radial bins, the angular power spectrum could be computed instead, which could be used to constrain $b_{\rm e}$ alone (as galaxy bias has been constrained with e.g., weak lensing in \citealp{Dvornik2017}, or CMB lensing in \citealp{Alonso2021}). 

Therefore, our forecast for DESI in the following subsection will only focus on the $\Delta \langle L_{\rm Ly\alpha} \rangle$ SNR we expect for the optimal $r_{\rm patch}$ and $z_{\rm min\, QSO}$, as well as the upper bounds that may be placed on $\langle L_{\rm Ly\alpha} \rangle$ and $b_{\rm e}$ in the case of a non-detection. It is worth noting that, following the isolines in \cref{fig:lya_diff_map_with_isolines}, smaller $r_{\rm patch}$ values result in a $\Delta \langle L_{\rm Ly\alpha} \rangle$ estimator more sensitive to $b_{\rm e}$: a given variation in $b_{\rm e}$ results in a larger variation in $\Delta \langle L_{\rm Ly\alpha} \rangle$. Hence, if the SNR is high enough for a confident detection with the optimal $r_{\rm patch}=26.4$ arcmin, exploring smaller $r_{\rm patch}$ may yield better constraints in the $\langle L_{\rm Ly\alpha} \rangle$-$b_{\rm e}$ space.

\subsubsection{Additional remarks}\label{sec:additional remarks}

Before moving on, here we shortly comment on two relevant validations we performed while evaluating the behaviour of $\Delta \langle L_{\rm Ly\alpha} \rangle$. First, for a number of QSOs smaller than the number of pixels of our Ly$\alpha$ simulated image (\cref{fig:bass_im_lya_img}), selecting adjacent QSO pixels (in a "compact" configuration") or sparse pixels from the whole image (which would be closer to a real QSO distribution) did not make a significant difference for the $\Delta \langle L_{\rm Ly\alpha} \rangle$ value or SNR. Therefore, an arbitrarily wide survey with an arbitrary large number of QSOs can be simulated by just drawing enough Ly$\alpha$ patch-noise patch combinations, with the only limit being the cosmic variance intrinsic to the hydrodynamic simulation.

Second, the BASS intensity map we use to model all noise sources together might actually lead to a SNR underestimation. The real Ly$\alpha$ LSS signal is already contained in that map, but uncorrelated to the simulated Ly$\alpha$ signal of our hydrodynamic simulation, so it is an additional source of noise that will not be present in an eventual observational study. To evaluate this effect, we computed  $\Delta \langle L_{\rm Ly\alpha} \rangle$ and its SNR by adding the flux of an uncorrelated Ly$\alpha$ patch to each Ly$\alpha$ patch (i.e., from a randomly drawn QSO that does not correspond to the QSO used to extract the Ly$\alpha$ fores data). This uncorrelated Ly$\alpha$ signal decreased SNR by 30\%-40\% when no extra noise sources were added. Nevertheless, when a fraction of the BASS noise was added to our forecast, the SNR decrease became negligible, i.e.: the SNR was virtually the same for the 10\% BASS noise cases in \cref{fig:snr_vs_radius} and \cref{fig:snr_vs_z}. Hence, while having a real, uncorrelated Ly$\alpha$ signal in our noise model makes our forecast more conservative, the effect we expect in our SNR predictions with real BASS noise is negligible.

\subsection{Results for DESI/DESIx2}\label{sec:Results for DESI/DESIx2}

We computed our final forecast for two surveys: DESI, as originally laid out in \citet{DESICollaboration2016} (with 579,183 QSOs at $z>2.2$ considered in our study), and a hypothetical DESI phase II extension which doubles the amount of observed Ly$\alpha$ QSOs (1,158,366 for our study). The exact characteristics of DESI phase II are not disclosed/decided at the time of writing (to the best of the authors knowledge), hence we will simply dub our hypothetical extension as DESIx2. Simply increasing the number of QSOs does not consider other factors that influence our forecast. For example, the characteristics of the overlapping imaging surveys, or a likely change in the QSO redshift distribution towards higher redshift stemming from deeper target selection. Regardless, doubling the number of QSOs is a reasonable order-of-magnitude estimate of what an eventual DESI extension may yield. For example, \citealp{Schlegel2022} proposes observing an additional extra 1.1 million LAEs at $z > 2.3$, although it does not specify how well the Ly$\alpha$ forest could be extracted from the spectra of those sources.

\begin{figure*}
 	\includegraphics[width=\textwidth]{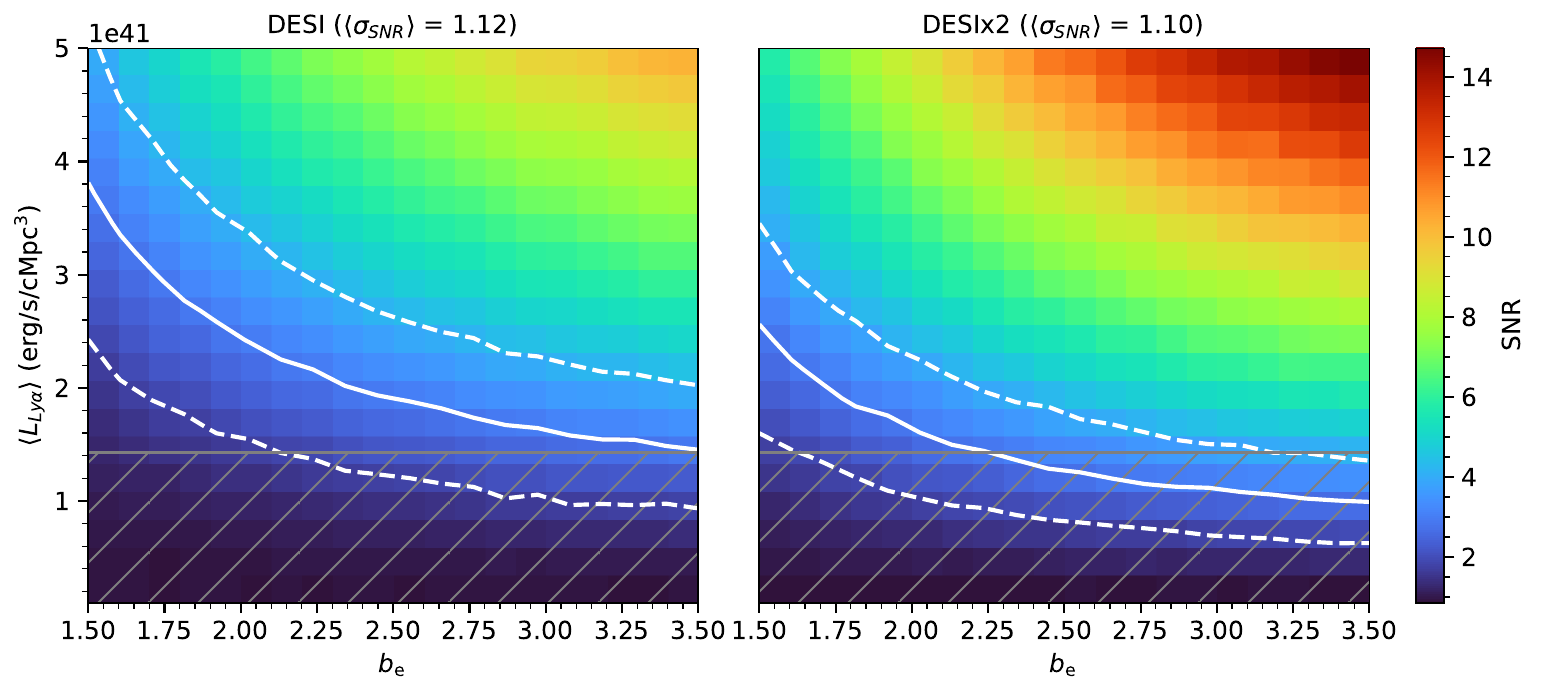}
     \caption{Average SNR of $\Delta \langle L_{\rm Ly\alpha} \rangle$ for our DESI forecast (left) and a hypothetical DESIx2 extension (right), versus $\langle L_{\rm Ly\alpha} \rangle$ and $b_{\rm e}$ of our Ly$\alpha$ emission model. The SNR is derived from 100 bootstrap resamplings, and for each $\langle L_{\rm Ly\alpha} \rangle$-$b_{\rm e}$ case 100 QSO realisations were computed. Solid white line is the isoline for average SNR = 3 (the average detection limit), while dashed white lines are the $1\sigma$ confidence interval of the detection limit. Grey hatched area is the $\langle L_{\rm Ly\alpha} \rangle$ range we deem realistic.}
     \label{fig:DESI_SNR_maps}
\end{figure*}

%For both DESI and DESIx2, our forecast will focus on the SNR of $\Delta \langle L_{\rm Ly\alpha} \rangle$: how realistic is to expect a detection of Ly$\alpha$ LSS (i.e.,  $\Delta \langle L_{\rm Ly\alpha} \rangle$ SNR > 3), and what upper bounds can be derived from a non-detection.
\Cref{fig:DESI_SNR_maps} shows the $\Delta \langle L_{\rm Ly\alpha} \rangle$ average SNR versus the parameters of our Ly$\alpha$ emission model ($\langle L_{\rm Ly\alpha} \rangle$, $b_{\rm e}$). We have explored a larger range of  $\langle L_{\rm Ly\alpha} \rangle$ than the limit we considered realistic (up to $5 \cdot 10^{\rm 41}$ erg/s/cMpc$^3$ instead of $1.5 \cdot 10^{\rm 41}$ erg/s/cMpc$^3$): the credible $\langle L_{\rm Ly\alpha} \rangle$ range (up to $1.5 \cdot 10^{\rm 41}$ erg/s/cMpc$^3$) is hatched in grey. For all simulated cases, the 100 realisations of the forecast were computed, with 100 bootstrap resamplings each.

The solid white line in \cref{fig:DESI_SNR_maps} represents the detection limit (i.e. average SNR = 3), while the two dashed white lines are its $\pm\sigma$ contours. By examining the colormaps on \cref{fig:DESI_SNR_maps}, the SNR dependence with Ly$\alpha$ tracer bias $b_{\rm e}$ becomes evident: for both DESI and DESIx2 a detection seems very unlikely for $b_{\rm e}=1.5$ (the SNR = 3 contour lies above the grey-hatched area by more than 1$\sigma$)). On the other hand, for $b_{\rm e} = 3.5$ a detection appears plausible for DESI (the SNR = 3 contour sits just at the limit of the realistic $\langle L_{\rm Ly\alpha} \rangle$ region). For DESIx2, a detection becomes highly likely, with the SNR = 3 line $1\sigma$ below the limit of the grey hatched area.

\begin{figure}
 	\includegraphics[width=\columnwidth]{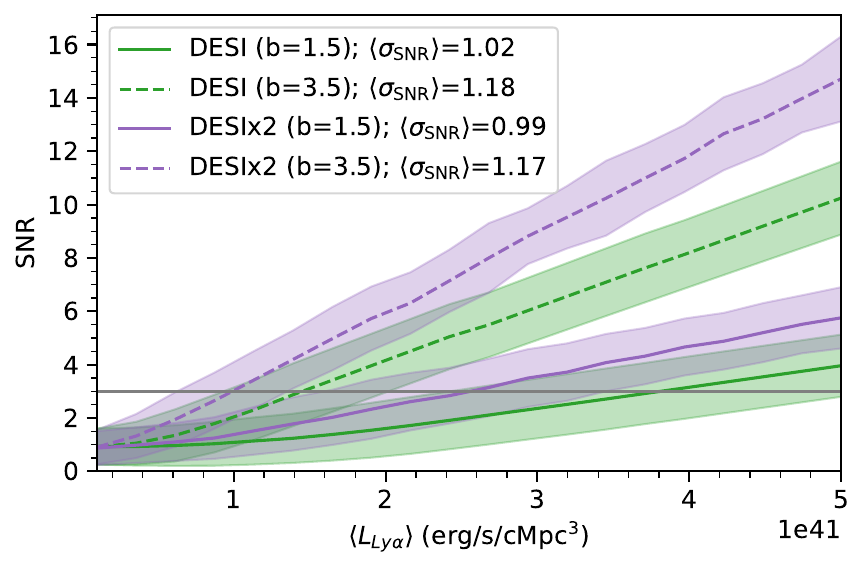}
     \caption{Mean SNR and the $1\sigma$ confidence interval (coloured area), versus $\langle L_{\rm Ly\alpha} \rangle$, for DESI (green) and DESIx2 (purple). Solid lines represent the cases with the minimum bias of our model ($b_{\rm e}=1.5$), and dashed lines the maximum bias ($b_{\rm e}=3.5$). Horizontal grey line represents the SNR = 3 detection threshold.}
     \label{fig:DESI_SNR_vs_lya_lum}
\end{figure}

To better understand the SNR variability between forecast realisations, in \cref{fig:DESI_SNR_vs_lya_lum} the average SNR and its $\pm\sigma$ range are displayed for both DESI and DESIx2, for the highest and lowest bias ($b_{\rm e}=3.5$ and $b_{\rm e}=1.5$ respectively). For all cases, the SNR variability (i.e. $\sigma_{\rm \rm SNR}$) remains approximately constant versus $\langle L_{\rm Ly\alpha} \rangle$, except in the lowest SNR regime (SNR $\sim$ 1), which stems from the SNR distribution becoming progressively more skewed (as SNR is strictly positive). Higher bias $b_{\rm e}$ increases the SNR and its variability (from $\sigma_{\rm \rm SNR} \sim 1$ at $b_{\rm e} = 1.5$ to $\sigma_{\rm \rm SNR} \sim 1.2$ at $b_{\rm e} = 3.5$), but a larger survey footprint or number of QSOs seems to increase the SNR while leaving $\sigma_{\rm SNR}$ largely unaffected. Regardless, the most relevant result of \cref{fig:DESI_SNR_vs_lya_lum} is that the SNR variability is only significant in the low SNR regime, namely: when a detection seems uncertain. For an average SNR = 3 we find $\sigma_{\rm SNR} \sim 1$, which means that a detection is plausible, yet unsure. Also, any upper bounds placed by a non-detection will have significant errors intervals. On the other hand, for a hypothetical $\Delta \langle L_{\rm Ly\alpha} \rangle$ SNR = 10, which could happen if the total $\langle L_{\rm Ly\alpha} \rangle$ was heavily underestimated in the literature, or if a deeper imaging survey was used, we still find $\sigma_{\rm SNR} \sim 1$.  On such a high SNR regime, the intrinsic variability between forecast realisations (and between QSO positions and imaging data with real observations) becomes subdominant.

Hence, for a given value $\langle L_{\rm Ly\alpha} \rangle$, there are two sources of variability for the expected SNR of our forecast: the (unknown) real value of the bias $b_{\rm e}$, and the intrinsic variability between forecast realisations of $\Delta \langle L_{\rm Ly\alpha} \rangle$. Nevertheless, we can still place upper bounds on $\langle L_{\rm Ly\alpha} \rangle$ for a non-detection of $\Delta \langle L_{\rm Ly\alpha} \rangle$. For each ($\langle L_{\rm Ly\alpha} \rangle_n$, $b_{\rm e \, n}$) case evaluated in \cref{fig:DESI_SNR_maps} (i.e. each pixel $n$ on the colormap), the fraction of forecast realisations with SNR < 3 is the probability of not detecting $\Delta \langle L_{\rm Ly\alpha} \rangle$ for that given ($\langle L_{\rm Ly\alpha} \rangle_n$, $b_{\rm e \, n}$). We can also interpret it as the probability of the real value of $\langle L_{\rm Ly\alpha} \rangle$, $b_{\rm e}$ being equal or lower to $\langle L_{\rm Ly\alpha} \rangle_n$, $b_{\rm e \, n}$, given a non-detection. In other words, the cumulative distribution function $P(\langle L_{\rm Ly\alpha} \rangle \leq \langle L_{\rm Ly\alpha} \rangle_n, b_{\rm e}\leq b_{\rm e \, n} \, | \, {\rm SNR} < 3)$. Therefore, by differentiating this cumulative distribution with respect of $\langle L_{\rm Ly\alpha} \rangle$ and $b_{\rm e}$, we obtain $P(\langle L_{\rm Ly\alpha} \rangle_n, b_{\rm e \, n}\, | \, {\rm SNR} < 3)$, the probability distribution of the upper bounds of ($\langle L_{\rm Ly\alpha} \rangle$, $b_{\rm e}$).

\begin{figure}
 	\includegraphics[width=\columnwidth]{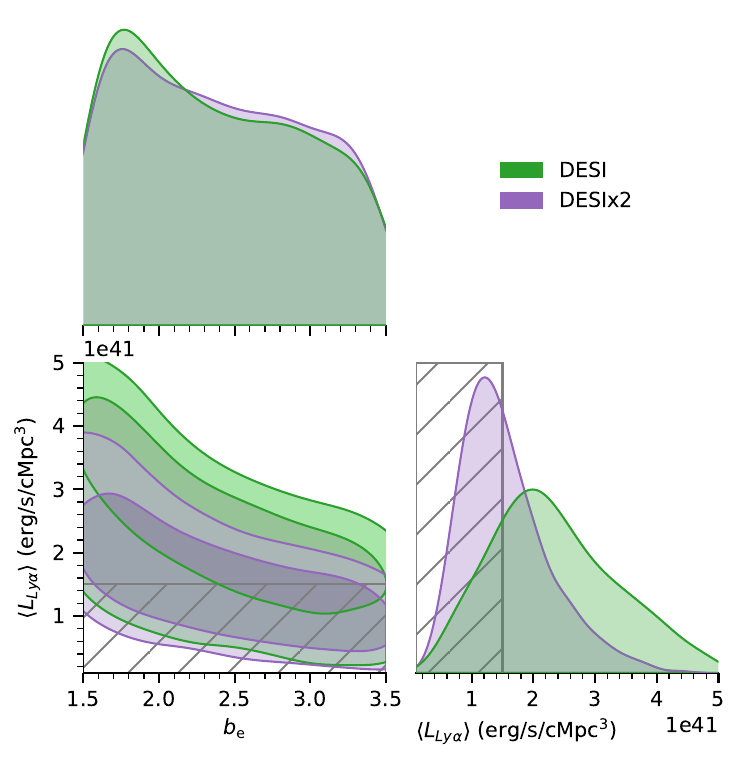}
     \caption{Triangle plot of the upper bounds on $\langle L_{\rm Ly\alpha} \rangle$ and $b_{\rm e}$ placed by a non-detection ($\Delta \langle L_{\rm Ly\alpha} \rangle$ SNR < 3) for DESI (green) and DESIx2 (purple). Contours in the bottom left panel correspond to $1\sigma$ and $2\sigma$ constraints, while the hatched areas indicate the range of $\langle L_{\rm Ly\alpha} \rangle$ we deem realistic.}
     \label{fig:DESI_triangle_plot}
\end{figure}

\Cref{fig:DESI_triangle_plot} displays this probability distribution of the upper bounds, together with the marginalised probability distributions for $\langle L_{\rm Ly\alpha} \rangle$ and $b_{\rm e}$. \Cref{tab:lya_upper_bounds} contains the mean and $\sigma$ of the marginalised upper bounds, together with the probability of detection for an actual observational study ( $\Delta \langle L_{\rm Ly\alpha} \rangle$  SNR > 3). This probability of detection was computed by integrating  $P(\langle L_{\rm Ly\alpha} \rangle, b_{\rm e}\, | \, {\rm SNR} < 3)$ over the whole bias range of our Ly$\alpha$ emission model, and over the credible range of $\langle L_{\rm Ly\alpha} \rangle$ values (assuming a flat prior) and then normalising by the integral of $P(\langle L_{\rm Ly\alpha} \rangle, b_{\rm e}\, | \, {\rm SNR} < 3)$ over the whole ($\langle L_{\rm Ly\alpha} \rangle$, $b_{\rm e}$) parameter space sampled in \cref{fig:DESI_triangle_plot}.

From both \cref{fig:DESI_triangle_plot} and \cref{tab:lya_upper_bounds} a clear conclusion can be drawn: our methodology can impose increasingly tight upper bounds on $\langle L_{\rm Ly\alpha} \rangle$, but it has little constraining power for the tracer bias $b_{\rm e}$. Indeed, compared to DESI the marginalised $\langle L_{\rm Ly\alpha} \rangle$ distribution for DESIx2 is significantly narrower and shifted towards lower $\langle L_{\rm Ly\alpha} \rangle$ values, while the marginalised $b_{\rm e}$ distribution remains effectively the same, namely: a rather flat distribution across the sampled $b_{\rm e}$ range, with some preference for lower $b_{\rm e}$ values (given that a detection is less likely for lower $b_{\rm e}$).% However, if the SNR was high enough to significantly detect $\Delta \langle L_{\rm Ly\alpha} \rangle$ in two or more radial bins, the shape of the angular power spectrum could be constrained instead, which would give an independent measurement of $b_{\rm e}$ alone.

\begin{table}
\centering
	\caption{Marginalised upper bounds for $\langle L_{\rm Ly\alpha} \rangle$ and $b_{\rm e}$ for a non-detection in our forecast, together with probability of detecting Ly$\alpha$ LSS (SNR $\Delta \langle L_{\rm Ly\alpha} \rangle> 3$).}
 	\label{tab:lya_upper_bounds}
 	\begin{tabular}{rccc}
 		\hline
 		 &  $\langle L_{\rm Ly\alpha} \rangle$ ($10^{\rm 40}$ erg/s/cMpc$^3$)&$b_{\rm e}$&$P_{\rm \rm detect}$ (\%)\\
 		\hline
 		DESI& $23 \pm 10$&$2.39 \pm 0.56$&23.95\\
        DESIx2& $15.5 \pm 7.0$&$2.40 \pm 0.56$&54.93\\
		\hline
 	\end{tabular}
 \end{table}

Regarding the $\langle L_{\rm Ly\alpha} \rangle$ upper bounds in \cref{tab:lya_upper_bounds}, a non-detection in DESI would predict them to be approximately a factor of $\sim$ 1.5 times larger than the most optimistic literature estimates ($2.3 \pm 1.0 \cdot 10^{\rm 41}$ erg/s/cMpc$^3$ versus $1.5 \cdot 10^{\rm 41}$ erg/s/cMpc$^3$). On the  other hand, for the hypothetical DESIx2, the upper bounds are extremely similar to said estimates ($1.55 \pm 0.73 \cdot 10^{\rm 41}$ erg/s/cMpc$^3$). This is also reflected in the probability of a detection: going from just 23.95\% in DESI to 54.93\% in DESIx2. Consequently, we can claim that even with DESI alone the upper  bounds on $\langle L_{\rm Ly\alpha} \rangle$ are of the same order of magnitude of the state-of-the-art LFs of faint LAEs \citep{Drake2017}, and with DESIx2 a proper detection of Ly$\alpha$ LSS in $g$ band images becomes quite plausible. In the following section, we will discuss more in-depth the implications of these results, and in \cref{sec:Forecast extrapolation to DESI-LSST} we show the results of a extrapolation of this forecast to the cross-correlation of DESI with LSST \citep{Ivezic2019}.

\section{Discussion}\label{sec:Discussion}

The results of our forecast are moderately optimistic for DESI: while a detection of Ly$\alpha$ LSS is uncertain, the upper bounds that we derive from a non-detection are competitive. Here we discuss the interpretation of the results presented here, as well as the possible ways to improve our analysis and results in future works, either by improving the SNR of an eventual study with real DESI-DECaLS/BASS data, or tightening the constraints on $\langle L_{\rm Ly\alpha} \rangle$ and $b_{\rm e}$. We also comment on the results and science that can be expected from upcoming imaging/spectroscopic surveys.

\subsection{Extrapolation to number of observed QSOs}\label{sec:Extrapolation to number of observed QSOs}

he forecast presented in this work is intentionally conservative at every stage; whenever a decision was necessary, we consistently opted for  a reasonable and cautious approach. For example, imaging data was modelled using BASS instead of DECaLS, which is slightly shallower, and thus noisier \citep[23.95 5$\sigma$ $g$-band PSF depth versus 23.65,][]{Dey2019}, and public SDSS BOSS mocks were used to model the Ly$\alpha$ forest noise \citep{Bautista2015}, which are bound to be noisier than DESI observations. In fact, DECaLS has a $g$-band magnitude limit 0.3 mag deeper than BASS \citep{Dey2019}; extrapolating the imaging noise map with a very simple approximation (\cref{eq:noise_factor_LSST}) we can expect a reduction in noise (and thus SNR increase) of $\sim$25\%. Given that DECaLS covers 9,000 deg$^2$ of the DESI footprint, and BASS 5,000 deg$^2$, a total SNR increase of $\sim$15\% can be expected (assuming the total SNR is the average of DECaLS and BASS, weighted by footprint size). This simple extrapolation confirms that our estimations are conservative, although a SNR increase of 15\% does not change the general conclusions of this work.

The image reduction procedure to mitigate photometric noise was also the most rudimentary we could perform while ensuring that the Ly$\alpha$ emission signal was preserved (subtraction of average sky per pointing and resolved foreground subtraction, \cref{sec:Photometric noise: BASS intensity map and image reduction}). We did not assume more sophisticated image reduction or foreground removal methods that might be developed in future work. Therefore, the results presented throughout \cref{sec:Results for DESI/DESIx2}, and especially in \cref{tab:lya_upper_bounds}, are to be taken as conservative estimates.

However, we can easily update our estimates for the actual number of Ly$\alpha$ QSOs contained in an eventual DESI Y5 data release. This number will exceed the baseline projection from \citet{DESICollaboration2016}. Moreover, a extension/second phase of DESI will not be an exact duplicate of its currently ongoing first phase, as assumed here; an extension reaching a total footprint of 18,000 deg$^2$ is is currently more likely (instead of 28,000 deg$^2$ assumed for DESIx2).

On the other hand, for the first phase of DESI there is reason for optimism: current QSO densities observed with DESI are approximately 20\% higher than the requirements in \citet{DESICollaboration2016} \citep{Chaussidon2022}. Regardless, the $\langle L_{\rm Ly\alpha} \rangle$ upper bounds scale well with the inverse of the square root of the number of QSOs: in \cref{tab:lya_upper_bounds} both the upper bound derived for DESIx2 and its error are approximately the DESI upper bounds divided by $\sqrt{2}$. Hence, the upper bounds on $\langle L_{\rm Ly\alpha} \rangle$ for a given number of observed DESI QSOs $N_{\rm \rm QSO\, obs}$ can be rescaled from our DESI forecast simply with

\begin{equation}\label{eq:upper_bounds_scaling}
    \langle L_{\rm Ly\alpha} \rangle_{\rm \rm UB\, obs} = (2.3 \pm 1.0) \cdot \sqrt{\frac{579,183}{N_{\rm \rm QSO\, obs}}} \cdot 10^{\rm 41} \, {\rm erg/s/cMpc^3},
\end{equation}

where the $\langle L_{\rm Ly\alpha} \rangle_{\rm \rm UB\, obs}$ is the observational upper bound, and the numerical values correspond to the upper bound and number of QSOs of our work. Following this extrapolation, just the increase in 20\% of QSO density shown in \citet{Chaussidon2022} already lowers the $\langle L_{\rm Ly\alpha} \rangle$ upper bounds of our forecast by $\sim$9\%. If we consider a DESI phase-II with a total footprint of 18,000 deg$^2$ in addition to the QSO density improvement from \citet{Chaussidon2022}, the total QSO number would increase by 54\% compared to our DESI forecast, lowering the $\langle L_{Ly\alpha} \rangle$ upper bounds by approximately 20\%.

The extrapolation presented in \cref{eq:upper_bounds_scaling} would only work as long as observations were to be performed in an area with $g$-band data of similar depth to DECaLS/BASS. In case a DESI-phase II target selection was to be carried out with a significantly deeper survey, our forecast would have to be recomputed with images from such survey.

\subsection{Future work}\label{sec:Future work}

This work demonstrates that detecting LSS in Ly$\alpha$ emission within a $g$-band intensity map is feasible with DESI, particularly with a potential phase II. Even if no LSS is detected, the upper bounds on Ly$\alpha$ emission that can be established are comparable to the most optimistic estimates in the literature. Nevertheless, two areas of further research could substantially enhance this projections: improved image reduction/foreground subtraction and more sophisticated simulations of the Ly$\alpha$ emission and absorption fields. We also expect that future work will use larger simulation boxes, computed with state-of-the-art cosmological models \citep[e.g, MillenniumTNG][]{Hernandez-Aguayo2023}.

\subsubsection{Foreground removal and image reduction}\label{sec:Foreground removal and image reduction}

Foreground subtraction has arguably been the most important observational challenge for 21 cm IM studies, as the resulting intensity maps are not dominated by 21 cm emission of HI LSS, but synchrotron and free-free emission of galactic origin \citep{Santos2005, Jelic2010}. Several methods have been developed to subtract this unwanted signal, from polynomial fitting \citep[e.g.,][]{Ansari2012} to more sophisticated statistical methods such as Principal Component Analysis \citep{Alonso2015, Zuo2023}, or machine learning approaches like Gaussian Processes \citep{Soares2022}. All these approaches rely on the foreground emission being smoother in frequency than the 21 cm cosmic signal; the foreground-contaminated region in Fourier space is usually called the "foreground wedge" \citep{Liu2014}.

In this study, it is evident that the Ly$\alpha$ intensity map derived from $g$-band images is predominantly dominated by components other than the Ly$\alpha$ signal. The chequered pattern in \cref{fig:bass_im_lya_img} roughly matches the CCD size of the BASS camera, which clearly points to the dominant component of the intensity map being of instrumental/atmospheric origin. Moreover, in \citet{Renard2021} it was already proven that cosmic foregrounds are subdominant for similar imaging depths: for a Ly$\alpha$ forest-narrow band imaging cross-correlation on 100 deg$^2$ (eBOSS/DESI-PAUS, with magnitude limit $i = 23$), cosmic foregrounds alone would still allow a clear detection of the 2PCF in several radial bins, while adding instrumental noise would make a detection unfeasible.

Hence, we are faced with a fundamentally different problem than foreground subtraction for 21 cm: instead of modelling and removing a foreground intensity map of mostly galactic origin (i.e., a scalar field defined in RA, dec, and z coordinates), we have to model and subtract a series of signals that will not depend on cosmic coordinates, but on observation time, atmospheric and instrumental conditions. It is worth noting that observations of the 21 cm line is also contaminated by noise of atmospheric origin \citep[ionospheric effects,][]{Shen2021} or even man-made, dependant on observation time \citep[radio-frequency intereference,][]{Wilensky2023}, but these remain subdominant.

The mitigation of instrumental systematics (mainly the CCD-to-CCD variability in \cref{fig:bass_im_lya_img}) and undesired emission of atmospheric/lunar/zodiacal origin (e.g., sky gradients remaining in \cref{fig:bass_im_lya_img}) could be tackled separately. For example, detailed sky emission models already exist for certain observatories \citep[e.g.,][]{Noll2012}, albeit they are more focused on modelling sky spectra. The development and subtraction of mock sky images based on similar models could remove part of the undesired sky foregrounds. Reducing the CCD-to-CCD variability, on the other hand, should be mostly attained via the improvement of image reduction pipelines.

%Another promising approach would be to both remove instrumental and sky emission by generating images without any cosmic component and subtracting them. This could be achieved with deep learning: training a neural network to generate telescope images using as a training dataset images of the considered survey (e.g., DECaLS and BASS ) downsampled to intensity map resolution ($\sim$1 arcmin in this work), but using as labels only the image metadata relevant to atmospheric/instrumental condition (e.g., airmass, moon phase/location, seeing, camera temperature). If the image metadata used as labels is carefully chosen, the neural network would effectively be unable to infer the RA and dec of a given image, and thus would be "blind" to any signal of cosmic origin. Hence, the generated images could not contain any Ly$\alpha$ emission, and subtracting them from real images could only suppose a net increase of the SNR of $\Delta \langle L_{\rm Ly\alpha} \rangle$ (or any other estimator employed), regardless of how well the network is able to predict real images based on metadata.

Therefore, we can confidently state that there is a margin for improvement on the SNR side and on the general previsions of our forecast. This is especially true if research is carried out on image reduction/foreground subtraction. However, we cannot quantify the extent of this improvement without further work.

\subsubsection{Modelling the Ly$\alpha$ emission field}\label{sec:Modelling the Lya emission field}

In this work, we have used a simple two-parameter Ly$\alpha$ emission model that allows tuning the total average cosmic Ly$\alpha$ emission, $\langle L_{\rm Ly\alpha} \rangle$, and its bias as a cosmological tracer, $b_{\rm e}$, while respecting within $\sim$1 dex the constraints imposed by the brightest QSOs and Ly$\alpha$ nebulae on the bright end (\cref{sec:Lya emission model} and \cref{sec:Lya emission model cutoff}). While this model generates a Ly$\alpha$ emission field with the desired features, it is agnostic when it comes to the relationship between $\langle L_{\rm Ly\alpha} \rangle$ and $b_{\rm e}$: it allows for any combination of both, as long as $b_{\rm e}$ is within the range it has been tuned to work.

Nevertheless, not all combinations of $\langle L_{\rm Ly\alpha} \rangle$ and $b_{\rm e}$ are plausible or allowed by a physically-motivated model. In \cref{fig:bias_values} we showed that QSOs are significantly more biased ($b_{\rm e} \gtrsim 3.5$) than observed LAE samples ($b_{\rm e} \sim 1.5$). However, the total Ly$\alpha$ luminosity contribution of QSOs is fairly well constrained, given their brightness \citep[e.g.,][]{Spinoso2020, Liu2022, Torralba-Torregrosa2023}, while the faint of the LAE luminosity function is just hypothesised from ultra-deep observations of cosmic filaments \citep{Bacon2021}. Hence, if the real value of $\langle L_{\rm Ly\alpha} \rangle$ corresponds to the most optimistic estimates ($1.5\cdot 10^{\rm 41}$ erg/s/cMpc$^3$, \cref{fig:lya_avg_values}), the reason for it will be that the bulk of Ly$\alpha$ luminosity is produced by the currently undetected faint end of the LAE LF of \citet{Drake2017}, plus diffuse fluorescent/resonant emission like the one sampled by \citet{Croft2016} and \citet{Lin2022}.

Therefore, a high value of $\langle L_{\rm Ly\alpha} \rangle$ will push the tracer bias of Ly$\alpha$ emission to the low end ($b_{\rm e}\sim1.5$). Conversely, a low value of $\langle L_{\rm Ly\alpha} \rangle$ will necessarily imply a higher bias, since QSOs, the most biased tracers, are making a significant contribution to the total Ly$\alpha$ luminosity budget. From this reasoning it follows that there must be a "forbidden" high-$\langle L_{\rm Ly\alpha} \rangle$ high-$b_{\rm e}$ zone in our parameter space, namely: a region where $\langle L_{\rm Ly\alpha} \rangle$ is high enough to be mostly produced by faint LAEs and diffuse emission, but a $b_{\rm e}$ so high to also require a significant contribution from QSOs.

Such "forbidden zone" would restrict our $\langle L_{\rm Ly\alpha} \rangle$-$b_{\rm e}$  parameter space, and would be represented by a masked area in the upper right corner of \cref{fig:DESI_SNR_maps}. This restriction has remarkable implications for our results, since we would not integrate over the whole $\langle L_{\rm Ly\alpha} \rangle$-$b_{\rm e}$  space sampled in \cref{fig:DESI_triangle_plot}. First, the upper bounds on  $\langle L_{\rm Ly\alpha} \rangle$ displayed in \cref{tab:lya_upper_bounds} would tighten, as we marginalise over a smaller $b_{\rm e}$ range. Second, the detection probabilities in \cref{tab:lya_upper_bounds} would be likely to decrease, given that we are removing the highest SNR region of our parameter space. Regardless, a physically-motivated Ly$\alpha$ emission model that considers all possible sources (from QSOs to fluorescent IGM emission) is out of the scope of this work, and without such a model, any speculation of the shape of this "forbidden zone" would be just a guess.

With a physically motivated (or at least more developed) Ly$\alpha$ emission model, we may also improve the SNR of an eventual Ly$\alpha$ IM study. In this work, we have integrated the $g$-band patches with uniform radial weighting —\cref{eq:patch_flux}—, but for a given Ly$\alpha$ emission model and a given radial length, there must be a radial kernel ($W(r)$ in \cref{eq:patch_flux_kernel}) that maximises the SNR of $\Delta \langle L_{\rm Ly\alpha} \rangle$. Convolution kernels that increase the SNR of a given statistic are already used in cosmology., i. e., in CMB lensing by weighting the redshift convolution by the inverse of the variance \citep[e.g.,][]{Qu2024}, or in CMB polarisation by weighting more the redshift ranges where a signal is expected \citep{Montero-Camacho2018}. However, since we have performed our forecast for the $r_{\rm patch}$ with maximum $\langle L_{\rm Ly\alpha} \rangle$ SNR, the gains we may expect from a more sophisticated $W(r)$ will be limited. 

Moreover, we do not expect a different $W(r)$ to break the $\langle L_{\rm Ly\alpha} \rangle$-$b_{e}$ degeneracy displayed in \cref{fig:lya_diff_map_with_isolines}. The isolines of $\Delta \langle L_{\rm Ly\alpha} \rangle$ for different $r_{\rm patch}$ display a radial symmetry in the $\langle L_{\rm Ly\alpha} \rangle$-$b_{\rm e}$ space (which is the cause of the degeneracy). A different $W(r)$ would be a weighted average of $\Delta \langle L_{\rm Ly\alpha} \rangle$ using step kernels of different $r_{\rm patch}$, i.e., a linear combination that would not break the radial symmetry of isolines in \cref{fig:lya_diff_map_with_isolines}.

There are already examples in the literature of physically-motivated Ly$\alpha$ emission models for the cosmic web, computed by post-processing hydrodynamic simulations with radiative transfer \citep[e.g.,][]{Cantalupo2005, Goerdt2010, Behrens2019, Byrohl2022}. Less computationally expensive approaches, such as log-normal mocks, have also been explored \citep{Niemeyer2024}.
However, the box sizes of these simulations are generally too small for a study on the scales we sample ($>1$ Gpc in redshift direction, and $> 30$ cMpc in transverse direction for our patch size). Some of these simulations \citep[e.g.,][]{Cantalupo2005, Behrens2019} are of halo scales or slightly above, and even the largest ones (to the best of the author's knowledge) are not large enough for the proper study of LSS. For example, in \citep{Byrohl2022} the simulation box size is of 50 cMpc, as it is post-processed on the Illustris TNG50 simulation \citep{Nelson2019}. Moreover, not all possible sources of Ly$\alpha$ emission are always considered (e.g., in \citealp{Behrens2019} only stellar emission as an ionising source is considered, and in \citealp{Byrohl2022} ionisation from local stellar sources is not considered).

Regardless, the implementation of a radiative transfer model, instead of the simple Ly$\alpha$ emission model presented in this work, would significantly improve the accuracy of our forecast, and tighten the $\langle L_{\rm Ly\alpha} \rangle$ upper bounds. State-of-the-art hydrodynamical simulations with larger sizes than our simulation box already exist \citep[e.g, MillenniumTNG][with a box size of 500 cMpc/h]{Hernandez-Aguayo2023}. Running radiative transfer simulations with different sets of parameters (e.g., UV background, ionisation from QSOs and stellar emission, cooling mechanisms) may result in different values of $\langle L_{\rm Ly\alpha} \rangle$ and $b_{\rm e}$ for the Ly$\alpha$ emission field, which would allow the Ly$\alpha$ IM studies like the one presented here to directly constrain physical properties of the IGM and/or cosmological parameters.

Simulating the Ly$\alpha$ emission field for different radiative transfer models may prove computationally too expensive; however, recent developments in machine learning may dramatically reduce computational costs. Moreover, the exploration of the cosmological parameter space with Ly$\alpha$ IM amplifies the need for inexpensive Ly$\alpha$ emission simulations, as the Ly$\alpha$ emission field will have to be computed on several boxes with different cosmologies.

The application of deep-learning models to extrapolate the results of small-scale radiative-transfer codes to larger simulated volumes is a very active field of research. Baryon inpainting generates baryon distributions from much faster N-body simulations without the need for hydrodynamical computations \citep{Wadekar2020, Dai2023}, and even HI density fields can be generated with similar methodologies \citep{Bernardini2022}. Physical properties of the IGM gas (e.g., temperature, pressure, electron density) can also be simulated with neural networks \citep{Mohammad2022, Andrianomena2023}; a similar neural network could be trained to generate Ly$\alpha$ emission fields from hydrodynamical or N-body simulations for different sets of IGM physical parameters. Another possible approach would be the use of emulators to interpolate statistics such as the power spectrum between simulations with different parameters. This approach has been widely explored for the matter power spectrum \citep[e.g.,][]{Heitmann2013, Arico2021}, and tracers such as galaxies \citep{Kwan2015} or the Ly$\alpha$ forest \citep{Bird2023}.

In fact, if next-generation spectroscopic and imaging surveys allow for Ly$\alpha$ IM studies with higher SNR than our expectations for DESI/DESIx2 (i.e., enough SNR to constrain the angular power spectrum in several bins of distance), a very clear synergy between simulations and observations may emerge. More sophisticated Ly$\alpha$ emission models than those currently available will give more constraining power to Ly$\alpha$ IM observations, and observations themselves will help rule out IGM physical models or place cosmological constraints. Moreover, resolved sources could be selectively removed from Ly$\alpha$ intensity maps before coadding (i.e., by applying colour cuts to remove QSOs, or general magnitude cuts), to better understand how diffuse emission versus different kinds of resolved emitters trace the cosmic web.

\subsection{Expectations for upcoming surveys}\label{sec:Expectations for upcoming surveys}

The Ly$\alpha$ IM methodology presented in this work requires both imaging data to generate the Ly$\alpha$ emission intensity maps, and spectroscopic data to extract the Ly$\alpha$ forest and cross-correlate it with the patches integrated around QSOs. Hence, future Ly$\alpha$ IM studies beyond DESI and its Legacy Imaging Surveys are contingent on the development of both imaging surveys in the optical spectrum, and spectroscopic surveys targeting QSOs at redshifts high enough to observe the Ly$\alpha$ forest. Moreover, if the SNR provided by future surveys is high enough to have significant detections of $\Delta \langle L_{\rm Ly\alpha} \rangle$ in several $r_{\rm patch}$, the 2PCF could be used instead of our estimator to cross-correlate Ly$\alpha$ emission with Ly$\alpha$ forest absorption. A more "conventional" IM study with the 2PCF would constrain the shape of the power spectrum in Ly$\alpha$ emission, which would have a much more direct link to the existing LSS theory \citep[e.g.,][]{Peebles1970, Bernardeau2002} than our custom estimator.

Two types of upcoming imaging surveys hold great promise for not only detecting Ly$\alpha$ IM, but even constraining the Ly$\alpha$ power spectrum: space telescopes and ground-based time-domain surveys.

\subsubsection{Broad-band imaging surveys}\label{sec:Broad-band imaging surveys}

As we have discussed and shown in \cref{fig:bass_im_lya_img}, the dominant pattern in our BASS intensity map, and thus our main source of noise, is from instrumental and atmospheric origin. Observations from space telescopes are free of any undesired emission of atmospheric origin, so an optical intensity map constructed from such images should have the chequered pattern displayed in \cref{fig:bass_im_lya_img} significantly mitigated, even with the exact same image reduction procedure carried out in this paper. The already-launched Euclid mission \citep{Scaramella2022} will observe over 15,000 deg$^s$ of the sky significantly overlapping with the DESI footprint \citep[over 9,000 deg$^2$,][]{Naidoo2023}. Its filter system consists of a very broad optical band $I_E$ (between 5500-9000 \AA), expected to reach a depth of $I_E = 26.2$, as well as three near-infrared bands up to similar depths. The Euclid $I_E$ band observes Ly$\alpha$ in a redshift range higher than most of DESI Ly$\alpha$ forest data ($3.5 \leq z_{\rm Ly\alpha} \leq 6.4$ versus $2.1 \leq z_{\rm Ly\alpha} \leq 4$ for the great majority of DESI Ly$\alpha$ QSOs), so deeper, next-generation spectroscopic surveys may be needed to cross-correlate with Euclid images \citep[e.g.,][]{Schlegel2022}.

Another space telescope projected to launch in the next years that holds even more promise for Ly$\alpha$ IM is the China Space Station Telescope \citep[CSST,][]{Miao2022, Liu2023}. Planned to carry out a imaging survey covering at least the entire Euclid footprint, CSST is designed to observe at bluer wavelength, using standard NUV+$ugrizy$ filters, with a approximate 5$\sigma$ magnitude limit of 25.5 for point sources \citep{Liu2023}. Reaching almost two magnitudes deeper than DECaLS/BASS, and free from any undesired atmospheric emission, Ly$\alpha$ IM with $g$-band CSST images may not only yield a very significant detection of our estimator $\Delta \langle L_{\rm Ly\alpha} \rangle$, but even allow to constrain the angular power spectrum in Ly$\alpha$ in several angular bins.

Regarding ground-based imaging surveys, the largest planned for the upcoming years is the Vera C. Rubin Observatory \citep[Rubin, formerly LSST,][]{Ivezic2019}, planned to observe at least 18,000 deg$^2$ in the $ugrizy$ filter system, and reaching a $g$-band 5$\sigma$ point-source magnitude depth of 25.6 and 26.84 in the first and tenth year of operations, respectively \citep{LSSTCollaboration2018}. Rubin is designed to be a time-domain survey, observing variable and transient phenomena (such as supernovae or AGN) over an unprecedented volume. Consequently, for each survey pointing, approximately 1,000 exposures are planned along its 10-year programme \citep{LSSTCollaboration2018}. For the interest of IM studies, averaging out $\sim$1,000 exposures instead of the 3 exposures per pointing taken by DECaLS/BASS \citep{Dey2019} will greatly mitigate any exposure-dependent systematic in the $g$-band intensity map, be it instrumental or atmospheric. Although Rubin will be based on the southern hemisphere, at least 3,200 deg$^2$ of overlap are expected with DESI \citep{Bolton2018}. In \cref{sec:Forecast extrapolation to DESI-LSST}, we perform a simple extrapolation of our forecast to this minimal DESI-LSST overlap and show that, just after a year of operations of LSST, a detection of the cosmic web in Ly$\alpha$ with our estimator seems extremely likely. Given this promising preliminary forecast, we expect future work to investigate the insight on IGM physical properties and cosmological parameters that can gained from Ly$\alpha$ IM with LSST.

\subsubsection{Spectroscopic surveys}\label{sec:Spectroscopic surveys}

The deeper and cleaner images from upcoming imaging surveys will be the main observational drive moving Ly$\alpha$ IM forward, but we can also briefly comment on upcoming spectroscopic surveys. While very preliminary plans have been laid out for a Stage-V successor to DESI \citep[MegaMapper,][]{Schlegel2022}, and telescopes with similar characteristics and science goals are under construction \citep[MUltiplexed Survey Telescope, MUST,][]{Zhang2024}, the improvement they may bring over DESI for Ly$\alpha$ IM is less certain. First, because their target selection is entirely contingent on available imaging data, and second, because only the targets that provide Ly$\alpha$ forest data can be used for Ly$\alpha$ IM. Regardless, we can state that increases in Ly$\alpha$ forest sightline densities can be expected from future spectroscopic surveys, due to both increased completeness and depth. An increase in the average redshift sampled by their Ly$\alpha$ forest and a shift towards higher redshifts of the redshift kernel for Ly$\alpha$ IM studies (\cref{fig:qso_vs_z_g_band}) is also to be expected, due to increased target selection depth.

\subsubsection{Narrow-band imaging surveys}\label{sec:Narrow-band imaging surveys}

Finally, we also have to mention multi-narrow-band photometric surveys, with filters of FWHM $\sim$ 100 \AA, instead of the typical FWHM $\gtrsim$ 1000 \AA , of standard broad-band filters. The increase of spectral resolution provided by these narrow-band surveys results in integration of the Ly$\alpha$ line in fine redshift bins of $\Delta z \sim 0.1$ instead of $\Delta z \sim 1$. This finer binning results in a much stronger correlation of the underlying Ly$\alpha$ emission field than broad-band surveys, given that the smoothing in the redshift direction is smaller than the scale of homogeneity \citep[just $\sim$50 cMpc/h,][]{Renard2021}. While this would make narrow-band imaging ideal for Ly$\alpha$ IM, narrow-band surveys are significantly scarcer, and most of them present important caveats. For example, some are devoted to studying specific emission lines in fields of few deg$^2$ \citep[e.g., SILVERRUSH,][]{Ouchi2018}, while wider ones reach relatively shallow depths in a limited set of filters \citep[e.g., $r<21$ in J-PLUS and S-PLUS][]{Cenarro2019, DeOliveira2019}. Narrow-band imaging surveys with depths similar to DECaLS/BASS ($r$ or $i$ magnitude up to 23) covering a significant fraction of the optical spectrum are generally not wide enough for Ly$\alpha$ IM, e.g., PAUS \citep{Benitez2009} covers $\sim$ 50 deg$^2$ \citep{Navarro-Girones2023}, and \citet{Renard2021} already showed it was insufficient for Ly$\alpha$ IM.

However, the recently started J-PAS survey \citep{Benitez2014} will be deep \citep{Bonoli2021} and wide enough for Ly$\alpha$ IM: with a projected total footprint of 8,000 deg$^2$, reaching an approximate magnitude limit of $i = 23$. Approximately $\sim$6,000 deg$^2$ are expected to overlap with the spectroscopic survey WEAVE \citep{Jin2024}, reaching almost 100\% completeness in its Ly$\alpha$ QSO observation program \citep{Pieri2016}. Cross-correlation of J-PAS narrow-band intensity maps with WEAVE Ly$\alpha$ forest data may yield a detection of the transversal Ly$\alpha$ power spectrum in several redshift bins  (Renard et al., in prep.)

\section{Conclusions}\label{sec:Conclusions}

In this work, we propose a novel approach for IM using the Ly$\alpha$ line, involving the cross-correlation of broad-band imaging data with Ly$\alpha$ forest observations. While this method is applicable to various imaging-spectroscopic surveys, we focus on a detailed forecast for DESI and its associated Legacy Imaging Surveys, DECaLS and BASS. In our analysis, we have especially focused on how image reduction and coadding of the intensity maps should be carried out to minimise noise while preserving most of the Ly$\alpha$ emission signal. Furthermore, we introduce a simple analytical model of Ly$\alpha$ emission, enabling us to explore a wide range of key parameters characterising Ly$\alpha$ emission as a tracer of LSS, including its average emission per volume unit ($\langle L_{\rm Ly\alpha} \rangle$) and its linear bias ($b_{\rm e}$) relative to the dark matter distribution.

The results of our forecast show that, while a detection of LSS in Ly$\alpha$ emission is not certain, competitive upper bounds on the average Ly$\alpha$ emission can still be placed for DESI-DECaLS/BASS and for a hypothetical DESI extension with twice the number of Ly$\alpha$ QSOs (DESIx2). For DESI, this upper bound on $\langle L_{\rm Ly\alpha} \rangle$ is $23\pm10\cdot 10^{\rm 40}$ erg/s/cMpc$^3$ ($\sim$ 50\% higher than the most optimistic literature estimate of $15\cdot 10^{\rm 40}$ erg/s/cMpc$^3$from integration of the Ly$\alpha$ LF in \citealp{Drake2017}). In contrast, for DESIx2, the upper bounds are lowered to $15.5\pm7.0\cdot 10^{\rm 40}$ erg/s/cMpc$^3$ (very similar to the most optimistic literature estimates). Moreover, a simple extrapolation of our forecast to the overlap between DESI and LSST (\cref{sec:Forecast extrapolation to DESI-LSST}) points to an almost certain detection of Ly$\alpha$ LSS, even after just a year of LSST observations.

Consequently, in this work we provide one of the very first practical examples and guidelines for performing IM with the Ly$\alpha$ line using optical imaging. We prove that currently ongoing surveys will provide, at the very least, valuable constraints in the absence of a detection. Moreover, our discussion of upcoming surveys and extrapolation to LSST points at next generation surveys ushering the beginning of Ly$\alpha$ IM as a fully-fledged cosmological/astrophysical probe, that would help disentangling how baryons distribute with respect of the underlying dark matter large-scale structure, and how the IGM physical properties determine the scattering of Ly$\alpha$ emission across the cosmic web. Its redshift range ($2.2 < z < 3.4$ for the $g$-band) is higher than typical galaxy surveys, providing a complementary cosmological probe to the Ly$\alpha$ forest that may help constrain dark matter candidates, probe the matter power spectrum, or even study the impacts of reionisation at later redshifts \citep[e.g.,][]{Long2023}.

\section*{Acknowledgements}

PR and DS acknowledge the support by the Tsinghua Shui Mu Scholarship. PR, DS and ZC acknowledge funding of the National Key R\&D Program of China (grant no. 2023YFA1605600), the science research grants from the China Manned Space Project with No. CMS-CSST2021-A05, and the Tsinghua University Initiative Scientific Research Program (No. 20223080023). PR acknowledges additional funding from the National Science Foundation of China (grant no. 12350410365). HZ acknowledges the supports from National Key R\&D Program of China (grant Nos. 2023YFA1607800, 2023YFA1607804, 2022YFA1602902) and the National Natural Science Foundation of China (NSFC; grant Nos. 12120101003, and 12373010). PMC was supported by the Major Key Project of PCL. We acknowledge the fruitful scientific discussions with Rupert A.C. Croft, Enrique Gaztanaga and Richard Grumitt. We would also like to thank the referee for their insightful comments.

%%%%%%%%%%%%%%%%%%%%%%%%%%%%%%%%%%%%%%%%%%%%%%%%%%
\section*{Data Availability}

The data and code underlying this article will be shared on reasonable request to the corresponding author.



\bibliographystyle{mnras}
\bibliography{main} % if your bibtex file is called example.bib

%\begin{thebibliography}{99}
%\end{thebibliography}

%%%%%%%%%%%%%%%%%%%%%%%%%%%%%%%%%%%%%%%%%%%%%%%%%%

%%%%%%%%%%%%%%%%% APPENDICES %%%%%%%%%%%%%%%%%%%%%

\appendix

\section{Validation of forest smoothing assumptions}\label{sec:Validation of forest smoothing assumptions}

The probability of a given QSO sightline having a Ly$\alpha$ forest absorption field convolved with the whole $g$ band $\delta^{\rm g\, full}_{\rm F}$ for a given observed Ly$\alpha$ forest $\delta^{\rm g\, obs}_{\rm F}$ is a key component of our correlation estimator $\Delta \langle L_{\rm Ly\alpha} \rangle$ (\cref{sec:Correlation estimator: Lya average fluctuation}). The determination of this probability $P(\delta^{\rm g\, full}_{\rm F} | \delta^{\rm g\, obs}_{\rm F})$ in a model-independent manner, as laid out in \cref{sec:Spectroscopic data: Convolved forest probabilities}, relies on a key assumption: we can smooth the forest data on scales above the scale of homogeneity $\chi_{\rm h}$, without altering the statistical distribution of $\delta^{\rm g\, full}_{\rm F}$.

\begin{figure}
 	\includegraphics[width=\columnwidth]{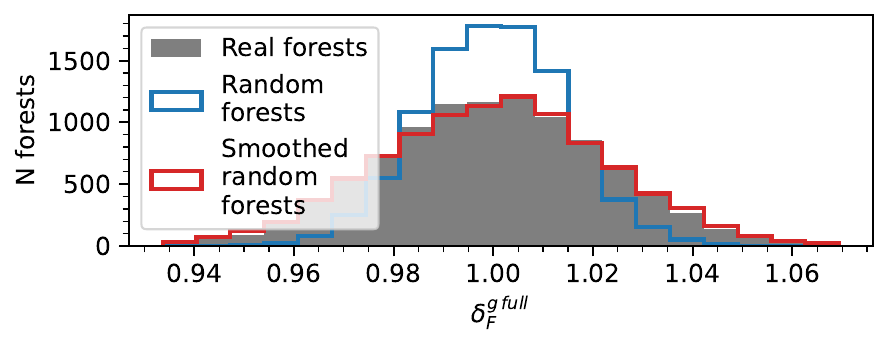}
     \caption{Distribution of $\delta_{\rm F}^{\rm g\, full}$ for 10,000 simulated forests. In grey, forest data as extracted from our simulation. In blue, randomly generated forests by resampling forest pixels, without smoothing (so the correlation at scales $\chi < \chi_{\rm h}$ is broken). In red, randomly generated smoothed forests, resampling forests pixels smoothed to $\Delta \chi = \chi_{\rm h}$.}
     \label{fig:forests_histogram_comparison}
\end{figure}

\Cref{fig:forests_histogram_comparison} shows the distribution of $\delta^{\rm g\, full}_{\rm F}$ values for our simulation (grey), as well as for resampled pixels of our simulation, without smoothing (blue), and resampled pixels, with smoothing at a homogeneity scale $\chi_{\rm h} = 150$ cMpc (red). The resampled sample without smoothing is significantly narrower than the original distribution of $\delta^{\rm g\, full}_{\rm F}$, while the resampled sample with smoothing is almost identical to the original. For a more quantitative assessment of this claim, \cref{tab:smoothed_forest_percentiles} shows the values of $\delta_{\rm F}^{\rm g\, full}$ for the percentiles corresponding to the 1$\sigma$ and 2$\sigma$ intervals. The percentiles for the random smoothed forests agree with the real forest distribution one order of magnitude better than the random forests without smoothing.

\begin{table}
\centering
	\caption{1$\sigma$ (16th-84th) and 2$\sigma$ (2.5th-97.5th) percentiles for the distributions of real, random and random smoothed forests in \cref{fig:forests_histogram_comparison}.}
 	\label{tab:smoothed_forest_percentiles}
 	\begin{tabular}{rcccc}
 		\hline
 		Percentile &  2.5th & 16th & 84th & 97.5th\\
 		\hline
 		Real&  0.9565 & 0.9780 & 1.0225 & 1.0439 \\
        Random&  0.9717 & 0.9858 & 1.0144 & 1.0283 \\
		Random smoothed & 0.9538 & 0.9770 & 1.0233 & 1.0458 \\
        \hline
 	\end{tabular}
 \end{table}

When resampling at the scales of our simulation pixels (2.23 cMpc), any correlation is broken, which results in a Ly$\alpha$ absorption field without over/underdensities at scales significantly larger than 2.23 cMpc, but a random succession of equally likely pixel values. The result is a distribution of $\delta^{\rm g\, full}_{\rm F}$ with significantly less variability than the original simulation. On the other hand, if the simulation is smoothed in redshift direction to $\Delta\chi = \chi_{\rm h} = 150$ cMpc, most of the correlation (large over/underdensities) is preserved inside these smoothed pixels. Hence, when we resample from these smoothed pixels to generate new Ly$\alpha$ forests, these pixels display a much larger variability per length unit than the 2.23 cMpc pixels, to the point where the distribution of $\delta^{\rm g\, full}_{\rm F}$ closely resembles the original. Therefore, from \cref{fig:forests_histogram_comparison} we can safely rely on the assumption that smoothing the Ly$\alpha$ forest data to  $\Delta\chi = 150$ cMpc and resampling generates new forests that are statistically identical after applying the $g$ band smoothing, and thus valid for our study.

\begin{figure}
    \includegraphics[width=\columnwidth]{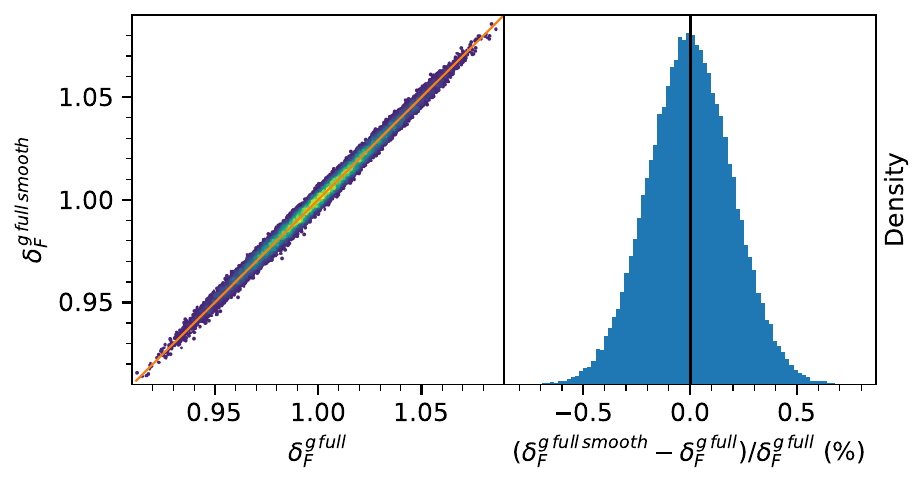}
     \caption{\textit{Left panel:} Scatter plot of $\delta_{\rm F}^{\rm g\, full}$ versus $\delta_{\rm F}^{\rm g\, full\, smooth}$. \textit{Right panel:} Histogram of relative error of $\delta_{\rm F}^{\rm g\, full\, smooth}$ with respect to $\delta_{\rm F}^{\rm g\, full}$, in \%.}
     \label{fig:forest_scatter_error_hist}
\end{figure}

The $\delta^{\rm g\, full}_{\rm F}$ from the unaltered simulation forests and the $\delta^{\rm g\, full}_{\rm F}$ from the smoothed and resampled forests follow a virtually identical distribution. However, we also need to examine if the smoothing on scales $\Delta \chi = 150$ cMpc significantly affects $\delta^{\rm g\, full}_{\rm F}$ for the same forest: we do so in \cref{fig:forest_scatter_error_hist}. The left panel shows that the values $\delta^{\rm g\, full\, smooth}_{\rm F}$ from smoothed forests are unbiased with respect to the unsmoothed counterpart  $\delta^{\rm g\, full}_{\rm F}$, with a small scatter. The right panel shows that this scatter introduces a relative error well within $\pm0.5 \%$ in  $\delta^{\rm g\, full\, smooth}_{\rm F}$ with respect to $\delta^{\rm g\, full}_{\rm F}$. Consequently, the error introduced in $\delta^{\rm g\, full}_{\rm F}$ by smoothing the forests is negligible. (Note that in \cref{sec:Spectroscopic data: Convolved forest probabilities}  $\delta^{\rm g\, full}_{\rm F}$ refers to a smoothed forest, since smoothing is necessary before resampling. Only in the context of \cref{fig:forest_scatter_error_hist} we refer to $\delta^{\rm g\, full}_{\rm F}$ as the $g$-band convolution of an unsmoothed forest).

\section{Ly$\alpha$ emission model cutoff}\label{sec:Lya emission model cutoff}

As specified in \cref{sec:Lya emission model}, if $L_{\rm Ly\alpha} \leq L_{\rm Ly\alpha\, cut}$, our Ly$\alpha$ emission model follows the power law

\begin{equation}\label{eq:lya_emission_cosmo_annex}
    L_{\rm Ly\alpha}(\delta) = C_L \cdot (1+\delta)^{b_{\rm e}},
\end{equation}

which results in a cosmological field with a given bias $b_{\rm e}$ \cref{eq:lya_emission_cosmo}. $C_L$ in \cref{eq:lya_emission_cosmo_annex}  is a normalisation constant so the total average luminosity in our simulation matches the input value $\langle L_{\rm Ly\alpha} \rangle$, i.e.,

\begin{equation}\label{eq:lya_average_condition}
    \frac{1}{N_{\rm voxels} V_{\rm voxel}}\sum_{\rm i=0}^N L_{\rm Ly\alpha\, i} = \langle L_{\rm Ly\alpha} \rangle,
\end{equation}

where $N_{\rm voxels}$ and $V_{\rm voxel}$ are the number of voxels and the voxel volume in the simulation box respectively, and $L_{\rm Ly\alpha\, i}$ the Ly$\alpha$ luminosity of voxel $i$. However, for voxels with $L_{\rm Ly\alpha} > L_{\rm Ly\alpha\,cut}$, the Ly$\alpha$ luminosity is computed with another power law,

\begin{equation}\label{eq:lya_emission_bright}
    L_{\rm Ly\alpha}(r) = \frac{L_{\rm Ly\alpha\, max} - L_{\rm Ly\alpha\, cut}}{(\delta_{\rm max} - \delta_{\rm cut})^\eta}(\delta - \delta_{\rm cut})^\eta + L_{\rm Ly\alpha\, cut},
\end{equation}

which has been determined \textit{ad hoc} to ensure that the bright end of the Ly$\alpha$ luminosity distribution matches within $\sim$1 dex the LFs for bright QSOs presented in \citet{Spinoso2020}. In other words, we assume that the dominant component of Ly$\alpha$ emission for the densest voxels $(\delta > \delta_{\rm cut})$ comes from very bright, resolved QSOs, as well as their surrounding nebulae. Given that the physical volume of our cells is $\sim$613 pkpc at $\langle z \rangle=2.64$, and that the largest Ly$\alpha$ nebulae detected so far have similar sizes \citep[e.g., 442 pkpc in][]{Cai2017}, this is a perfectly reasonable assumption. The exponent $\eta$ and maximum voxel luminosity $L_{\rm Ly\alpha\, max}$ of \cref{eq:lya_emission_bright} are

\begin{gather}
    \eta =  1.55 - 0.3 b_{\rm e}\label{eq:lya_cutoff_eta}\\
    L_{\rm Ly\alpha\, max} = 10^4 \langle L_{\rm Ly\alpha} \rangle \cdot V_{\rm voxel} \label{eq:lya_cutoff_max},
\end{gather}

These parameters have been chosen to ensure that 1) the bright end of the $L_{\rm Ly\alpha}$ histogram is within $\sim$1 dex of \citet{Spinoso2020} for the range of $L_{\rm Ly\alpha}$ specified in \cref{sec:Lya emission model}, and 2) the model converges for the range of bias in \cref{sec:Lya emission model} ($1.5 \leq b_{\rm e}\leq 3.5$). Regarding the cutoff luminosity, $L_{\rm Ly\alpha\,cut}$, it is derived from the condition of continuity in the $L_{\rm Ly\alpha}$ distribution, i.e., forcing the derivatives of \cref{eq:lya_emission_cosmo_annex} and \ref{eq:lya_emission_bright} to be equal,

\begin{equation}\label{eq:lya_cutoff_lya_cut}
    L_{\rm Ly\alpha\, cut} = L_{\rm Ly\alpha\, max} - \frac{b_{\rm e}}{\eta} C_L \cdot (1 + \delta_{\rm cut})^{b_{\rm e}-1} \cdot (\delta_{\rm max} - \delta_{\rm cut})^{\rm \eta},
\end{equation}

and $\delta_{\rm cut}$ is derived from evaluating \cref{eq:lya_emission_cosmo_annex} at $L_{\rm Ly\alpha\, cut}$,

\begin{equation}\label{eq:delta_cut}
    \delta_{\rm cut} = \left(\frac{L_{\rm Ly\alpha\, cut}}{C_L}\right)^{1/b_{\rm e}} - 1.
\end{equation}

Consequently, our Ly$\alpha$ emission model has just two entry parameters: $\langle L_{\rm Ly\alpha} \rangle$ and $b_{\rm e}$. For a given $( \langle L_{\rm Ly\alpha} \rangle, b_{\rm e})$ pair, $C_L$ and $L_{\rm Ly\alpha\, cut}$ are solved numerically from \cref{eq:lya_average_condition} and \cref{eq:delta_cut}. A numerical solution is needed because \cref{eq:lya_average_condition} is evaluated over all simulations voxels, and not only over those voxels under $L_{\rm Ly\alpha\, cut}$ that follow \cref{eq:lya_emission_cosmo_annex}.

\begin{figure}
 	\includegraphics[width=\columnwidth]{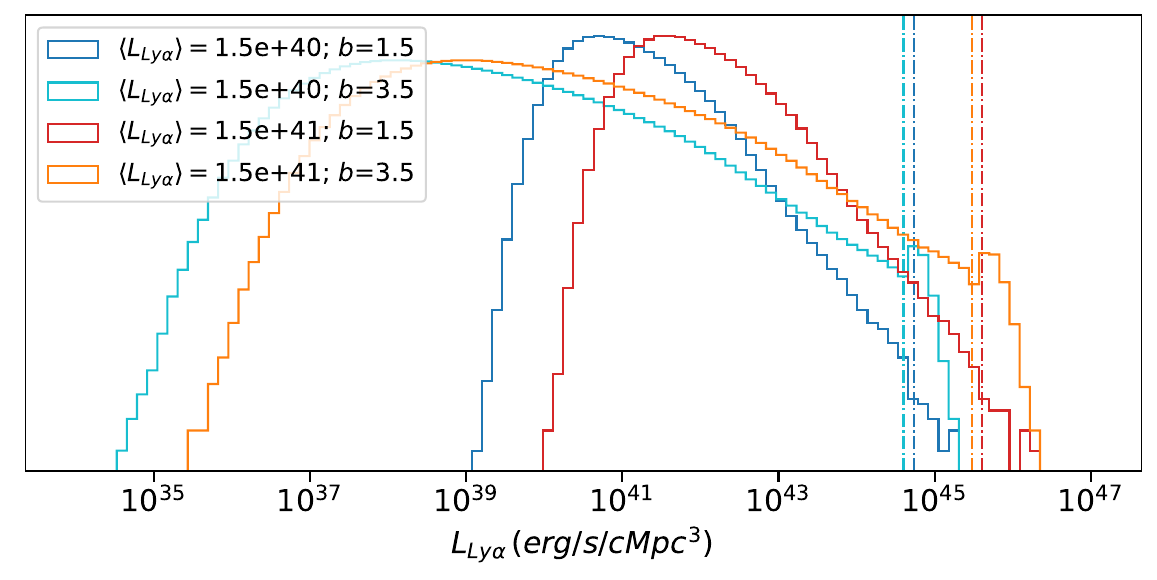}
 	\includegraphics[width=\columnwidth]{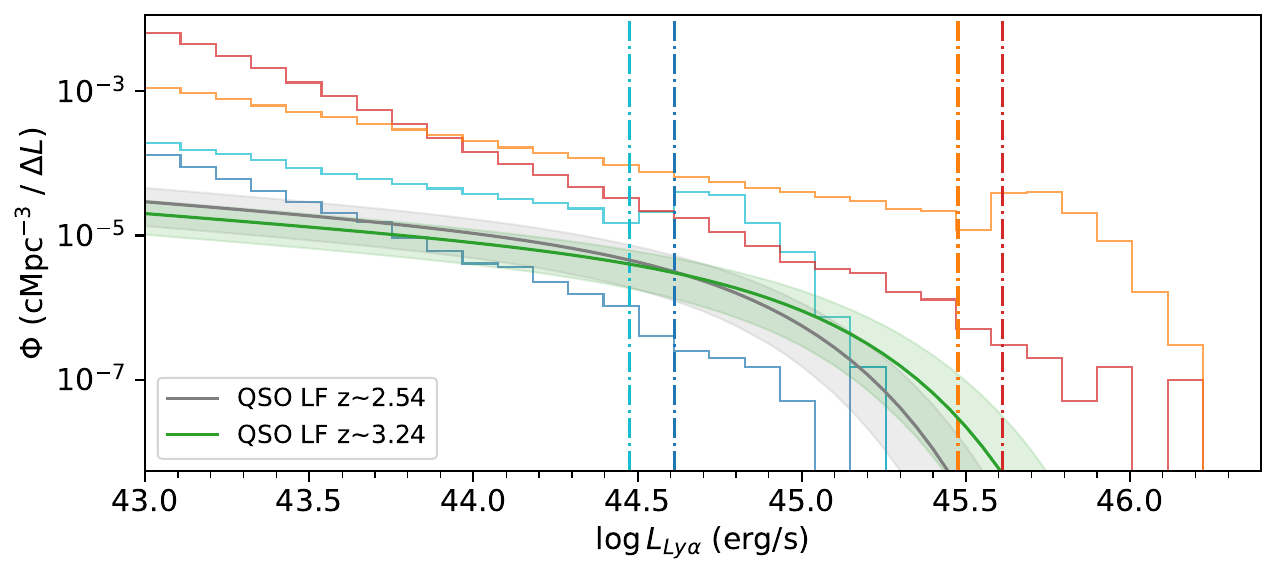}
     \caption{\textit{Upper panel:} Histogram of Ly$\alpha$ luminosity per volume unit of our simulation box, for the 4 possible combinations of maximum/minimum $\langle L_{\rm Ly\alpha} \rangle$ and $b_{\rm e}$. The dash-dotted lines are the values of $L_{\rm Ly\alpha\,cut}$ for each case. \textit{Lower panel:} Detail of the histogram for the bright end, plotted together with two of the QSO luminosity functions of \citet{Spinoso2020} (i.e. those compatible in redshift with our analysis). Units have been modified to match those of luminosity functions (number density per unit volume versus total luminosity per voxel).}
     \label{fig:lya_emission_hists}
\end{figure}

\Cref{fig:lya_emission_hists} displays histograms of our Ly$\alpha$ emission model applied to our simulation box, for the 4 possible maximum/minimum combinations of $\langle L_{\rm Ly\alpha} \rangle$, $b_{\rm e}$, together with a comparison with the QSO LFs of \citet{Spinoso2020} at the bright luminosity end, recomputed after removing all QSOs with $g < 19$ (as this is the foreground cut we apply in \cref{sec:Photometric noise: BASS intensity map and image reduction}). For this comparison, we have considered each voxel as a point source emitting in Ly$\alpha$, and compared their distributions to the QSO LFs. We find that, except for the faintest, less biased case ($\langle L_{\rm Ly\alpha} \rangle = 1.5\cdot10^{\rm 40}$ erg/s/cMpc$^3$, $b_{\rm e}=1.5$), and the brightest, most biased case ($\langle L_{\rm Ly\alpha} \rangle = 1.5\cdot10^{\rm 41}$ erg/s/cMpc$^3$, $b_{\rm e}=3.5$), the bright end of the histograms is at most $\sim$1 dex and $\sim$0.5 mag brighter than the LFs. This difference is consistent with the QSO LFs only considering the resolved Ly$\alpha$ QSO emission in a small aperture, but our simulation including all Ly$\alpha$ luminosity enclosed in a $\sim$560 pkpc$^3$ volume around a bright QSO \citep[e.g., in the extremely large Ly$\alpha$ nebula described in][, only 4\% of detected Ly$\alpha$ emission comes from a resolved source.]{Cai2017}. The extreme cases of $\langle L_{\rm Ly\alpha} \rangle$, $b_{\rm e}$ show a luminosity distribution which may be deemed unrealistic, as it deviates from the QSO LFs of \citet{Spinoso2020} by being approximately half an order of magnitude fainter or an order of magnitude brighter. However, since these represent the extremes of our parameter space, they are the most likely to be far away from the $\langle L_{\rm Ly\alpha} \rangle$, $b_{\rm e}$ values of the real Ly$\alpha$ emission field, so a larger deviation from the real QSO LFs is to be expected. 

One interesting feature of the model that can be observed in \cref{fig:lya_emission_hists} is that $\langle L_{\rm Ly\alpha} \rangle$ and $b_{\rm e}$ are independent entry parameters. Increasing/decreasing $\langle L_{\rm Ly\alpha} \rangle$ is equivalent to shifting right/left the histogram in \cref{fig:lya_emission_hists}, while increasing/decreasing $b_{\rm e}$ just modifies the shape of the histogram (making it become wider/narrower). $L_{\rm Ly\alpha\,max}$ is proportional to $\langle L_{\rm Ly\alpha} \rangle$ in \cref{eq:lya_cutoff_max}, and $\eta$ only depends on $b_{\rm e}$ in \cref{eq:lya_cutoff_eta}, this means that the convergence of the model only depends on $b_{\rm e}$. If the model has converged for a given value of $b_{\rm e}$ (i.e., $L_{\rm Ly\alpha}(\delta)$ is a continuous function for a given histogram shape), the histogram shape will be the same regardless of the value of $\langle L_{\rm Ly\alpha} \rangle$, as $L_{\rm Ly\alpha\,max}$ will shift accordingly.

The independent behaviour of $\langle L_{\rm Ly\alpha} \rangle$ and $b_{\rm e}$ has some useful consequences. First, even though we restrict our study to the values of  $\langle L_{\rm Ly\alpha} \rangle$ specified in \cref{sec:Lya emission model}, our model can simulate any value of  $\langle L_{\rm Ly\alpha} \rangle$. Likewise, the model can be solved (i.e., the values of $C_L$ and $L_{\rm Ly\alpha\, cut}$ can be found) for a single pair of ($\langle L_{\rm Ly\alpha} \rangle$, $b_{\rm e}$), and then simply renormalised to any other value of $\langle L_{\rm Ly\alpha} \rangle$. Moreover, the model can be extrapolated to different bias ranges, or even to other hydrodynamic simulations with similar voxel sizes, simply by slightly tuning \cref{eq:lya_cutoff_eta}.

\section{Forecast extrapolation to DESI-LSST}\label{sec:Forecast extrapolation to DESI-LSST}

Here, we present a simple extrapolation of our forecast for a hypothetical DESI-LSST Ly$\alpha$ IM study. The area for of the DESI-LSST overlap has been assumed to be 3,200 deg$^2$ \citep[the baseline in][]{Bolton2018}. By using the same QSO density distribution used throughout this work \citep[from ][]{DESICollaboration2016}, we expect a total of 132,384 QSOs in the overlap in the $g$-band redshift range. The same BASS $g$-band IM as in \cref{fig:bass_im_lya_img} has been used for imaging noise, but multiplying its flux by a factor $C_{\rm noise}$ determined by the magnitude limits as

\begin{equation}\label{eq:noise_factor_LSST}
    C_{\rm noise} = 10^{\rm (g_{\rm lim\, BASS} - g_{\rm lim\, LSST}) / 2.5},
\end{equation}

where $g_{\rm lim\, BASS}$ and $g_{\rm lim\, LSST}$ are the 5$\sigma$ PSF detection limits for the respective surveys. With this simple scaling, and assuming that all flux contained in our BASS intensity map is noise from non-cosmic origin, the SNR of any hypothetical source of magnitude $g_{\rm lim\, LSST}$ in LSST would be the same of a hypothetical source of magnitude $g_{\rm lim\, BASS}$ in BASS. We simulated two different stages of LSST with two different depth limits \citep{LSSTCollaboration2018}: LSST Y1, with $g_{\rm lim\, LSST} = 25.5$ ($C_{\rm noise} = 0.182$) and LSST Y10, with $g_{\rm lim\, LSST} = 26.9$ ($C_{\rm noise} = 0.050$).

By applying this scaling factor $C_{\rm noise}$, however, we are also artificially reducing the emission of cosmic foregrounds (i.e., the flux of all real galaxies in the intensity map is also downscaled). Nevertheless, in \citet{Renard2021} the cross-correlation between Ly$\alpha$ forest and narrow-band imaging ($i < 23$) could still clearly be detected for 100 deg$^2$ with cosmic foregrounds alone. Hence, it seems unlikely that cosmic foregrounds on their own could hide the cross-correlation between DESI Ly$\alpha$ forest data and LSST $g$-band images. Moreover, since we are just rescaling the flux of the BASS intensity map, we also neglect the mitigation of the sky and instrument variability noise components by the very high number of exposures of LSST, as discussed in \cref{sec:Expectations for upcoming surveys}.

\begin{figure}
 	\includegraphics[width=\columnwidth]{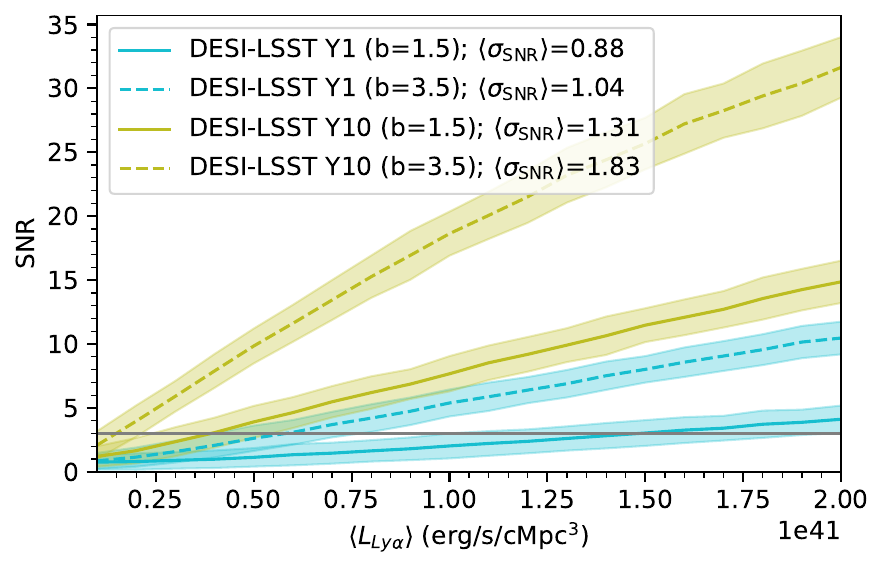}
     \caption{Mean SNR and the $1\sigma$ confidence interval (coloured area), versus $\langle L_{\rm Ly\alpha} \rangle$, for DESI-LSST Y1 (cyan) and DESI-LSST Y10 (olive). Solid lines represent the cases with the minimum bias of our model ($b_{\rm e}=1.5$), and dashed lines the maximum bias ($b_{\rm e}=3.5$).}
     \label{fig:DESI-LSST_SNR_vs_lya_lum}
\end{figure}

\begin{figure}
 	\includegraphics[width=\columnwidth]{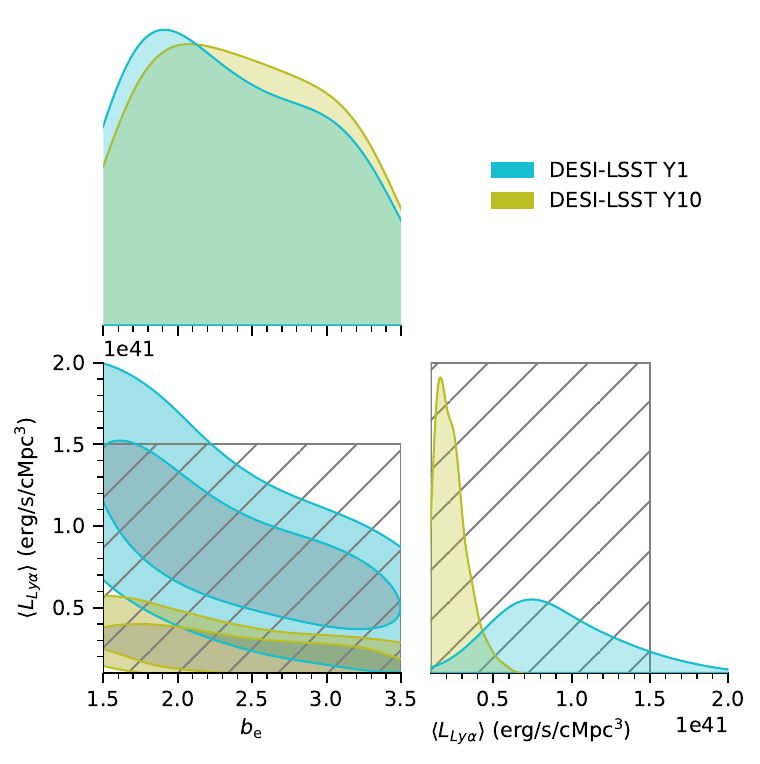}
     \caption{Triangle plot of the upper bounds on $\langle L_{\rm Ly\alpha} \rangle$ and $b_{\rm e}$ placed by a non-detection ($\Delta \langle L_{\rm Ly\alpha} \rangle$ SNR < 3), for DESI-LSST Y1 (cyan) and DESI-LSST Y10 (olive). Contours on the bottom left panel correspond to $1\sigma$ and $2\sigma$ constraints, hatched areas to the range of $\langle L_{\rm Ly\alpha} \rangle$ we deem realistic.}
     \label{fig:DESI-LSST_triangle_plot}
\end{figure}

In an analogous manner to the results presented in  \cref{sec:Results for DESI/DESIx2}, \cref{fig:DESI-LSST_SNR_vs_lya_lum} displays the evolution of $\Delta \langle L_{\rm Ly\alpha} \rangle$ SNR versus $\langle L_{\rm Ly\alpha} \rangle$ for $b_{\rm e} = 1.5$ and $b_{\rm e} = 3.5$, for both DESI-LSST Y1 and DESI-LSST Y10. \Cref{fig:DESI-LSST_triangle_plot} shows the triangle plot of the upper bounds on $\langle L_{\rm Ly\alpha} \rangle$ and $b_{\rm e}$, and \cref{tab:lya_upper_bounds_LSST} the marginalised upper bounds, together with the probability of detection. However, for this case we have only sampled $\langle L_{\rm Ly\alpha} \rangle$ up to $2\cdot 10^{\rm 41}$ erg/s/cMpc$^3$, since at this level the $\Delta \langle L_{\rm Ly\alpha} \rangle$ SNR is already so high that a detection seems certain ($\Delta \langle L_{\rm Ly\alpha} \rangle$ SNR = 4$\pm$1 for DESI-LSST Y1 and $b_{\rm e} = 1.5$).

\begin{table}
\centering
	\caption{Marginalised upper bounds for $\langle L_{\rm Ly\alpha} \rangle$ and $b_{\rm e}$ for a non-detection in our forecast extrapolated to DESI-LSST, together with probability of detecting Ly$\alpha$ LSS (SNR$_{\rm \Delta \langle L_{\rm Ly\alpha} \rangle}> 3$).}
 	\label{tab:lya_upper_bounds_LSST}
 	\begin{tabular}{rccc}
 		\hline
 		 &  \makecell{$\langle L_{\rm Ly\alpha} \rangle$ \\ ($10^{\rm 40}$ erg/s/cMpc$^3$)} &$b_{\rm e}$&$P_{\rm \rm detect}$ (\%)\\
 		\hline
 		DESI-LSST Y1& $8.8 \pm 3.6$&$2.38 \pm 0.56$&93.30\\
        DESI-LSST Y10& $2.8\pm 1.0$&$2.36 \pm 0.55$&100.00\\
		\hline
 	\end{tabular}
 \end{table}

%%%%%%%%%%%%%%%%%%%%%%%%%%%%%%%%%%%%%%%%%%%%%%%%%%

% Don't change these lines
\bsp	% typesetting comment
\label{lastpage}
\end{document}